\documentclass[
 reprint,
superscriptaddress,
nofootinbib,
bibnotes,
 amsmath,amssymb,
 aps,
prb,
]{revtex4-2}
\usepackage[plainpages = false, pdfpagelabels, 
                 bookmarks,
                 bookmarksopen = true,
                 bookmarksnumbered = true,
                 breaklinks = true,
                 linktocpage,
                 colorlinks = true,
                 linkcolor = blue,
                 urlcolor  = blue,
                 citecolor = blue,
                 anchorcolor = green,
                 hyperindex = true,
                 hyperfigures
                 ]{hyperref} 
\usepackage{enumerate}
\usepackage{makecell, multirow}
\usepackage{tikz}
\usetikzlibrary{patterns}
\usepackage{slashed}
\usepackage{braket}
\usepackage{bbold}
\usepackage{bm}
\usepackage{booktabs}
\usepackage[caption=false,singlelinecheck=off]{subfig}
\usepackage{blindtext}
\usepackage{bbm}
\usepackage{graphicx}
\usepackage{dcolumn}
\usepackage{bm}
\usepackage[mathlines]{lineno}
\definecolor{ballblue}{rgb}{0.13, 0.67, 0.8}

\begin{document}
\title{Does delta-\textit{T} noise probe quantum statistics?}
\author{Gu Zhang}
\affiliation{Beijing  Academy  of  Quantum  Information  Sciences, 100193 Beijing, China}
\affiliation{\mbox{Institute for Quantum Materials and Technologies, Karlsruhe Institute of Technology, 76021 Karlsruhe, Germany}}
\affiliation{\mbox{Institut f\"{u}r Theorie der Kondensierten Materie, Karlsruhe Institute of Technology, 76128 Karlsruhe, Germany}}
\author{Igor V. Gornyi}
\affiliation{\mbox{Institute for Quantum Materials and Technologies, Karlsruhe Institute of Technology, 76021 Karlsruhe, Germany}}
\affiliation{\mbox{Institut f\"{u}r Theorie der Kondensierten Materie, Karlsruhe Institute of Technology, 76128 Karlsruhe, Germany}}
\affiliation{Ioffe Institute, 194021 St.~Petersburg, Russia}
\author{Christian Sp\r{a}nsl\"{a}tt}
\affiliation{\mbox{Department of Microtechnology and Nanoscience (MC2), Chalmers University of Technology, S-412 96 G\"oteborg, Sweden}}
\affiliation{\mbox{Institute for Quantum Materials and Technologies, Karlsruhe Institute of Technology, 76021 Karlsruhe, Germany}}
\affiliation{\mbox{Institut f\"{u}r Theorie der Kondensierten Materie, Karlsruhe Institute of Technology, 76128 Karlsruhe, Germany}}
  
\begin{abstract}
We study delta-$T$ noise---excess charge noise at zero voltage but finite temperature bias---for weak tunneling in one-dimensional interacting systems. We show that the sign of the delta-$T$ noise is generically determined by the nature of the dominating tunneling process. More specifically, the sign is governed by the leading charge-tunneling operator's scaling dimension. We clarify the relation between the sign of delta-$T$ noise and the quantum exchange statistics of tunneling quasiparticles. We find that, for infinite systems hosting chiral channels with local interactions (e.g., quantum Hall or  quantum spin Hall edges), when the delta-$T$ noise is negative, the tunneling particles are boson-like, revealing their tendency towards bunching. However, the opposite is not true: Boson-like particles do not necessarily produce negative delta-$T$ noise. Importantly, the bosonic nature of particles generating the negative delta-$T$ is not necessarily intrinsic, but can be induced by the interactions. This, in particular, implies that negative delta-$T$ noise for tunneling between the edge states cannot serve as a smoking gun for detecting ``intrinsic anyons''. We also establish a connection between the delta-$T$ noise and the temperature derivative of the Nyquist-Johnson (thermal) noise in interacting systems, both governed by the same scaling dimensions.
As a demonstration of the above statements, we study tunneling between two interacting quantum spin Hall edges. With bosonization and renormalization-group techniques, we find that many-body interactions can generate negative delta-$T$ noise for both direct tunneling through a point contact and in Kondo exchange tunneling via a localized spin. In both these setups, we show that the noise can become negative at sufficiently low temperatures, when interactions renormalize the tunneling to favor boson-like pair-tunneling of electrons rather than single-electron tunneling. Our findings show that delta-$T$ noise can be used to probe the nature of collective excitations in interacting one-dimensional systems.
\end{abstract}
\date{\today}
\maketitle

\section{Introduction}
\subsection{Background and motivation}
Non-equilibrium noise in nanoscale electronic conductors refers traditionally to the partition noise in a current induced by a voltage bias. This  distinguishes it from the thermal equilibrium (Nyquist-Johnson) noise~\cite{BlanterButtikerPhysRep00}.
In recent decades, partition noise has successfully been used as a sensitive tool~\cite{Kobayashi2021} to probe collective properties in systems of strongly correlated electrons. Examples include shot-noise measurements of fractional charges and non-trivial scaling dimensions in the fractional quantum Hall (FQH)~\cite{WenPRB91,KaneFisherPRL94,FendleyLudwigSaleurPRL95,ChamonPRB95,ChamonPRB96,SaminadayerPRL97,PicciottoPhysB98,KyryloVadimPRB15} and Kondo~\cite{Gogolin1997,Sela2006,Mora2008,Zarchin2008,Mora2009,Sakano2011,Yamauchi2011,Ferrier2016} effects, observation of anyonic statistics via noise correlations~\cite{SafiPRL01,KanePRL03,VishveshwaraPRL03,KimPRL05,LawFeldmanGefenPRB06,GrosfeldPRL06,FeldmanPRB07,CampagnanoPRL12,HalperinPRL16,LeeHanSimPRL19,BartolomeiScience20,NakamuraX20,SafiPRB20}, and detecting neutral modes in complex FQH states~\cite{Bid2012,Gross2012,Inoue2014,Park2019,Spanslatt2019,Spanslatt2020QPC,Spanslatt2020,Dutta2022,Kumar2022}.

Accurate control of local heating and readout of temperatures have in recent years spurred studies of a novel type of non-equilibrium noise, the so-called delta-$T$ noise~\cite{Lumbroso2018}. By definition, this quantity is an excess charge noise, on top of the equilibrium noise, which is induced only by a temperature difference $\delta T\neq0$ at vanishing voltage bias, $\delta V=0$. In contrast to the voltage-induced noise, the delta-$T$ noise does not require a net flow of a non-equilibrium charge current. Since initial theoretical~\cite{Sukhorukov1999,Mu2019,Zhitlukhina2020,RechMartinPRL20,Hasegawa2021,duprezX2021,Popoff2021,Schiller2021} and experimental~\cite{TikhonovSciRep16,Lumbroso2018,Sivre2019,TikhonovPRB2020,Larocque2020,Rosenblatt2020,Melcer2022} work, it has been shown that non-equilibrium noise in the absence of currents is, in fact, a more general phenomenon than delta-$T$ noise~\cite{ErikssonX21}. Still, what properties of nanoscale conductors---and, in particular, facets of their strongly-correlated nature---this class of zero-current non-equilibrium noise might reveal, 
remains an open and interesting question.

In this context, a recent work~\cite{RechMartinPRL20} theoretically investigated delta-$T$ noise in the FQH effect. It was found that the tunneling of fractionally charged quasiparticles in a quantum point contact (QPC) comes with an associated \textit{negative} delta-$T$ noise. In other words, the non-equilibrium noise induced by the temperature difference in a correlated state turns out to be, quite counter-intuitively, smaller in magnitude than that at equilibrium. By contrast, when the tunneling is dominated by electrons, the delta-$T$ noise is positive. The same work argued that tunneling between two infinite non-chiral Luttinger liquids (LLs) was only capable of generating positive delta-$T$ noise. 

The emergence of a negative delta-$T$ noise is rather surprising, since one could expect that, at least for energy-independent tunneling~\cite{Lesovik1993}, the non-equilibrium noise should be additive to the equilibrium noise. Given these intriguing results for the strongly correlated nature of FQH tunneling, the authors in Ref.~\cite{RechMartinPRL20} further speculated that the emergence of negative delta-$T$ noise could be related to the anyonic statistics of the tunneling quasiparticles. It has remained, however, not clear if the negative delta-$T$ noise is a quantum statistical effect unique to anyons, or if many-body interactions alone suffice to produce such a signature. 

Moreover, the results of Ref.~\cite{RechMartinPRL20} hinted that the delta-$T$ noise depends explicitly on the scaling dimension $\Delta_{\mathcal{T}}$ of the leading charge-tunneling operator. Further, a very recent work, Ref.~\cite{Schiller2021}, specifically emphasized the role of scaling dimension in the delta-$T$ noise. In particular, the authors proposed to use delta-$T$ noise to extract scaling dimensions of particles tunneling between complex FQH edges. The sign change of the noise occurs precisely when $\Delta_{\mathcal{T}}=1/2$. It is thus natural to ask if the scaling dimension alone determines the sign of the delta-$T$ noise, 
and whether quantum statistics of the tunneling particles
plays a role here. 

In this Paper, we address the above questions by exploring the connection between delta-$T$ noise and quantum statistics in a more generic setting of strongly interacting one-dimensional (1D) systems. In this context, we define the quantum statistics of 1D excitations (we will use the words ``excitations'', ``quasiparticles'', or ``particles'' interchangeably in this work) as the phase $\Theta$ generated upon exchanging the order of the corresponding field operators. To be more precise, consider operators $\psi(x)$ and $\psi(x')$ describing excitations of the same kind at positions $x$ respectively $x'\neq x$ (both at the same time instant). The statistical phase $\Theta$ (mod $2\pi$) can then be expressed as
\begin{align}
\label{eq:CBH_relation_Intro}
    \psi(x)^{}\psi(x')^{} = e^{i\Theta}\psi(x')^{}\psi^{}(x).
\end{align}

For non-interacting particles, $\Theta$ takes discrete values. For $\Theta=0$, the order of the operators does not matter and the excitations are boson-like.  For $\Theta=\pi$, exchanging the order introduces a negative sign, which reflects the Pauli-principle of  fermions. For FQH Laughlin states, the hosted anyons carry fractional phases $\Theta = \pi/(2 n + 1)$, with $n$ being positive integers~\cite{WenPRB91}. The value $\Theta=\pi/3$ was recently measured experimentally in a Fabry-P\'{e}rot interferometer at FQH filling $\nu=1/3$~\cite{NakamuraX20}.
In these three situations, $\Theta$ is fixed by the free-particle statistics or a bulk topological order. However, in 1D interacting systems $\Theta$ may instead depend explicitly on the many-body interactions. Then, by changing the interaction strength, the statistics of the excitations continuously interpolates between being more boson-like ($|\Theta| < \pi/2$ modulo $2\pi$) or fermion-like ($|\Theta| > \pi/2$). This is the case, e.g., for tunneling between LLs, where interaction-influenced tunneling operators for electrons share similar long-time correlations as those for quasiparticles tunneling between FQH edges~\cite{StoneFisherJMPB94,FendleyLudwigSaleurPRB95}. It is therefore interesting to understand delta-$T$ noise also in this generic situation.

\subsection{Overview of models and results}
\label{sec:overview}

A central result in this Paper is that if two interacting 1D electron systems are connected at a single tunneling bridge,
the sign of the delta-$T$ noise is uniquely determined by the scaling dimension $\Delta_\mathcal{T}$ of the leading charge tunneling operator(s). More specifically, the delta-$T$ noise can only become negative if $\Delta_{\mathcal{T}} < 1/2$ (cf. Ref.~\cite{Schiller2021}). The determination of the sign of the noise by $\Delta_{\mathcal{T}}$ indicates that negative delta-$T$ noise is not a unique feature of anyonic systems (e.g., Laughlin FQH edges), but can be generated by interactions in many 1D electrons systems.

At the same time, we demonstrate that, if the two
infinite chiral systems with local interactions
are identical~\cite{IdenticalNote}, negative delta-$T$ noise implies that the statistical phase $\Theta$ [see Eq.~\eqref{eq:CBH_relation_Intro} above] of the tunneling particles must be boson-like: \begin{equation}
  |\Theta|\leq \pi/2.  
\end{equation}
The opposite does not hold, however, as boson-like particles may generate positive delta-$T$ noise. Thus, delta-$T$ noise, while not being uniquely determined by the quantum statistics, can still serve for exploring statistical properties of excitations in correlated systems.

We anticipate our conclusion for the relation of the delta-$T$ noise to the statistical phase to be valid for Hong-Ou-Mandel (HOM) geometries~\cite{HOM1987} (see  Refs.~\cite{idrisov2022,Taktak2022} for very recent examples of studies for HOM setups based on quantum-Hall edges), involving only local interactions, and other setups that are equivalent.
These are geometries relevant to experiments on quantum Hall and spin quantum Hall edges. However, the above relation does not apply to the setup of weak tunneling between two semi-infinite Luttinger liquid leads, which generates non-local interactions upon mapping to the HOM geometry~\cite{GiamarchiBook}. Nevertheless, a universal relation between delta-$T$ noise and scaling dimension $\Delta_\mathcal{T}$ of the dominant tunneling operator holds in this case as well.

As a concrete test of the relation between delta-$T$ noise, scaling dimensions, and quantum statistics in 1D, we study tunneling between interacting edges of a quantum spin Hall (QSH) insulator (see Fig.~\ref{fig:qsh_kondo}). A QSH edge may realize the so-called ideal (``$S_z$-conserving'') helical LL, in which electron spins are locked to their momentum, i.e., to their propagation direction (chirality).  Initially proposed for graphene~\cite{KaneMelePRL05,Kane2005b}, the QSH states were first realized in HgTe ~\cite{BernevigZhangPRL06,konig2007} and later in InAs/GaSb~\cite{Knez2011,Knez2014,Du2015,Li2015} quantum wells. Recently, a practical way of inducing QSH states has been suggested also for graphene, by exploiting heavy ad-atom deposition~\cite{WeeksPRX11,SantosPRB18,AvsarRMP20}. While previous works~\cite{Maciejko2009,DelMaestro2013,Tikhonov2015,Aseev2016,Vayrynen2017,Nagaev2018,KurilovichPRB19} analyzed ordinary shot noise due to magnetic impurities on the QSH edge, investigations of delta-$T$ noise in this system has so far been lacking.

Our choice of the QSH edge system is motivated by the possibility to study non-trivial tunneling behavior in a fairly simple, strongly correlated \cite{K-QSH} setup:

\begin{itemize}
\item[(i)] The electron tunneling between two interacting QSH edges, Fig.~\ref{fig:qsh_kondo}\textcolor{blue}{(a)}, depends continuously on the interaction strength. This allows us to study tunneling of particles with statistics that continuously interpolate between being more boson-like to being fermion-like. This is in contrast to FQH edges where the tunneling behaviour is fixed by the underlying topological order. 

\begin{figure}[t!]
  \centering
    \includegraphics[width=0.8\columnwidth]{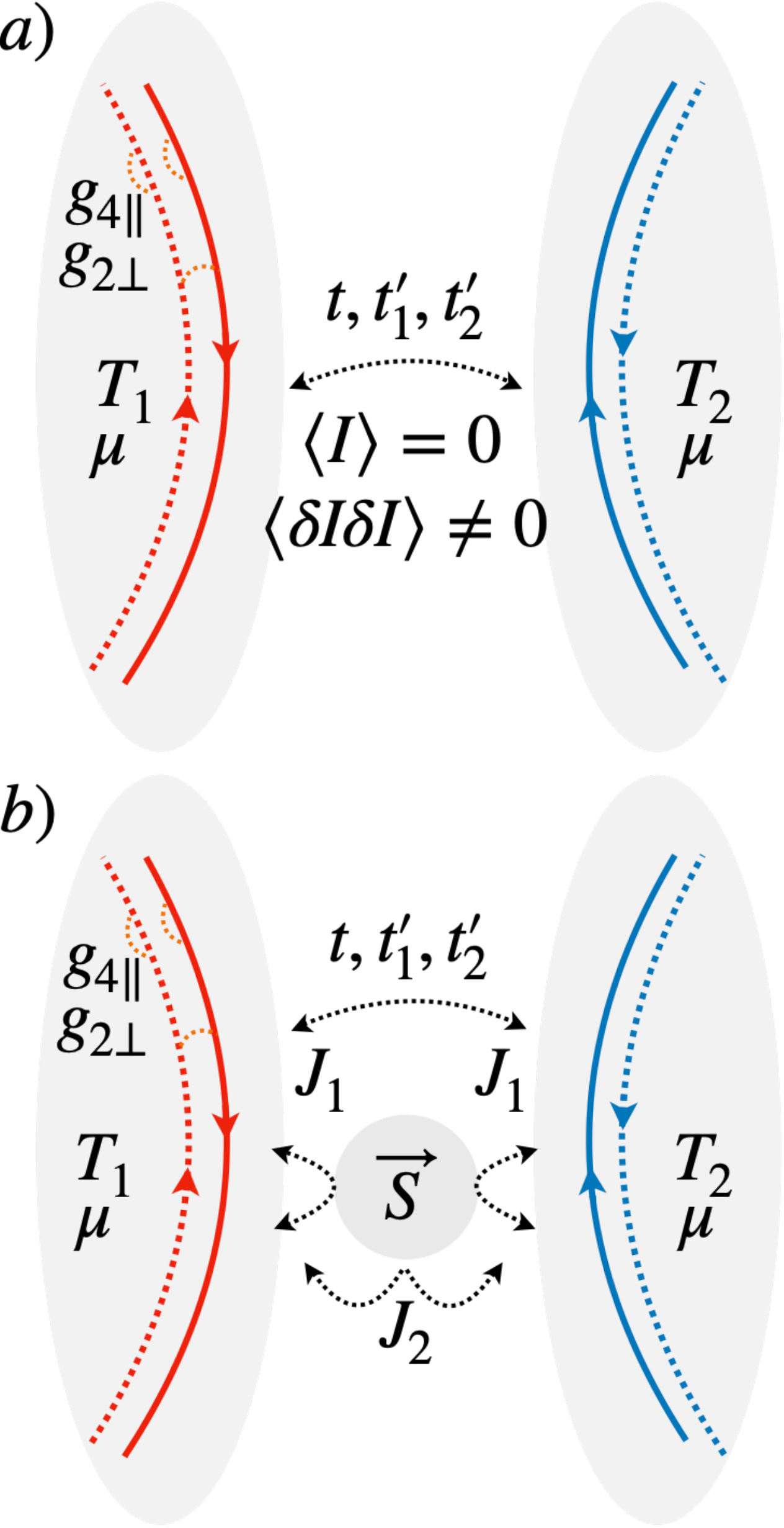}
    \caption{Two interacting QSH edges with clockwise (solid lines) and counter-clockwise (dashed lines) propagating edge modes. The modes have opposite spin projections because of spin-momentum locking. The couplings $g_{4\parallel}$ and $g_{2\perp}$ determine, respectively, forward (same chirality and, hence, same spin) and dispersive (opposite chiralities and spins) scattering amplitudes with a small momentum transfer. The edges are at the same chemical potential $\mu$ but at temperatures $T_1\neq T_2$ and are bridged by a single tunneling point. This results in a vanishing tunneling current $\langle I \rangle =0$, but finite excess charge noise $S\propto \langle \delta I \delta I \rangle $, called delta-$T$ noise. In panel (a), the particles transfer between the edges is described by the single-electron tunneling amplitude $t$, as well as by the coherent pair tunneling amplitudes $t_1'$ and $t'_2$ for spin preserving and spin-flipping tunneling, respectively. In panel (b), in addition to the above processes, tunneling is also possible via the Kondo exchange mediated by the spin $\vec{S}$, with backward (intra-edge) and forward (inter-edge) scattering amplitudes $J_1$ and $J_2$, respectively. }
    \label{fig:qsh_kondo}
\end{figure}

\item[(ii)] While point (i) is true also for the standard LL, spin-momentum locking on the ideal helical QSH edge allows several simplifications.
First, it reduces the number of degrees of freedom by a factor of two in comparison to the spinfull LL [see Fig.~\ref{fig:ll_structure}\textcolor{blue}{(a)}]. Second, it forbids the $g_{1\perp}$ (backscattering) interaction that normally prevents full diagonalization of the spinful LL system. Third, the spin-momentum locking also restricts the set of possible tunneling processes in a QSH system.

\begin{figure}[t!]
  \centering
    \includegraphics[width=1.0\columnwidth]{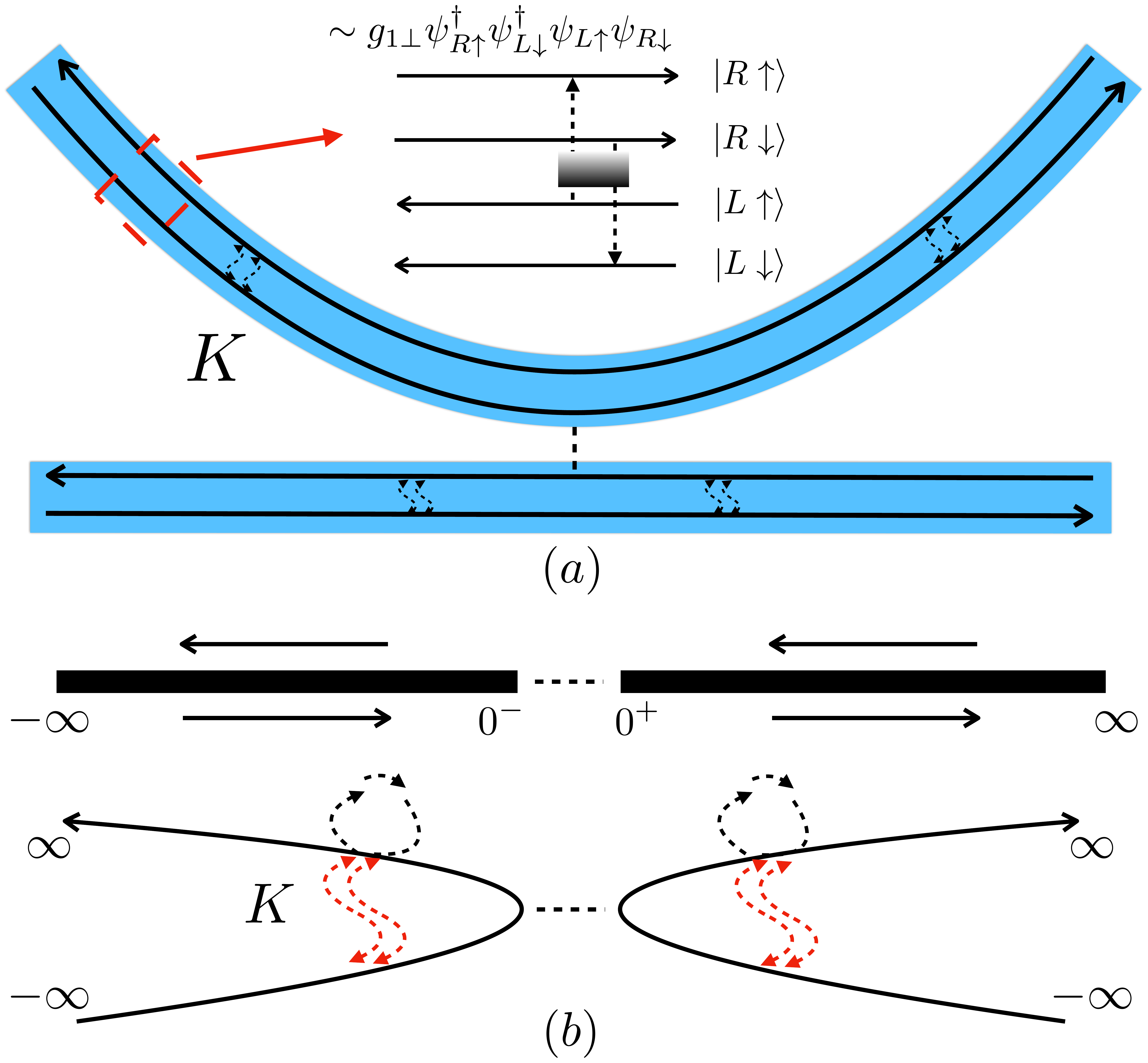}
    \caption{Two distinct setups for tunneling between two conventional Luttinger-liquid wires. The interaction strength is quantified by the Luttinger parameter $K$ (dashed arrows). (a) Tunneling between two adjacent wires. Each wire can be described by four independent bosonic fields. The degrees of freedom are doubled in comparison to Fig.~\ref{fig:qsh_kondo}. Inset: Zoomed-in view of the channel structure. The spinful LL model contains the $g_{1\perp}$ interaction consisting of two coherent sub-processes involving opposite spins (indicated by the two arrows connected by the shaded box). This interaction term can not be diagonalized with bosonization~\cite{GiamarchiBook}. On the ideal helical QSH edge, this term is forbidden by the time-reversal symmetry. (b) Tunneling across a weak link (between points $0^-$ and $0^+$) connecting two semi-infinite LL wires. Each semi-infinite wire (thick black lines) hosts particles of both chiralities, but can be transformed to host only a single chirality [see Eq.~\eqref{eq:chiral_fields}] by ``unfolding'' that extends the half-wire to be fully infinite (lower panel). However, in addition to local interactions (black, dashed lines) this unfolding transformation introduces non-local interactions (red, wiggly lines).}
    \label{fig:ll_structure}
\end{figure}

\item[(iii)] In contrast to the setup for tunneling between two semi-infinite LL wires [Fig.~\ref{fig:ll_structure}\textcolor{blue}{(b)}], the source and drain contacts attached to a QSH edge can be geometrically separated. This allows a wider range of experimentally accessible noise measurements. It is also true that tunneling between two semi-infinite LLs can map to infinite chiral channels with non-local interactions.
In these cases tunneling operators with scaling dimension $\Delta_\mathcal{T}<1/2$ does not necessarily indicate a boson-like tunneling.
\end{itemize}

Our study comprises an analysis of the delta-$T$ noise in two different setups of inter-edge QSH tunneling. In the first setup, displayed in Fig.~\ref{fig:qsh_kondo}\textcolor{blue}{(a)}, electrons tunnel only directly between the two edges. By the well-known mapping of this problem to the FQH system considered in Ref.~\cite{RechMartinPRL20}, one might expect that the direct tunneling should only be able to generate positive delta-$T$ noise. This is, however, the case when only single-electron tunneling is included. Here, we show that electron coherent pair tunneling, which for sufficiently strong interactions becomes the most relevant process, is sufficient to generate negative delta-$T$ noise. This result reveals that negative delta-$T$ noise in electronic systems can arise without any notion of anyon exchange statistics. However, we argue that one may, in fact, interpret the coherent pair tunneling as the tunneling of bosons, whose delta-$T$ noise, in contrast to fermions, is negative at low temperatures.

As our second setup, depicted in Fig.~\ref{fig:qsh_kondo}\textcolor{blue}{(b)}, we study a QSH-Kondo model, in which tunneling may also occur via a single spin, e.g., a localized electron or a quantum dot tuned to the Kondo valley (the regime where the dot is singly occupied). At sufficiently low temperatures, we show that both charge and spin fluctuations in the dot become frozen. Under these conditions, we find that negative delta-$T$ noise emerges only for attractive interactions. In comparison to the direct tunneling model, the negativity of the noise in the QSH-Kondo model originates from a combination of intra-edge and dot-edge interactions. 

In the QSH-Kondo model we find that the emergence of the negative noise is accompanied with the renormalization group (RG) flow of the system towards the single-channel Kondo fixed point. At this fixed point (see e.g., Ref.~\cite{SchillerHershfieldPRB98}), the leading RG relevant operators describe boson-like tunneling between the edges. This behavior can be contrasted with the situation of repulsive interactions, where the system instead flows towards the two-channel Kondo fixed point~\cite{LawPRB10}, inducing a positive delta-$T$ noise. This follows from the fact that the leading relevant operators close to this fixed point describe fermion-like tunneling. The intimate connection between the quantum criticality and the sign of the delta-$T$ noise, suggests that delta-$T$ noise probes the quantum phase transition of the QSH-Kondo model.

\subsection{Outline}
The remainder of this Paper is organized as follows.
We start in Sec.~\ref{sec:Free_Particles} by studying delta-$T$ noise of free fermions and bosons. Here, we establish relations between the free particle statistics, their delta-$T$ noise and their bunching/anti-bunching properties. 
In Sec.~\ref{sec:Discussion}, we move on to interacting 1D systems. We outline a bosonization approach to emergent excitations in such systems, and present a connection between quantum statistics, scaling dimensions, and the sign of the delta-$T$ noise. In Sec.~\ref{sec:model}, we use the above approach to study tunneling between interacting QSH edges. In Sec.~\ref{sec:Results}, we discuss the conditions for emergence of the negative delta-$T$ in direct and exchange tunneling between the edges. A summary and conclusions are given in Sec.~\ref{sec:summary}. Complementary calculations have been relegated to Appendices~\ref{sec:interaction_and_selection}, \ref{sec:bosonization_details} \ref{sec:braiding_delta_t}, and \ref{sec:AppendixCoulomb}. Throughout the paper, we use units where $\hbar=k_B=1$.

\section{Delta-\textit{T} noise for free fermions and bosons}
\label{sec:Free_Particles}

As our starting point, in this Section, we consider delta-$T$ noise of free fermions and bosons scattering in a two-terminal setup as shown in Fig.\,\ref{fig:hom}\textcolor{blue}{(a)}.
In this system, the noise of the tunneling current $I_\mathcal{T}$ is defined as
\begin{equation}
    S =  \int dt \left[ \langle I_\mathcal{T}(t) I_\mathcal{T} (0) \rangle- \langle I_\mathcal{T}(t) \rangle \langle I_\mathcal{T}(0) \rangle \right].
    \label{eq:tunneling_noise}
\end{equation}
For energy independent transmission probability, we show in Sec.~\ref{sec:FreeElectrons} that the delta-$T$ noise is strictly positive for fermions, while we show in Sec.~\ref{sec:boson} that bosons can generate negative delta-$T$ noise at low temperatures. In Sec.~\ref{sec:bunching}, we further relate the sign of the delta-$T$ noise to the bunching/anti-bunching tendency in scattering processes involving two particles.

\subsection{Free fermions}
\label{sec:FreeElectrons}
Here, we review the computation of delta-$T$ noise of free fermions, e.g., electrons, following Ref.~\cite{Lumbroso2018}.
Consider a single spinless quantum channel with a scattering region connected to two Fermi reservoirs. 
Assuming only elastic scattering, the zero-frequency noise for fermions, $S^{(\text{f})}$, can be obtained within a scattering approach and reads~\cite{BlanterButtikerPhysRep00} 
\begin{equation}
\begin{aligned}
S^{(\text{f})} & = \frac{2e^2}{h} \int d\epsilon \left\{ D \left[ f_1 (1 - f_1) + f_2 (1 - f_2) \right] \right. \\
&\left. \qquad \qquad + D(1 -D) (f_1-f_2)^2 \right\}.
\end{aligned}
\label{eq:correlator_fermion}
\end{equation}
The noise is given in terms of two Fermi distribution functions
\begin{equation}
f_{\alpha}(\epsilon) = \frac{1}{\exp\left[(\epsilon-\mu_{\alpha})/T_{\alpha}\right] + 1},
\label{eq:fermionic_distribution}
\end{equation}
where $\alpha = 1,2$ for the two reservoirs and $\mu_{\alpha}$ and $T_{\alpha}$ are the corresponding chemical potentials and temperatures.
In Eq.~\eqref{eq:correlator_fermion}, $D$ is the transmission probability of the scatterer. For simplicity, we consider here only energy-independent $D$. Then, the net current between the reservoirs vanish in the absence of a voltage bias. The energy dependence can be included for a more general investigation of zero-current excess noise, see Ref.~\cite{ErikssonX21}. We also discuss the effect of energy dependence of the elastic transmission probability on the delta-$T$ noise in Appendix~\ref{sec:interaction_and_selection}.
It is worth noting that the positive sign in front of the second term in Eq.~\eqref{eq:correlator_fermion} is a direct consequence of Fermi statistics, i.e., a statistical phase $\Theta_\text{fermion} = \pi$ upon particle exchange [see Eq~\eqref{eq:CBH_relation_Intro}].

To study delta-$T$ noise, we choose $T_1>T_2$ and $\mu_1=\mu_2=0$ in Eq.~\eqref{eq:fermionic_distribution}.  We further assume that the temperature difference 
\begin{equation}
\label{eq:deltaTDef}
    \delta T\equiv T_1 - T_2
\end{equation} 
is much smaller than the average temperature $\bar{T}$: 
$$\delta T \ll \bar{T} \equiv (T_1 + T_2 )/2.$$ 
We then expand $S^{(\text{f})}$ to lowest order in $\delta T/\bar{T}$ and find
\begin{equation}
S^{(\text{f})} =S_{\text{thermal}}^{(\text{f})} \left[ 1 + \mathcal{C}^{(2)} \left( \frac{\delta T}{2 \bar{T}} \right)^2 \right].
\label{delta-T-fermions}
\end{equation}
Here, the thermal (Nyquist-Johnson) noise
\begin{equation}
S_{\text{thermal}}^{(\text{f})} = \frac{4e^2}{h} D \bar{T}, 
\label{eq:sthermal_fermion}
\end{equation}
and the non-equilibrium delta-$T$ noise correction is parametrized by the coefficient
\begin{equation}
\mathcal{C}^{(2)} = \frac{\pi^2 -6}{9} (1 -D) >0.
\label{eq:c2_fermion}
\end{equation}
The delta-$T$ noise contribution~\eqref{eq:c2_fermion} originates from the second line in Eq.~\eqref{eq:correlator_fermion}. Indeed, the first line after integration becomes $\propto \bar{T}$, and is independent of the temperature difference $\delta T$. As is clear from Eq.~\eqref{eq:c2_fermion}, for energy-independent transmissions $ 0\leq D\leq 1$, fermions can only generate positive delta-$T$ noise. 

\subsection{Free bosons}
\label{sec:boson}
Next, we compute the delta-$T$ noise for scattering of free bosons. In Sec.~\ref{sec:FreeElectrons}, we have seen that, for a small temperature bias, delta-$T$ noise produced by elastic scattering of non-interacting electrons is positive due to fermionic anti-commutation relations. If the electrons are replaced by charged bosons (for simplicity, we take their charges to be $e$), the particles emanating from the reservoirs follow instead Bose-Einstein distributions
\begin{equation}
b_{\alpha}(\epsilon) = \frac{1}{\exp\left\{[\epsilon-\mu_\alpha(T_{\alpha})]/T_{\alpha}\right\} - 1}.
\label{eq:bosonic_distribution}
\end{equation}
The bosonic charge noise then reads~\cite{BlanterButtikerPhysRep00}
\begin{equation}
\begin{aligned}
S^{(\text{b})} & = \frac{2e^2}{h} \int d\epsilon \left\{ D \left[ b_1 (1 + b_1) + b_2 (1 + b_2) \right] \right. \\
&\left. \qquad \qquad - D (1 - D) (b_1-b_2)^2 \right\}.
\end{aligned}
\label{eq:correlator_boson}
\end{equation}
where, in contrast to Eq.~\eqref{eq:correlator_fermion}, the second term has a negative sign. This sign originates from Bose-Einstein statistics~\cite{BlanterButtikerPhysRep00}, i.e., $\Theta_\text{boson}=0$ in Eq.~\eqref{eq:CBH_relation_Intro}. A further analysis of the sign difference between Eqs.~\eqref{eq:correlator_fermion} and~\eqref{eq:correlator_boson} and the corresponding particle statistics is presented in Sec.~\ref{sec:bunching}.

In contrast to free fermions, the chemical potential of free bosons depends significantly on the temperature. With $N$ particles in a bosonic reservoir with distribution $b(\epsilon)$, the chemical potential can be found by solving
\begin{equation}
\rho \int_0^\infty d\epsilon\,  b(\epsilon) = N,
\label{eq:number_conservation}
\end{equation}
where $\rho$ is the density of states (assumed for simplicity to be constant). Equation~\eqref{eq:number_conservation} has the solution
\begin{equation}
\begin{aligned}
\mu( T ) =T \ln \left( 1 - e^{-\gamma/T} \right)<0,
\end{aligned}
\label{eq:particle_conservation}
\end{equation}
where $\gamma \equiv N/\rho >0$.
In an isolated system with fixed $N$, this temperature-potential relation is crucial, and leads e.g., to Bose-Einstein condensation (BEC). Here, we fix
$N$ such that the two reservoirs, with generally different temperatures, have the same chemical potential and thus, there is no net charge flow between them. Otherwise, we cannot isolate the delta-$T$ contribution to the total noise. 

We next define $\bar{n}\equiv\gamma/T$ as the average bosonic occupation number. At low temperatures, $\bar{n}\gg 1$, and it eventually diverges at the critical BEC temperature $T_{\rm BEC}$. At higher temperatures, $\bar{n}\ll 1$ and the bosonic distribution is well approximated by the Boltzmann distribution. As follows, we consider only these two limiting cases and  assume that the temperature in the low-temperature scenario is still higher than $T_{\rm BEC}$.

In the low-temperature (LT) case, we use  Eq.~\eqref{eq:correlator_boson} and expand it to leading order in $\delta T$. We then find
\begin{equation}
S^{(\text{b})} = S_{\text{thermal}}^{\text{LT}} \left[ 1 +  \left( -1 + 2D \right) \left( \frac{\delta T}{2\bar{T}}\right)^2 \right],
\label{eq:result_1}
\end{equation}
where
\begin{equation}
S_{\text{thermal}}^{\text{LT}} = -\frac{4e^2}{h} \frac{\bar{T}^2 D}{\mu(\bar{T})}  >0
\end{equation}
is the thermal noise in the low-temperature limit.
From Eq.~\eqref{eq:result_1}, we extract the bosonic version of $\mathcal{C}^{(2)}$ in the low-temperature limit as
\begin{equation}
\mathcal{C}^{(2)} = -1 + 2D.
\label{eq:c2_lt}
\end{equation}
We see that $\mathcal{C}^{(2)}<0$ for $D<1/2$, which includes the weak tunneling limit $D\ll1$.

The negative sign of $\mathcal{C}^{(2)}$ is a consequence of bosonic statistics. The physical mechanism behind the negative noise is that bosonic equilibrium fluctuations are super-Poissonian because of bunching of bosons into the same state. Partitioning of the current tends to split up these bunches, lowering the total noise towards the classical Poissonian limit~\cite{BlanterButtikerPhysRep00}.

We next consider the the high temperature (HT) limit $\bar{n} \ll 1$. This condition implies that the reservoirs are diluted gases which are approximately described by Boltzmann distributions. Once again, we expand Eq.~\eqref{eq:correlator_boson} to leading order in $\delta T/\bar{T}$, where we now find
\begin{equation}
S_b = S_{\text{thermal}}^{\text{HT}} \left[ 1 + \frac{\mu^2}{2\bar{T}^2} \left( \frac{\delta T}{2\bar{T}}\right)^2  \right],
\label{eq:result_2}
\end{equation}
with
\begin{equation}
S_{\text{thermal}}^{\text{HT}} = \frac{4e^2}{h} e^{\mu(\bar{T} ) /\bar{T}} \bar{T} D>0
\end{equation}
being the high-temperature thermal noise. In the high-temperature limit, the coefficient $\mathcal{C}^{(2)}$ can be read off from Eq.~\eqref{eq:result_2} as
\begin{equation}
\mathcal{C}^{(2)} =  \frac{\mu^2}{2\bar{T}^2} >0,
\label{eq:c2_ht}
\end{equation}
indicating positive delta-$T$ noise in this limit.

By comparing Eqs.~\eqref{eq:c2_lt} and \eqref{eq:c2_ht} we see that the delta-$T$ noise of free bosons becomes negative at sufficiently low temperatures when the bosonic statistics is manifest. With increasing temperature, the bosonic features properties encoded in the delta-$T$ noise vanish. Hence, we conclude that, for energy independent transmissions $D$, free fermions can only generate positive delta-$T$ noise whereas bosons can produce both signs. We thus anticipate that observing negative delta-$T$ noise may imply the tunneling of bosons or, as we shall see in later sections, boson-like particles.

\subsection{Bunching/anti-bunching and quantum statistics}
\label{sec:bunching}
To further establish connections between delta-$T$ noise and particle statistics, we consider here noise correlations in an HOM interferometer. This interferometer is a well-known setup to reveal fundamental particle properties, including quasiparticle charges and statistics~\cite{SaminadayerPRL97,PicciottoNature97,DolevNature08,BocquillonScience13,MartinPRL14,BartolomeiScience20}.

\begin{figure}[t!]
  \centering
    \includegraphics[width=0.8\columnwidth]{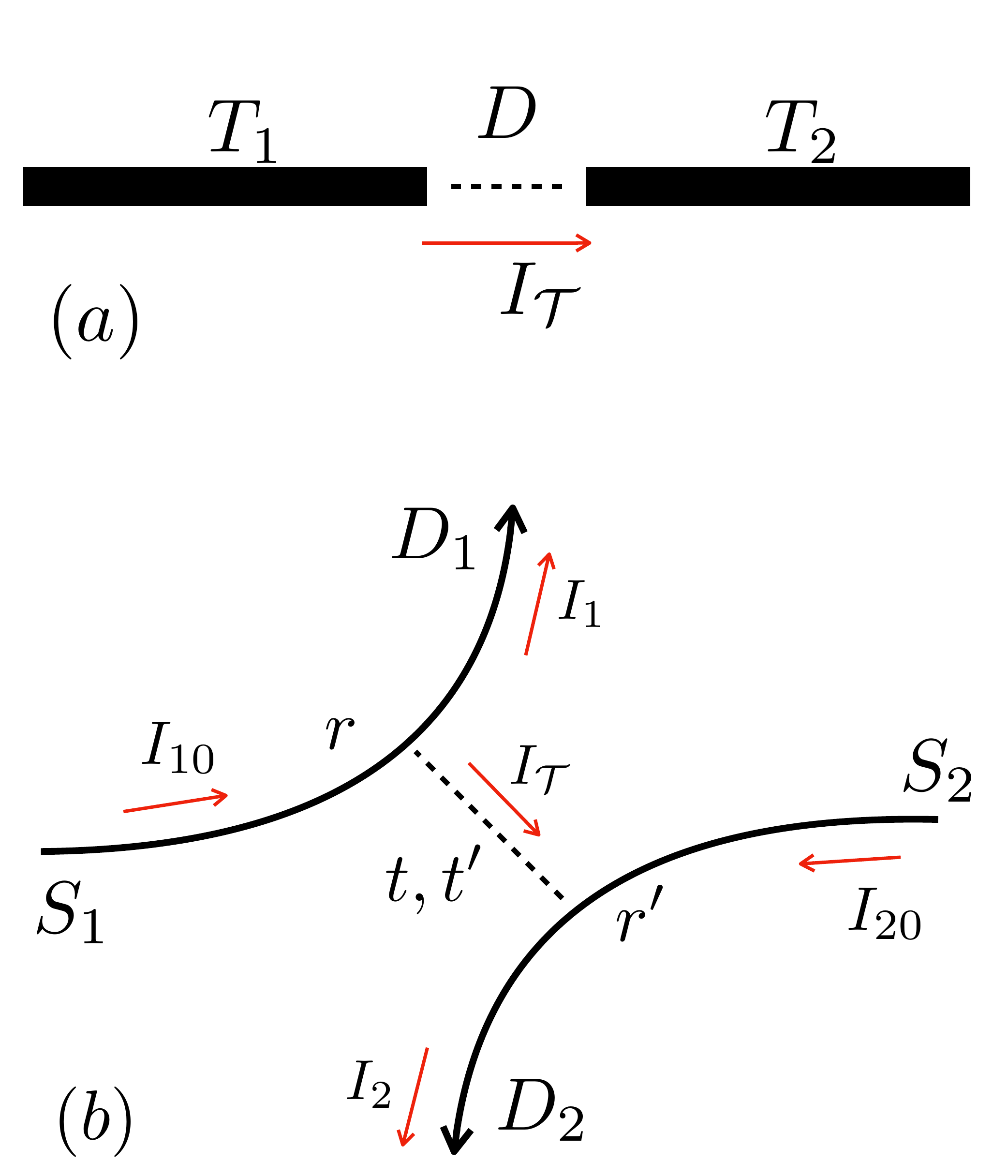}
    \caption{(a) A two-terminal setup where particles tunnel between two terminals with temperatures $T_1$ and $T_2$ through a structure (wire, single point contact, barrier or a weak link, etc.) characterized by the transmission probability $D$. The operator $I_\mathcal{T}$ (the red arrow) carries the current between two terminals. (b) Hong-Ou-Mandel interferometer with sources $S_1$ and $S_2$ and drains $D_1$ and $D_2$. Transmission and reflection amplitudes are denoted by $t,t'$ and $r,r'$ for particles emitted from $S_1$ and $S_2$, respectively. The current in the two drains and the tunneling current are denoted correspondingly by $I_1$, $I_2$, and $I_\mathcal{T}$.
    Current operators before the QPC are instead described by $I_{10}$ and $I_{20}$, respectively.
    }
    \label{fig:hom}
\end{figure}

A simple model of an HOM interferometer is depicted in Fig.~\ref{fig:hom}\textcolor{blue}{(b)}. It consists of two ``channels'' with four attached contacts: two source contacts where particles are injected ($S_1$ and $S_2$) and two drain contacts where particles are collected ($D_1$ and $D_2$) after scattering at a quantum point contact (QPC). Particles injected from source $S_i$ have the amplitude $t$ ($t'$) and $r$ ($r'$) to end up in the contact with the same ($D_i$) or different ($D_{j\neq i}$) labelling, respectively (for $i,j=1,2$). In this setup, we treat scattering of bosons and fermions on the same footing by considering a generic statistics angle $\Theta$ of the particles [see Eq.~\eqref{eq:CBH_relation_Intro}]. In fact, the presented approach works also for $\Theta\neq 0,\pi$, which is the case for, e.g., Laughlin anyons in the FQH state at filling $\nu=1/3$, where $\Theta=\pi/3$. Particles with generic $\Theta$ is discussed in more detail in Sec.~\ref{sec:Discussion}.

In the HOM interferometer, scattering generates noise by two mechanisms. First, noise is generated by partitioning of the injected currents. This mechanism is purely classical and reveals no quantum statistical information. Second, noise is also generated in two-particle scattering, i.e., when two particles arrive simultaneously at the scatterer. This noise depends directly on the particle statistics and can be interpreted in terms of bunching (the two particles tend to end up in the same drain) and anti-bunching (the two particles tend to end up in different drains) probabilities.

More specifically, when two particles (creation operators $a_1^\dagger$ and $a_2^\dagger$, from $S_1$ and $S_2$ respectively) arrive at the QPC, they are related to the operators in two drains $D_1$ (with the creation operator $b_1^\dagger$) and $D_2$ (with $b_2^\dagger$) via $a_1^\dagger = r b_1^\dagger + t b_2^\dagger$ and $a_2^\dagger = t' b_1^\dagger + r' b_2^\dagger$.
From these relations, we find the amplitude of anti-bunching as
\begin{equation}
\label{eq:PAB}
p_{\text{AB}} = tt' \exp(i\Theta) + r r' ,
\end{equation} 
where $\Theta$ is the statistical angle generated upon exchange of $b_1^\dagger$ and $b_2^\dagger$.
We note that Eq.~\eqref{eq:PAB} agrees with the result in Ref.~\cite{CampagnanoPRL12} for bunching Laughlin anyons.
We emphasize that the non-trivial phase $\Theta$ in front of $tt'$ in Eq.~\eqref{eq:PAB} originates from particle exchange enforced by the indistinguishability of quantum particles, not from anyonic braiding of particles at different positions. In Appendix~\ref{sec:braiding_delta_t} we show that braiding-induced delta-$T$ noise in fact vanishes at the second order of the tunneling amplitudes, and is thus negligible in our setups.

The probability for particles to end up in different drains $P_{\text{AB}}$, i.e., to anti-bunch, depends on the statistical phase as follows:
\begin{align}
P_{\text{AB}} = |p_{\rm AB}|^2 =  D^2 + (1 - D)^2 -2 D (1 - D)  \cos(\Theta),
\label{eq:anti-bunch_p}
\end{align}
where $D = |t|^2 = |t'|^2$ is the transmission probability at the point contact. Equation~\eqref{eq:anti-bunch_p} has two contributions, $P_{\rm AB}=P_{\rm class}+P_{\rm stat}$, where 
\begin{align}
\label{eq:p_class}
    P_\text{class}=D^2 + (1 - D)^2>0,
\end{align} 
is the classical partitioning contribution. The statistical contribution
\begin{align}
\label{eq:p_stat}
  P_\text{stat} = -2D (1 - D) \cos(\Theta), 
\end{align}
includes the statistical phase.
Equation~\eqref{eq:anti-bunch_p} indicates that for particles with $|\Theta| > \pi/2$ the probability $P_\text{AB}$ is increased compared to the classical result ($P_\text{stat} > 0$), while for $|\Theta | < \pi/2$ the statistical term reduces $P_\text{AB}$ ($P_\text{stat} < 0$) . We can call these two types of particles fermion-like respectively boson-like, with the borderline between them located at $|\Theta| = \pi/2$.

We further elaborate on the relation between delta-$T$ noise and bunching/anti-bunching, by showing that the tendency for particles to bunch or anti-bunch is revealed in the current cross-correlation function~\cite{BlanterButtikerPhysRep00}.
We consider the system where two channels are connected via a single point contact, as shown in Fig.~\ref{fig:hom}\textcolor{blue}{(b)}.
We take $\mu_1=\mu_2$ in the source contacts $S_1$ and $S_2$, but a small temperature bias $\delta T\ll \bar{T}$, similar to previous sections. We introduce the current operator 
\begin{equation}
   I_\mathcal{T}=I_{10}-I_1=I_2-I_{20} 
   \label{IT-I1-I2}
\end{equation}
that transfers charges between two channels. Its irreducible correlation function defines the noise, $S$,
Eq.~\eqref{eq:tunneling_noise}.
Similarly to Eq.~\eqref{eq:anti-bunch_p}, we split $S$ into two parts
\begin{align}
\label{eq:S12Div}
    S = S_\text{class}(\bar{T},\delta T) + S_\text{stat}(\Theta,\bar{T},\delta T),
\end{align}
where $S_\text{class}$ and $S_\text{stat}$ refer to the statistics-independent and statistics-induced contributions, respectively.

The classical term, $S_\text{class}$, contains the noise from the partitioning of classical particles at the QPC. 
As we will see below, the classical contribution to delta-$T$ noise is negligible in the weak-tunneling limit.
To demonstrate this explicitly, we consider a model that hosts distinguishable particles in both channels, with distribution functions $n_1$ and $n_2$, respectively, and the same charge $q$.
The first contribution to the classical noise, upon Fourier transforming 
\begin{equation}
    \int dt \langle I_\mathcal{T}(t) 
I_\mathcal{T} (0) \rangle
\to q^2 D \int d\epsilon (n_1 + n_2),
\label{II-class}
\end{equation}
is linear in transmission probability $D$. Importantly, the energy integrals of the distribution functions are temperature independent,
and, hence, this term does not contribute to the delta-$T$ noise.
The second term in Eq.~\eqref{eq:tunneling_noise} is proportional to $D^2$, as the average tunneling current 
$$ 
\langle I_\mathcal{T}\rangle 
= q D \int d\epsilon (n_1 - n_2)
$$
is itself linear in $D$. Thus, this contribution to the delta-$T$ noise is negligible in the weak-tunneling limit, $D\ll 1$.

For indistinguishable quantum particles, an additional noise term $S_\text{stat}$ emerges, which depends on the statistics of the particles undergoing two-particle scattering.
Indeed, when particles anti-bunch $(|\Theta|>\pi/2)$, the effective tunneling probability for two-particle scattering events is suppressed: when arriving simultaneously at a contact from different sources, the particles avoid tunneling in order not to meet in the same channel/drain. 
By contrast, for bunching particles $(|\Theta|<\pi/2)$, statistics enhances the tunneling probability and thus the noise is increased. Hence, in the absence of a voltage bias, the statistics-induced correction to the noise depends on the statistical angle $\Theta$: 
\begin{equation}
\begin{aligned}
\label{eq:Arg1}
& S_\text{stat}(|\Theta|<\pi/2, \bar{T},\delta T) > 0 \quad \text{(bunching)},\\
& S_\text{stat}(|\Theta|>\pi/2,\bar{T},\delta T) < 0 \quad \text{(anti-bunching)}.
\end{aligned}
\end{equation}
We note in passing that the statistical contribution is finite for $\delta T=0$, i.e., it affects the thermal noise.

By construction, we have $S_\text{stat}\propto -P_\text{stat}$, where the statistical contribution
$P_\text{stat}$ to the anti-bunching probability is given by Eq.~\eqref{eq:p_stat}. 
This result is consistent with the semiclassical result in Ref.~\cite{HalperinPRL16}, which predicts two-particle scattering noise stemming from particle statistics. At the same time, since this statistical term involves two-particle scattering for particles emitted from the two uncorrelated sources, the noise is determined by a product of two distribution functions, $n_1n_2$. 

For instance, for a system with free fermions 
($\Theta=\pi$), the statistical contribution of the noise takes the form
\begin{equation}
\begin{aligned}
S^{(\text{f})}_\text{stat} & = -\frac{2e^2}{h} 2 D (1 - D) \int d\epsilon\ f_1 f_2.
\end{aligned}
\label{eq:stat-S}
\end{equation}
The classical part of the noise combines the partitioning term of distinguishable particles and the term related to the classical part of $P_\text{AB}$, Eq.~\eqref{eq:p_class}. For fermions, this yields
\begin{equation}
\begin{aligned}
S^{(\text{f})}_\text{class} & = \frac{2e^2}{h} \int d\epsilon [D (f_1 + f_2) - D^2 (f_1^2 + f_2^2)].
\end{aligned}
\label{eq:class-S}
\end{equation}
The sum of Eqs.~\eqref{eq:stat-S} and \eqref{eq:class-S} reproduces the result of Eq.~\eqref{eq:correlator_fermion}.
To linear order in the tunneling probability $S^{(\text{f})}_\text{class}$ 
is determined by the single-particle partitioning, cf. Eq.~(\ref{II-class}), and does not depend on temperature (see discussion above; this holds even when the distribution functions of quasiparticles are not well-defined, as for anyons). The delta-$T$ noise for the tunneling current $I_\mathcal{T}$ in the weak-tunneling limit $D\ll 1$ is thus fully determined by the statistics-induced noise that reflects the anti-bunching/bunching preference of particles.
The sign of this contribution is governed by $\cos(\Theta)$ in Eq.~\eqref{eq:p_stat} and changes at $\Theta=\pi/2$.

Now, comparing the energy integrals in Eq.~(\ref{eq:stat-S}) for $\delta T=0$ and $\delta T\neq 0$, we observe that the temperature bias \textit{reduces} the statistical contribution to the noise by reducing the overlap between the two fermion distribution functions. For the anti-bunching particles, this reduction of the negative statistical contribution to the noise, Eq.~\eqref{eq:Arg1}, implies \text{positive} delta-$T$ noise, cf. Eq.~\eqref{delta-T-fermions}. For bosons in the low-$T$ regime, the same consideration yields negative delta-$T$ noise, Eq.~\eqref{eq:result_1} for $D\ll 1$. 

Before concluding this qualitative discussion, it is 
worth mentioning that the above consideration assumed
(i) elastic tunneling with (ii) an energy-independent transmission probability, and (iii) absence of interparticle interaction over the tunneling bridge.
In Appendix \ref{sec:interaction_and_selection}, we will address the first two conditions in some more detail. As for condition (iii), it is natural to expect that the interaction between the charged particles that belong to the two channels connected by the tunneling bridge may mimic effects of statistics-induced anti-bunching and bunching (see Refs.~\cite{MartinPRL14,BellentaniPRB19} for interaction-induced bunching for fermions). As a result, such inter-channel interactions may also affect the sign of the delta-$T$ noise. In what follows, we will, for simplicity, neglect these effects, relegating their study to future work. Even in this situation, as we will see, the interactions within the subsystems may effectively lead to an emergence of a nontrivial statistical phase that is manifested as a bunching tendency.

To summarize this Section, we have seen that delta-$T$ noise can disclose statistical features of fermions and bosons. For an energy independent transmission, free fermions come with positive delta-$T$ noise, whereas free bosons can produce either sign. Further, we associated the sign of the noise for these particles to anti-bunching respectively bunching tendencies and extended this analysis to particles with generic statistical phases $\Theta$. In the following Section, we will derive rigorous statements about the the sign of the delta-$T$ noise for interacting particles in 1D.

\section{Bosonization, quantum statistics, and delta-\textit{T} noise}
\label{sec:Discussion}

In this Section, we analyze delta-$T$ noise for particles with statistical phases $\Theta$ between those of bosons ($\Theta=0$) and fermions ($\Theta=\pi$). Such particles emerge naturally in various strongly interacting 1D electron systems. To describe these particles, we use in Sec.~\ref{sec:stat_and_ops} the bosonization framework and discuss in detail the relation between particle statistics and scaling dimensions $\Delta_\mathcal{T}$ of tunneling operators. In particular, we specify the conditions under which $\Delta_\mathcal{T}$ can be connected to the quantum statistics of the tunneling particles. Examples of the general formalism are provided in Sec.~\ref{sec:Examples}.
In Sec.~\ref{sec:deltaTnoiseFormula}, we derive a generic formula for the delta-$T$ noise due to a weak temperature difference between the two subsystems. We show that, for tunneling between two identical subsystems, the scaling dimension $\Delta_\mathcal{T} = 1/2$ is precisely the point where the delta-$T$ noise changes sign. As a consequence, negative delta-$T$ noise for tunneling between two infinite subsystems with local interactions can only be generated for particles that are boson-like, i.e., for $|\Theta|<\pi/2$.

\subsection{Tunneling operators, quantum statistics, and scaling dimensions}
\label{sec:stat_and_ops}

We consider a general bosonized (an introduction to bosonization can be found, e.g., in Ref.~\cite{GiamarchiBook}) Hamiltonian
\begin{align}
\label{eq:GenBosHam}
    H_\text{1D} = \frac{1}{4\pi} \sum_{i=1}^N \frac{1}{\lambda_i} \int_{-\infty}^\infty dx\, {v_i} (\partial_x \phi_i)^2.
\end{align}
It describes a set of $N$ 1D, free, chiral bosons $\phi_i$ propagating along infinite channels with velocities $v_i>0$. Here, $\lambda_i>0$ are real numbers which may depend on topological order and/or short-range density-density interactions. The Hamiltonian~\eqref{eq:GenBosHam} is supplemented with the equal-time commutation relations
\begin{align}
\label{eq:comm_gen}
\left[ \phi_i(x), \phi_j (x') \right] = i\pi\lambda_i\chi_i \delta_{ij}\, \text{sgn}(x - x'),
\end{align}
where $\text{sgn}(x-x')$ is the sign function and $\chi_i=\pm 1$ specifies the propagation direction of $\phi_i$. The usefulness of Eqs.~\eqref{eq:GenBosHam} and \eqref{eq:comm_gen} is that a large class of interacting 1D systems, including FQH edges~\cite{WenPRB91}, LL wires~\cite{KaneFisherPRB92,KaneFisherPRL94}, or QSH edges~\cite{Wu2006,Xu2006,Maciejko2009,TanakaFurusakiMatveevPRL11}, can be described in this form after suitable basis transformations.

We further define total and partial local charge densities by introducing the charge vector $\vec{t}=\{t_i\}$ according to
\begin{align}
\label{eq:rho_operator}
    \rho(x)=\sum_j \chi_j\rho_j(x) = \sum_j \chi_j t_j\frac{\partial_x \phi_j(x)}{2\pi}.
\end{align}
The basis dependent charge vector specifies the contribution of each bosonic mode to the charge transport. A common and convenient choice is $\vec{t}=(1,1,\ldots,1)^T$, but we keep $\vec{t}$ general to conform with the notation in some previous works~\cite{WenZee1992,KaneFisher1995Contact}. By combining Eqs.~\eqref{eq:comm_gen} and~\eqref{eq:rho_operator}, we obtain the additional commutation relations
\begin{align}
\label{eq:comm_gen2}
&\left[ \rho_i(x), \phi_j (x') \right] = it_i\lambda_i\chi_i \delta_{ij}\, \delta(x - x'),\\
& \label{eq:comm_gen3} \left[ \rho_i(x), \rho_j (x') \right] = \frac{i}{2\pi}t_i^2\lambda_i\chi_i \delta_{ij}\, \partial_x\delta(x - x').
\end{align}

Creation and annihilation of particle excitations are described by so-called vertex operators~\cite{Francesco2012}
\begin{align}
\label{eq:vertex_operator}
    \psi_j(x) = \frac{1}{\sqrt{2\pi a}} e^{im_j\phi_j(x)},
\end{align}
where $a$ is a short-distance cutoff, and $m_j$ is a real number. In principle, Eq.~\eqref{eq:vertex_operator} should be multiplied by a Klein factor $F_j$ to ensure commutation relations for vertex operators with $\phi_i\neq \phi_j$. For example, for fermionic vertex operators, the Klein factors obey $\{F^\dagger_i,F_j\}=2\delta_{ij}$, $\{F^\dagger_i,F^\dagger_j\}=\{F_i,F_j\}=0$. Klein factors are, however, not important for the calculations in this work (apart from Appendix~\ref{sec:braiding_delta_t}), and we ignore them in this Section.

We define the physical charge $q_j$ associated to $\psi_j(x)$ by computing the commutator
\begin{align}
\left[ \rho(x), \psi_j (x') \right] = \left[ \chi_j\rho_j(x), \psi_j (x') \right] = q_j\delta(x-x')\psi_j(x'),
\end{align}
and obtain
\begin{align}
\label{eq:charge_qj}
    q_j = -t_jm_j\lambda_j.
\end{align}
Whether $\psi_j$ describes the creation or annihilation of a charge $|q_j|$, depends on the sign of $m_jt_j$.

Next, we compute the scaling dimension $\Delta_j$ of $\psi_j$, defined by
\begin{align}
\label{eq:Delta_jDef}
\langle \psi_j(x)\psi^\dagger_j(x') \rangle \sim (x-x')^{-2\Delta_j}. 
\end{align}
The average is performed with respect to the diagonal Hamiltonian $H_{1D}$. Since the scaling dimension parametrizes correlation function, it influences various observable properties (see Sec.~\ref{sec:deltaTnoiseFormula} below). To proceed, we use the Baker-Campbell-Hausdorff identity
\begin{align}
\label{eq:BCH}
    e^A e^B = e^B e^A e^{[A,B]},
\end{align}
which is valid whenever two operators $A$ and $B$ commute with their commutator $[A,B]$. We combine this identity with the chiral boson propagators
\begin{align}
\label{eq:single_boson_propagator}
     &\langle \phi_i(x,t)\phi_j(0,0) \rangle  = -\lambda_{j}\delta_{ij} \notag\\
     &\qquad \times \begin{cases} \ln\dfrac{a+i(\chi_ix+ v_i t)}{a}, &\text{ for } T=0,\\[0.2cm]
    \!\ln \dfrac{\sin\!\left[\! \frac{\pi T}{v_i}\! \left(a\! + \! i(\chi_i x \!+ \!v_i t)\right)\right]}{a \pi T/v_i}, &\text{ for }T\neq 0.
    \end{cases}
\end{align}
By inserting the result in Eq.~\eqref{eq:Delta_jDef}, we arrive at the scaling dimension of $\psi_j$ as
\begin{align}
\label{eq:Local_Scaling}
    \Delta_j = \frac{\lambda_j m_j^2}{2}.
\end{align}

The statistical phase of excitations is defined as that generated upon exchanging vertex operators at different locations.  Applying~\eqref{eq:BCH} to the vertex operator~\eqref{eq:vertex_operator} and using the commutation relations~\eqref{eq:comm_gen}, we find
\begin{align}
    \psi_j(x)\psi_j(x') = e^{i\Theta_j}\psi_j(x')\psi_j(x).
\end{align}
Here, the statistical phase is given by
\begin{align}
\label{eq:theta_j}
    \Theta_j = m_j^2\pi\lambda_j\chi_j.
\end{align}
Above, we have defined $\Theta_j$ for $x'>x$. For $x<x'$, we have $\Theta_j\rightarrow-\Theta_j$.
From Eq.~\eqref{eq:theta_j}, we see that if $\Theta_j=0$, we recover bosonic commutation of the particles, whereas $|\Theta_j|=\pi$ gives the fermionic anti-commutation relation. However, since $\Theta_j$ can depend on interactions (encoded in $\lambda_j m_j$), the corresponding excitations can interpolate between being more boson-like ($|\Theta_j|<\pi/2$) or more fermion-like ($\pi/2<|\Theta_j|<\pi$).

We next describe tunneling of particles. We consider a local tunneling operator $\mathcal{T}_{\vec{m}}(x)$ which transfers particles between bosonic species. It consists of a product of vertex operators~\eqref{eq:vertex_operator} and is therefore specified by the string of numbers $\vec{m}=\{m_j\}$ according to
\begin{align}
\label{eq:TunnelingOperators}
    \mathcal{T}_{\vec{m}}(x) & = \mathcal{N} \prod_{j=1}^M e^{i m_j\phi_j(x)}.
    \end{align}
Here, $\mathcal{N}=(2\pi a)^{M/2}$ is a normalization constant with $M\leq N$ being the total number of involved vertex operators. 
The tunneling operator $\mathcal{T}_{\vec{m}}$ does not create any net charge $Q$, which means that $\vec{m}$ must be fixed such that the following constraint is fulfilled
\begin{align}
\label{eq:TotalChargeQ}
    Q \equiv -\sum_j m_j\lambda_jt_j = 0.
\end{align}
By appropriate choices of $\vec{m}$, the tunneling operators describes single- or multiple-electron and quasiparticle tunneling. Examples are given below in Sec.~\ref{sec:Examples}.

The scaling dimension of the tunneling operator is defined similarly to~\eqref{eq:Delta_jDef} as
\begin{align}
\label{eq:Delta_TDef}
\langle \mathcal{T}_{\vec{m}}(x) \mathcal{T}^\dagger_{\vec{m}}(x') \rangle \sim (x-x')^{-2\Delta_{\mathcal{T}}}.
\end{align}
By repeated use of the identity~\eqref{eq:BCH}, we have for a chain of vertex operators
\begin{align}
\label{eq:BCH_Long}
    \langle e^{A_1}\ldots e^{A_n} \rangle \propto e^{\sum_{i<j} \langle A_i A_j \rangle}.
\end{align}
Using Eqs.~\eqref{eq:BCH_Long} and~\eqref{eq:single_boson_propagator} in Eq.~\eqref{eq:Delta_TDef}, we find
\begin{align}
\label{eq:primary_field_scaling}
    \Delta_{\mathcal{T}} = \frac{1}{2}\sum_j \lambda_j m_j^2 = \sum_j \Delta_j,
\end{align}
where in the final equality, we used Eq.~\eqref{eq:Local_Scaling}.

With the tunneling operators~\eqref{eq:TunnelingOperators}, we can describe point tunneling (here chosen without loss of generality to occur at $x=0$) of particles by adding  the term
\begin{align}
\label{eq:HIntroTunnel}
    H_\mathcal{T} =\int dx\, \delta(x)\left[ t_{\mathcal{T}}\mathcal{T}_{\vec{m}}(x) + \text{H.c.\,} \right]
\end{align}
to $H_{1D}$.
Here, the type and number of particles tunneling is captured by ${\vec{m}}$. The tunneling strength is quantified by the constant $t_{\mathcal{T}}$ and we assume that the energy dependence of $t_\mathcal{T}$ is negligible. The terms in the Hamiltonian density in Eq.~\eqref{eq:HIntroTunnel} take the form $\delta(x)\times t_\mathcal{T} \times \mathcal{T}$. With decreasing temperature (i.e., for an RG flow towards lower energies), the amplitude $t_\mathcal{T}$ is renormalized, to leading order, as follows (see, e.g., Ref.~\cite{GiamarchiBook}):
\begin{align}
\label{eq:renormalizationEquation}
    \frac{dt_\mathcal{T}}{d\ell} = \left( 1-\Delta_\mathcal{T}\right)t_\mathcal{T},
\end{align}
where $\ell$ is the running logarithmic length scale and $\Delta_\mathcal{T}$ is given in Eq.~\eqref{eq:primary_field_scaling}. For $\Delta_\mathcal{T}>1$, the tunneling decreases with lower energies (increasing $\ell$) and $H_{\mathcal{T}}$ is an irrelevant perturbation with respect to $H_{1D}$. For $\Delta_{\mathcal{T}}<1$ the tunneling strength instead increases at lower energies and $H_{\mathcal{T}}$ is a relevant perturbation. 
For $\Delta_\mathcal{T}=1$, $H_{\mathcal{T}}$ is called a marginal perturbation. In this situation, one has to analyze the RG flow of $t_\mathcal{T}$ by taking into account next orders in the beta-function.
In what follows, we will consider the situations where 
the renormalization of $t_\mathcal{T}$ is terminated by temperature that sets the infrared length cutoff at the thermal length $v/T$, such that the tunneling remains weak at all scales.

 We now consider tunneling between two sub-systems $A$ and $B$. To this end, we write Eq.~\eqref{eq:TunnelingOperators} as
\begin{align}
\label{eq:TAB}
\mathcal{T}_{\vec{m}}(x) = \Psi_\text{A} \Psi_\text{B},
\end{align}
where 
\begin{align}
   & \Psi_A = (2\pi a)^{-N_A/2}\ \prod_{j=1}^{N_A}\ \exp(i\,m_j \phi_j), \\
   &\Psi_B = (2\pi a)^{-N_B/2}\! \prod_{j = N_A + 1}^{M} \! \exp(i\,m_j \phi_j)
\end{align}
are products of the vertex operators in subsystems $A$ respectively $B$. The number of vertex operators in each subsystem is $N_A$ respectively $N_B$, with $N_A+N_B=M$. We assume in the following that subsystems A and B are each in separate thermal equilibria, so that Eq.~\eqref{eq:single_boson_propagator} at finite $T$ is applicable. 

At this point, we emphasize that the possible tunneling processes are constrained by the type of vacuum between $A$ and $B$. For example, if these subsystems are bridged by a topologically trivial vacuum, only electrons can tunnel. This happens in the strong backscattering regime of a QPC in the FQH regime. For tunneling across a topologically ordered region such as the FQH bulk in the weak backscattering regime of the FQH QPC, fractional charges can tunnel. 

We next define statistical phases also for the $A$ and $B$ subsystems as
\begin{align}
\label{eq:tunneling_statAB}
     \Psi_{A,B}(x) \Psi_{A,B}(x') = e^{i\Theta_{A,B}}\Psi_{A,B}(x') \Psi_{A,B}(x).
\end{align}
Without loss of generality, we use the convention $x' > x$ when studying the statistical angle following Eq.~\eqref{eq:tunneling_statAB} to avoid an additional negative sign.
Noteworthily, the definition of $\Theta_{A,B}$ contains two groups of operators [$\Psi_{A,B}(x)$ and $\Psi_{A,B}(x')$] that are only different in the spatial arguments ($x$ and $x'$). Possible Klein factors, if included, are the same for each group. These Klein factors are thus irrelevant to the statistical angle $\Theta_{A,B}$ defined in Eq.~\eqref{eq:tunneling_statAB}, and safely neglected in the definition Eq.~\eqref{eq:vertex_operator} and the rest of the paper.
Again, by using Eq.~\eqref{eq:BCH}, we find
\begin{align}
    & \Theta_{A,B} = \pi\sum_{j\in A,B}\chi_j\lambda_j m_j^2.
    \label{eq:theta_ab}
\end{align}
The statistical phases are bounded as
\begin{align}
\label{eq:bound_of_phase}
    |\Theta_{A,B}| \leq 2\pi \Delta_{A,B},
\end{align}
where
\begin{align}
    \Delta_{A,B} = \frac{1}{2}\sum_{j\in A,B} \lambda_j m_j^2
\end{align}
are the scaling dimensions of $\Psi_{A,B}$. They satisfy $\Delta_A+\Delta_B=\Delta_{\mathcal{T}}$. 

We now consider a special, but frequent, situation where subsystems $A$ and $B$ are identical. Then, we have $\Theta_A=\Theta_B$ and from Eq.~\eqref{eq:primary_field_scaling},  $\Delta_{A}+\Delta_{B}=2\Delta_{A}=\Delta_{\mathcal{T}}$. In this case, from  Eq.~\eqref{eq:bound_of_phase} we get the bound
\begin{align}
\label{eq:Scaling_and_statistics}
    |\Theta_A| \leq \pi \Delta_{\mathcal{T}}.
\end{align}
From this equation, we see that $\Delta_\mathcal{T}=1/2$ is precisely the dividing point for $\Psi_A=\Psi_B$  describing boson-like or fermion-like particles, i.e.,
\begin{align}
\label{eq:Theta_Delta_bound}
    |\Theta_A| < \frac{\pi}{2}, \ \text{if} \  \Delta_{\mathcal{T}}<\frac{1}{2}.
\end{align}
Equation~\eqref{eq:Theta_Delta_bound} is a key observation in this work. It states that, if $\Delta_{\mathcal{T}}<1/2$ in a system governed by Eq.~(\ref{eq:GenBosHam}), the particles corresponding to $\Psi_A$ must be boson-like. The opposite is, however, not necessarily true. Boson-like operators $\Psi_A$ may very well produce $\Delta_{\mathcal{T}}>1/2$ (see Sec.~\ref{sec:direct_tunneling}). In Sec.~\ref{sec:deltaTnoiseFormula}, we show that the above observations allow us to connect the sign of the delta-$T$ noise to the statistics of the tunneling particles. Before that, we clarify the general formalism above, with a few simple examples.

\subsection{Examples of the relation between statistics and scaling dimension}
\label{sec:Examples}
As our first example, consider the operator
\begin{align}
  \psi_{e} \sim e^{i\phi}
\end{align}
in a theory with a single boson $\phi$ obeying
\begin{align}
    \left[ \phi(x), \phi (x') \right] = i\pi\, \text{sgn}(x - x').
\end{align}
According to Eqs.~\eqref{eq:comm_gen} and \eqref{eq:vertex_operator}, we have $\lambda=m=1$, and, by combining the formulas for the statistical phase~\eqref{eq:theta_j} and the scaling dimension~\eqref{eq:Local_Scaling}, we find $\Theta_e=\pi$ and $\Delta_e=1/2$. The charge is read off as $q=1$ from Eq.~\eqref{eq:charge_qj}. In other words, the operator $\psi_e$ describes a non-interacting, chiral fermion. This identification is the foundation of the bosonization technique~\cite{GiamarchiBook}.

As a slightly more involved example, consider the Hamiltonian of a spinless LL wire with Luttinger parameter $K$ and velocity $u$~\cite{GiamarchiBook}. In the chiral basis, the LL Hamiltonian can be put on the form~\eqref{eq:GenBosHam} as
\begin{align}
    H_\text{LL}  = \frac{u}{4\pi K } \int dx \left[(\partial_x \phi_{1})^2 + (\partial_x \phi_{2})^2 \right],
    \label{eq:free_LL_hamiltonian}
\end{align}
with the accompanying commutation relations
\begin{align}
\label{eq:KComm}
    \left[ \phi_i(x), \phi_j (x') \right] = i\pi \chi_i K\delta_{ij}\, \text{sgn}(x - x'), \quad i=1,2,
\end{align}
where $\chi_i = \pm 1$, for the right- and left-moving chiral channels, respectively.
In this case, we read off $\lambda_1=\lambda_2=K$. The characteristics of particles produced by
\begin{align}
    \psi_{1,2}\sim e^{i\phi_{1,2}},
\end{align}
are found from Eqs.~\eqref{eq:theta_j} and \eqref{eq:Local_Scaling} as $\Theta_1=\Theta_2=\pi K$ and $\Delta_1=\Delta_2=K/2$, respectively. These values show that the particle statistics and scaling dimensions of creation and annihilation operators in a LL depend continuously on $K$. For sufficiently strong repulsive interactions, $0<K<1/2$, the excitations $\psi_{1,2}$ are boson-like ($\Theta_{1,2}$ closer to $0$), while for weaker interaction $1/2<K<1$ they are more fermion-like ($\Theta_{1,2}$ closer to $\pi$). 

As our last example, we consider the chiral LL model describing a single-channel FQH edge at filling factor $\nu$~\cite{WenPRB91}. This theory includes a single boson $\phi$. In this case, 
$\lambda^{}=\nu^{}=1/(2 n + 1)$, with $n$ being a positive integer. The operators
\begin{align}
    & \psi_{e}\sim e^{i\phi/\nu},\\
    & \psi_\text{qp}\sim e^{i\phi}
\end{align}
describe electronic and anyonic excitations, respectively. By using Eqs.~\eqref{eq:theta_j} and~\eqref{eq:Local_Scaling}, one may check that $q_e=1$, $\Theta_e =\pi$ (mod $2\pi$), and $\Delta_e=1/(2\nu)$. On the other hand, for anyons one finds: $q_\text{qp}=\nu$, $\Theta_\text{qp} =\pi\nu$ and $\Delta_\text{qp}=\nu/2$. The Laughlin FQH edge and the spinless LL are addressed in more detail in Appendix~\ref{sec:bosonization_details}. 

One observes that the statistical angles and scaling dimensions for excitations in the spinless LL and in Laughlin FDH edges are related by Eq.~\eqref{eq:Scaling_and_statistics} with the equality sign.
However, for more complex FQH edges with counter-propagating edge channels (e.g., at filling $\nu=2/3$), the relation~\eqref{eq:Scaling_and_statistics} holds in the sense of inequality, since the parameters $\chi_i$ have different signs for different modes (cf. Ref.~\cite{Schiller2021}).

\subsection{Delta-\textit{T} noise in interacting 1D systems}
\label{sec:deltaTnoiseFormula}

We now compute the delta-$T$ noise for weak tunneling in interacting 1D electron systems. The total Hamiltonian reads as
\begin{equation}
    H=H_{1D}+H_\mathcal{T},
    \label{eq:HO}
\end{equation}
where $H_{1D}$ is given in Eq.~\eqref{eq:GenBosHam} and
$H_\mathcal{T}$ in Eq.~\eqref{eq:HIntroTunnel} with the tunneling operator in Eq.~\eqref{eq:TAB}. The type of tunneling is specified by the choice of $\vec{m}$ and the most relevant type of tunneling is further determined by Eqs.~\eqref{eq:primary_field_scaling} and~\eqref{eq:renormalizationEquation}.
In what follows, we consider $H_\mathcal{T}$ as a weak perturbation.

We further assume that, in the absence of tunneling, the two subsystems $A$ and $B$ [see the discussion below Eq.~\eqref{eq:Theta_Delta_bound}] are at the same electrochemical potential $\mu$, which we set to zero. By contrast, $A$ and $B$ are chosen to have distinct temperatures, $T_1$ and $T_2$, respectively. We restrict ourselves to the case of weak thermal bias, i.e., $\delta T \equiv |T_1 - T_2|$ is assumed to be much smaller than the average temperature $\bar{T} \equiv (T_1 + T_2)/2$. Under these conditions, we are interested in noise effects for small $\delta T/\bar{T}$.

To find the delta-$T$ noise, we introduce the charge tunneling current operator:
\begin{equation}
\begin{aligned}
    I_\mathcal{T}&\equiv i \int d x[H_\mathcal{T}, \rho(x) ] 
    = -it_\mathcal{T} q\left[ \mathcal{T}_{\vec{m}}(0) - \mathcal{T}_{\vec{m}}^\dagger(0) \right]. 
\end{aligned}
\end{equation}
Here, $q\equiv|q_A|=|q_B|$ is the transferred charge, with $Q=q_A+q_B=0$ according to the charge conservation condition~\eqref{eq:TotalChargeQ}. The density $\rho(x)$ is given in Eq.~\eqref{eq:rho_operator}.

Next, we follow the standard Keldysh approach (outlined, e.g., in Ref.~\cite{RechMartinPRL20}), and compute the average tunneling current to leading order in the perturbation $H_\mathcal{T}$. We find 
\begin{align}
\label{eq:It}
   I=\langle I_\mathcal{T} \rangle = 2i q|t_\mathcal{T}|^2\int dt \sin\left(\omega_0 t\right) \langle \mathcal{T}^{}_{\vec{m}}(t,0)\mathcal{T}_{\vec{m}}^\dagger(0,0) \rangle,
\end{align}
where the time-evolution of $\mathcal{T}_{\vec{m}}(t,0)$ is described in the interaction picture with $H_{\mathcal{T}}$ as the interaction Hamiltonian. The ``Josephson frequency'' $\omega_0=q\delta V$ includes a possible voltage difference $\delta V$, to be put to zero below.
From $I$, we find the differential conductance $$g_\mathcal{T}\equiv \partial_{\delta V} I_\mathcal{T}|_{\delta V=0}$$ 
as
\begin{align}
    \label{eq:gt}
    g_\mathcal{T} = 2i q^2|t_\mathcal{T}|^2\int dt \;t \langle \mathcal{T}_{\vec{m}}^{}(t,0)\mathcal{T}_{\vec{m}}^\dagger(0,0) \rangle.
\end{align}

The key quantity of interest to us is the zero-frequency charge noise, defined as
\begin{align}
\label{eq:ST}
   S \equiv 2\int dt \left[ \langle I_\mathcal{T}(t) I_\mathcal{T}(0) \rangle - \langle I_\mathcal{T}(t)\rangle\langle I_\mathcal{T}(0)\rangle \right]. 
\end{align}
Since the current~\eqref{eq:It} is proportional to $|t_{\mathcal{T}}|^2$, the noise includes, to lowest order in the tunneling probability, only the reducible current correlator $\langle I_\mathcal{T}(t) I_\mathcal{T}(0) \rangle$. Then the noise is found as
\begin{align}
\label{eq:S}
   S = 4 q^2|t_\mathcal{T}|^2\int dt \cos\left(\omega_0 t\right) \langle \mathcal{T}_{\vec{m}}^{}(t,0)\mathcal{T}_{\vec{m}}^\dagger(0,0) \rangle. 
\end{align}
It is worth noting that the leading-order contributions to both the average tunneling current
(\ref{eq:It}) and the noise of the tunneling current (\ref{eq:S}) are quadratic in $|t_\mathcal{T}|$ and involve the same correlation function, $\langle \mathcal{T}_{\vec{m}}^{}(t,0)\mathcal{T}_{\vec{m}}^\dagger(0,0) \rangle$, quadratic in the tunneling operators.

From Eq.~\eqref{eq:It} we see that for vanishing voltage bias $\delta V$, we have $\omega_0=0$ and the current vanishes: $I=0$. Still, the noise $S$ contains non-equilibrium contributions due to the temperature difference. To extract this delta-$T$ noise, we choose zero bias $\omega_0=0$. To proceed, we use the vertex operator formula~\eqref{eq:BCH_Long} with the thermal propagators~\eqref{eq:single_boson_propagator} in Eq.~\eqref{eq:S}, and arrive at
\begin{align}
\label{eq:SwrittenOut}
    S&=\frac{4q^2|t_{\mathcal{T}}|^2}{(2\pi a)^M}
    \frac{(\pi a )^{2\Delta_{\mathcal{T}}} T_1^{2\Delta_A}T_2^{2\Delta_B}}{\prod_{i=1}^M v_i^{2\Delta_i}} \notag \\
    &\times \int dt \Big[ \frac{1}{\cosh \left(\pi T_1 t \right)} \Big]^{2\Delta_A} \Big[ \frac{1}{\cosh \left(\pi T_2 t \right)} \Big]^{2\Delta_ B}.
\end{align}
We see that the integrand is split into a product of the two sectors $A$ and $B$, with $\Delta_{A,B}$ being the scaling dimensions of $\Psi_{A,B}$. They are related to the total  scaling dimension of the tunneling operator as $\Delta_A+\Delta_B=\Delta_{\mathcal{T}}$.
Importantly, Eq.~(\ref{eq:SwrittenOut}) requires that each subsystem is in its thermal equilibrium state.

Next, we expand $S$ in small $\delta T/\overline{T}$, and find
\begin{equation}
\label{eq:deltaTS}
S \approx S_{\text{thermal}} \left[ 1 + \mathcal{C}^{(1)}\left(\frac{\delta T}{2\bar{T}}\right)+ \mathcal{C}^{(2)} \left( \frac{\delta T}{2 \bar{T}} \right)^2 \right].
\end{equation}
Here, 
\begin{align}
    S_{\text{thermal}}= 4g_\mathcal{T}(\bar{T})\bar{T}
    \label{eq:sth}
\end{align}
is the equilibrium thermal noise, expressed in terms of the average temperature $\bar{T}$ and the equilibrium tunneling conductance
\begin{align}
\label{eq:gtauGen}
g_\mathcal{T}(\bar{T}) \equiv q^2\frac{|t_{\mathcal{T}}|^2 \left(2\pi a \bar{T} \right)^{2\Delta_\mathcal{T}-2}}{ \prod_{i=1}^M v_i^{2\Delta_i}}  \frac{|\Gamma(\Delta_\mathcal{T})|^{2}}{\Gamma(2\Delta_\mathcal{T})}, 
\end{align}
with $\Gamma(z)$ being the Gamma function. For scattering of non-interacting electrons (having equal speeds $v$), we have $\Delta_{\mathcal{T}}=1$, $q=1$, and we recover Eq.~\eqref{eq:sthermal_fermion} upon identifying the transmission probability $D=|t_{\mathcal{T}}|^2 /v^2$. Note that the derivative of $S_\text{thermal}$ with respect to $\bar{T}$
vanishes at $\Delta_\mathcal{T}=1/2$, see Appendix~\ref{app:NJ} for a discussion.

In Eq.~\eqref{eq:deltaTS}, the leading-order coefficient is given as
\begin{align}
\mathcal{C}^{(1)} = (\Delta_A-\Delta_B)\times \frac{2\Delta_{\mathcal{T}}-1}{\Delta_{\mathcal{T}}}.
\end{align}
Importantly, since $\mathcal{C}^{(1)}\propto (\Delta_A-\Delta_B)$, it is non-zero only when the scaling dimensions of the two subsystems are distinct. Hence, no such term is present for the free bosons and fermions in Sec.~\ref{sec:Discussion}. However such a term is present, e.g., in inter-mode tunneling on a single FQH edge with counter-propagating channels. Such delta-$T$ noise was recently measured in Ref.~\cite{Melcer2022}. We note also that $C^{(1)}$ vanishes at $\Delta_{\mathcal{T}}=1/2$ in accordance with Ref.~\cite{RechMartinPRL20}. Physically, $\mathcal{C}^{(1)}$ reflects an asymmetry of the setup upon reversing the bias: $\delta T\rightarrow -\delta T$ [notice that the $\mathcal{C}^{(n)}$ coefficients are defined with the convention of $\delta T$ as in Eq.~\eqref{eq:deltaTDef}]. 

To avoid such sign effects from the bias direction, we assume in the following a symmetric setup with identical subsystems, so that $\Delta_A=\Delta_B$ and $\mathcal{C}^{(1)} = 0$. In this case, the leading-order contribution in Eq.~\eqref{eq:deltaTS} is $\mathcal{C}^{(2)}$ which is found as~\cite{RechMartinPRL20}
\begin{equation}
\mathcal{C}^{(2)} = \Delta_{\mathcal{T}} \left\{ \frac{\Delta_{\mathcal{T}}}{2 \Delta_{\mathcal{T}} + 1}  \left[  \frac{\pi^2}{2} - \psi'(\Delta_{\mathcal{T}}+1) \right] -1 \right\}.
\label{eq:c2}
\end{equation}
Here, $\psi'(z)$ is the derivative of the digamma function $\psi(z)$. It can readily be checked that $\mathcal{C}^{(2)}<0$ for $\Delta_{\mathcal{T}}<1/2$ and $\mathcal{C}^{(2)}>0$ for $\Delta_{\mathcal{T}}>1/2$.
In addition, the higher-order term $|\mathcal{C}^{(4)}|\ll| \mathcal{C}^{(2)}|$ is negligible when the tunneling is weak. These are the main observations in Ref.~\cite{RechMartinPRL20} for the case of a QPC device at FQH filling $\nu=1/(2n+1)$, for $n$ a positive integer. There, $\Delta_{\mathcal{T}}=\nu$ and $\Delta_{\mathcal{T}}=1/\nu$ for anyon respectively electron tunneling. Consequently, $\mathcal{C}^{(2)}$ is negative for anyon tunneling in the entire Laughlin series, whereas for electron tunneling  $\mathcal{C}^{(2)}>0$ in accordance with the result in Sec.~\ref{sec:FreeElectrons}. This highly non-trivial behavior of negative noise due to the strongly correlated nature of FQH states was speculated to be related to the anyonic statistics of the tunneling quasiparticles~\cite{RechMartinPRL20}.

At this point, we note that the $\mathcal{C}^{(n)}$ coefficients can be viewed as corresponding Fano factors for the delta-$T$ noise. Where the Fano factor for weak, voltage biased tunneling, reveals the charges of the tunneling particles, the $\mathcal{C}^(n)$ coefficients can be used to extract scaling dimensions\cite{Schiller2021} and (in some cases) the quantum statistics of the particles. 

We emphasize that Eq.~\eqref{eq:c2} is expressed in terms of the scaling dimension of the full tunneling operator~\eqref{eq:TunnelingOperators} (see also Ref.~\cite{Schiller2021}). A major observation in the present work is therefore that, in the weak tunneling limit, negative delta-$T$ noise will, in fact, be generated when two identical subsystems are bridged by a leading relevant inter-edge tunneling operator satisfying two conditions:
\begin{enumerate}[i)]
\item It transfers a finite charge $q\neq0$ between the two edges;
\item It has a scaling dimension $\Delta_{\mathcal{T}}<1/2$. 
\end{enumerate}
In particular, this derivation is applicable to non-chiral, infinite LLs as considered in Fig.~\ref{fig:ll_structure}\textcolor{blue}{(a)} with $g_{1\perp}=0$. There $\Delta_{\mathcal{T}}=(K+K^{-1})/2\geq 1$ and generates only positive delta-$T$ noise, in agreement with Ref.~\cite{RechMartinPRL20}. For the analysis of noise for tunneling between two semi-infinite LLs, see Appendix~\ref{sec:interacting_delta_T}.

We now view this result for infinite subsystems in light of Eq.~\eqref{eq:Theta_Delta_bound}. 
The special scaling dimension $\Delta_{\mathcal{T}}=1/2$ is, under the conditions of identical subsystems, each of which comprises co-propagating channels, precisely the point where the tunneling particles change from being more boson-like to fermion-like~\cite{BosonScaling}. 
The emergence of a negative delta-$T$ noise $\mathcal{C}^{(2)} < 0$ always accompanies a boson-like statistical angle $|\Theta_A|=|\Theta_B| < \pi/2$, following the bound given by Eq.~\eqref{eq:Theta_Delta_bound}.
A transition from positive to negative delta-$T$ noise should therefore be expected whenever the most relevant tunneling operator changes from being more fermion-like to more boson-like. This result is consistent with the delta-$T$ noise for free fermions and bosons (see Secs. \ref{sec:FreeElectrons} and \ref{sec:boson}), where the former only have positive delta-$T$ noise but the latter can have negative noise. We emphasize again that boson-like particles can generate delta-$T$ noise of both signs. However, for negative sign, the particles must be boson-like. 

It should be noted that the above correspondence only holds for setups with infinite channels with local interactions [cf. Fig.~\ref{fig:ll_structure}{\color{blue}(a)}]. In Appendix~\ref{sec:interacting_delta_T}, we analyze the delta-$T$ noise for semi-infinite LL wires, Fig.~\ref{fig:ll_structure}{\color{blue}(a)}, and show that the negative value of $\mathcal{C}^{(2)}<0$,
while still being achieved for $\Delta_\mathcal{T}<1/2$ according to Eq.~(\ref{eq:c2}), does not imply a boson-like statistics of the tunneling particles (they are still electrons transferred through a trivial vacuum in the weak-link setup). In the unfolded geometry, this setup is characterized by effectively nonlocal interactions, and the above consideration of the statistical phases becomes obscured. In this paper, however, we concentrate on the case of infinite channels as relevant to the quantum Hall and helical edges, where the correspondence following from Eq.~\eqref{eq:Theta_Delta_bound} takes place.

In the next Section, we verify and illustrate these results through a comprehensively study of delta-$T$ noise in QSH tunneling. As we shall see, during renormalization, the delta-$T$ noise may change its sign when the dominating tunneling processes changes from being more fermion-like to boson-like.

\section{Tunneling between quantum spin Hall edges}
\label{sec:model}

In this Section, we describe two models (Sec.~\ref{sec:ModelSetup}) of tunneling between QSH edges, depicted in Fig.~\ref{fig:qsh_kondo}. By bosonizing (Sec.~\ref{sec:transformations}) and a suitable basis transformation (Sec.~\ref{sec:chiralFields}) we prepare for a perturbative RG analysis of the noise, which we describe in Sec.~\ref{sec:Results}.

\subsection{Setups and Models}
\label{sec:ModelSetup}
Electron motion on QSH edges is helical: clockwise and counter-clockwise moving electrons have opposite spin projections due to strong spin-orbit coupling. Without loss of generality, we label clockwise-moving particles in leads 1 and 2 as ``$L$'' and ``$R$'', respectively. At low energies, interacting electrons on the QSH edge are described by the Luttinger Hamiltonian \cite{Wu2006,Xu2006,Maciejko2009,TanakaFurusakiMatveevPRL11}
\begin{equation}
\begin{aligned}
H_{\alpha} & = iv_F \int  dx \left[ \psi^{\dagger}_{\alpha,R} \partial_x \psi_{\alpha,R}^{}- \psi^{\dagger}_{\alpha,L} \partial_x \psi_{\alpha,L}^{} \right]\\
& + \frac{g_{4\parallel}}{2} \int dx \left[ \rho_{\alpha,L}\, \rho_{\alpha,L} + \rho_{\alpha,R}\, \rho_{\alpha,R} \right]\\
& +g_{2\perp} \int dx \left[ \rho_{\alpha,L}\, \rho_{\alpha,R}   \right]+\text{H.c.\,}.
\end{aligned}
\label{eq:disconnected_leads}
\end{equation}
Here, we have for simplicity neglected possible bulk Rashba spin-orbit interaction that induces spin rotation in chiral channels (for the discussion of effects of this rotation on transport in helical edges, see, e.g., Refs.~\cite{Schmidt2012, Kainaris2014, Kainaris2017,yevtushenko2022,YevtushenkoPRB21}).
In Eq.~\eqref{eq:disconnected_leads}, $\alpha = 1,2$ labels the two edges, $v_F$ is the Fermi velocity, $\rho_{\alpha,L}=\psi^{\dagger}_{\alpha,L}\psi^{}_{\alpha,L}$ and $\rho_{\alpha,R}=\psi^{\dagger}_{\alpha,R}\psi^{}_{\alpha,R}$ are the left- and right-moving density operators on edge $\alpha$. For a simpler notation, we leave the position-dependence of the operators implicit. Only two types of interactions $g_{4\parallel}$ and $g_{2\perp}$ are included, as dictated by time-reversal symmetry in this $S_z$-conserving model. The absence of $g_{2\parallel}$ and $g_{4\perp}$ in the standard ``g-ology'' framework~\cite{GiamarchiBook} is a key difference between the ideal interacting QSH edge and that of the standard Luttinger liquid.

We are interested in delta-$T$ noise in tunneling between these edges for the two types of tunneling. The first type is depicted in Fig.~\ref{fig:qsh_kondo}\textcolor{blue}{(a)}, where the edges are coupled at a single point, $x=0$, at which electrons may directly tunnel. This situation describes, e.g., a minimal model of a QSH QPC~\cite{Strunz2020}. The full Hamiltonian is then given by
\begin{equation}
H_{\text{direct}} = H_1 + H_2 + H_\mathcal{T},
\label{eq:h_former}
\end{equation}
with 
\begin{equation}
\begin{aligned}
H_\mathcal{T} & =\int dx\; \delta(x) \Bigg[ t \left( \psi^{\dagger}_{1,L} \psi_{2,R}^{} + \psi^{\dagger}_{1,R} \psi_{2,L}^{} + \text{H.c.} \right)\\
 & +  t_1'\left( \psi^{\dagger}_{1,L} \psi^{\dagger}_{1,R} \psi_{2,L} \psi_{2,R} + \text{H.c.}  \right) \\
& +  t_2'\left( \psi^{\dagger}_{1,L} \psi^{\dagger}_{2,L} \psi_{1,R}^{} \psi_{2,R}^{} + \text{H.c.}  \right) \Bigg].
\end{aligned}
\label{eq:direct_tunneling}
\end{equation}
In Eq.~\eqref{eq:direct_tunneling}, only spin-conserving operators are included. The operators in the first line, $\propto t$, describe single particle tunneling between the two edges, while operators in the last two lines, $\propto t_1', t_2'$, describe coherent tunneling of two electrons. 

As shown in Sec.~\ref{sec:direct_tunneling} below,  the two-particle tunnelings $t_1', t_2' \ll t$ are negligible in comparison to the single particle tunneling at high energies (still below the UV cutoff, where the Luttinger model is valid). However, for sufficiently strong interactions, the two-particle tunneling operators can become more RG-relevant [i.e., have a faster growth according to Eq.~\eqref{eq:renormalizationEquation}] at low energies. These processes then dominate over single-particle tunneling, which we will show changes the sign of the delta-$T$ noise.

As a second type of inter-edge tunneling, we consider the setup in Fig.~\ref{fig:qsh_kondo}\textcolor{blue}{(b)}. Here, in addition to the direct tunneling, the edges are also bridged by a single, localized, spin-$1/2$ impurity, through which electrons can be transferred between the edges by Kondo exchange. This setup describes tunneling via a localized electron or a quantum dot tuned to the Kondo valley. As follows, we refer to the localized spin as the dot-spin. The full Hamiltonian including this exchange tunneling reads
\begin{equation}
H_{\text{exchange}} = H_1 + H_2 + H_{\text{Kondo}},
\label{eq:h_latter}
\end{equation}
where
\begin{align}
H_{\text{Kondo}} & = \int\; dx \delta(x) \Bigg(J_1 \left[ \vec{s}_1 \cdot \vec{S} + \vec{s}_2 \cdot \vec{S} \right] + J_2 \vec{s}_{12} \cdot\vec{S}\Bigg) \notag \\
&+H_\mathcal{T}
\label{eq:ham_kondo}
\end{align}
is the Kondo Hamiltonian, where $J_1$ and $J_2$ have the same dimension: energy (coupling strength) $\times$ length (effective size of the impurity). The first term in Eq.~\eqref{eq:ham_kondo} describes backscattering (with coupling strength $J_1$) and forward scattering (coupling strength $J_2$) via the dot spin $\vec{S}$, in terms of the spin density operators
\begin{equation}
\begin{aligned}
  &\vec{s}_{1}=\sum_{j=L,R} \psi^{\dagger}_{1,j}\vec{\sigma} \psi_{1,j} + \text{H.c.\,},\\
  & \vec{s}_{2}=\sum_{j=L,R} \psi^{\dagger}_{2,j}\vec{\sigma} \psi_{2,j} + \text{H.c.\,},\\
  & \vec{s}_{12} =\sum_{j=L,R} \psi^{\dagger}_{1,j}\vec{\sigma} \psi_{2,j} + \text{H.c.\,}.
\end{aligned}
\label{eq:spin_density_ops}
\end{equation}
Here, $\vec{\sigma}$ is the vector of Pauli matrices. 

In Eq.~\eqref{eq:ham_kondo}, we assume for simplicity equal forward scattering and backscattering coupling constants $J_1 = J_2$ at high energies. It is, however, known (see, e.g., Ref.~\cite{FurusakiNagaosaPRB93}) that $J_1$ and $J_2$ depart at lower energies, because of different scaling behavior during the RG flow~\eqref{eq:renormalizationEquation}.
We have also assumed an ideal spin-momentum locking, i.e., an isotropic impurity-lead Kondo exchange.
As will be shown in detail in Sec.\,\ref{sec:konod}, an anisotropy will not change the delta-$T$ noise in the regime of interest, as $J_1^z$ flows to its fixed-point value independently of its initial value.

\subsection{Bosonization and chiral representation}
\label{sec:transformations}

As with other Luttinger liquid systems, a powerful approach to study  Hamiltonians of the type~\eqref{eq:h_former} and~\eqref{eq:h_latter} is bosonization. Within this formalism, we express fermions via bosonic fields with the following bosonization convention~\cite{GiamarchiBook}:
\begin{equation}
\begin{aligned}
\psi^{\dagger}_{\alpha, L} & = \frac{F_{\alpha, L}}{\sqrt{2\pi a}} e^{-i( \phi_{\alpha} +  \theta_{\alpha} - k_F x)},\\
\psi^{\dagger}_{\alpha, R} & = \frac{F_{\alpha,R}}{\sqrt{2\pi a}} e^{i ( \phi_{\alpha} - \theta_{\alpha} - k_F x) }.
\end{aligned}
\label{eq:bosonization_convention}
\end{equation}
Here, $a$ is a short distance cutoff, $k_F$ is the Fermi wave vector, $F_{\alpha,L,R}$ are Klein factors, and $\phi_\alpha$ and $\theta_{\alpha}$ are bosonic fields obeying the equal time commutation relations
\begin{align}
\label{eq:Can_Comm}
\left[ \phi_{\alpha}(x), \partial_x \theta_{\alpha'} (x') \right] = i\pi \delta_{\alpha,\alpha'} \delta (x - x').
\end{align}
With~\eqref{eq:bosonization_convention}, the free Hamiltonians~\eqref{eq:disconnected_leads} become quadratic in terms of bosonic modes:
\begin{equation}
H_1 \!+\! H_2\! =\! \frac{u}{2\pi} \int\! dx\! \sum_{\alpha = 1,2 } \left[ K\left( \partial_x \theta_{\alpha} \right)^2 \!+\! \frac{1}{K} \left( \partial_x \phi_{\alpha} \right)^2 \right],
\end{equation}
where
\begin{equation}
u = v_F K
\end{equation}
is the dressed velocity, and
\begin{equation}
\label{eq:LuttingerK}
K= \sqrt{\frac{v_F - (2 g_{2\perp} - g_{4\parallel})/\pi}{v_F + (2 g_{2\perp} + g_{4\parallel})/\pi}}
\end{equation}
is the Luttinger parameter. We assume $K$ to be equal for the two helical edges. For repulsive and attractive interactions, $K<1$ and $K>1$, respectively, whereas vanishing interactions corresponds to $K=1$.

The bosonization rule~\eqref{eq:bosonization_convention} defines two pairs of canonical bosons, as manifested by the relations~\eqref{eq:Can_Comm}. In the present context, it is however more convenient to work with two chiral fields, defined as
\begin{equation}
\begin{aligned}
\tilde{\phi}_{\alpha,L} (x) & = K\theta_{\alpha} (-x) + \phi_{\alpha} (-x),\\
\tilde{\phi}_{\alpha,R} (x)& = K\theta_{\alpha} (x) - \phi_{\alpha} (x),
\end{aligned}
\label{eq:chiral_fields}
\end{equation}
obtained by a reflection $x\to -x$ for the left-moving fields. Then, all fields propagate to the right. This reflection is only permissible if two conditions are fulfilled. (i) The edges are coupled only at a single point~\cite{GiamarchiBook}. This condition is satisfied by construction in our model. (ii) All chiral fields are independent. This can be readily checked by using Eqs.~\eqref{eq:Can_Comm}~and~\eqref{eq:chiral_fields} to compute the commutators
\begin{equation}
\left[ \tilde{\phi}_{\alpha, \sigma} (x) ,\tilde{\phi}_{\alpha', \sigma'}(x') \right] =\delta_{\alpha,\alpha'} \delta_{\sigma,\sigma'} i\pi K \text{sgn} (x - x'),
\label{eq:commutator_chiral}
\end{equation}
showing that the fields are indeed independent \cite{FieldRotation}.
Importantly, this reflection changes the statistical angle of tunneling operators (see Sec.~\ref{sec:Discussion}). Since after the reflection, all channels now share the same chirality, the statistical angles of the tunneling operators are fixed at its maximum bound~\eqref{eq:Theta_Delta_bound}. We will revisit this point in more detail below.

In terms of the chiral bosonic operators~\eqref{eq:chiral_fields}, the free Hamiltonian becomes
\begin{equation}
\begin{aligned}
H_1 + H_2 & = \frac{u}{4\pi K } \int dx \left[(\partial_x \tilde{\phi}_{1,L})^2 + (\partial_x \tilde{\phi}_{2,L})^2 \right. \\
&\ \ \ \ \ \ \ \  \left. + (\partial_x \tilde{\phi}_{1,R})^2 + (\partial_x \tilde{\phi}_{2,R})^2\right],
\end{aligned}
\label{eq:ham_before_rotation}
\end{equation}
which is precisely the Hamiltonian of two spinless, non-chiral LL wires.
Equation~\eqref{eq:ham_before_rotation} indicates that $\lambda_i = K$ for all four fields, following the formalism described in Sec.~\ref{sec:stat_and_ops}.

The difference between a description of such wires and QSH edges emerges when one considers electron tunneling: since particles with the same spin travel in opposite directions on the two edges, only backward scattering $L\leftrightarrow R$ is allowed between the QSH edges.
For tunneling between spinless LL wires, forward scattering processes $L\leftrightarrow L$ and $R\leftrightarrow R$ are possible.

\subsection{Physical chiral fields}
\label{sec:chiralFields}

In Eq.~\eqref{eq:chiral_fields}, we defined four independent bosonic chiral fields. In terms of these fields, the edge Hamiltonian bears a close resemblance to that of a FQH system, with an effective ``filling factor'' depending on $K$. However, these fields are not very convenient for an RG procedure that is used to determine what  operators are most relevant. To simplify the following discussion, and render the physics more transparent, we define a new set of chiral fields:
\begin{equation}
\begin{aligned}
 \Phi_\text{c} \! &=\! \frac{1}{2K} [\tilde{\phi}_{1,L} \!+\! \tilde{\phi}_{1,R} \!+\! \tilde{\phi}_{2,L} \!+\! \tilde{\phi}_{2,R}],\\ 
\Phi_\text{cf} \!&= \!\frac{1}{2K} [\tilde{\phi}_{1,L}\! +\! \tilde{\phi}_{1,R}\! -\! \tilde{\phi}_{2,L}\! -\! \tilde{\phi}_{2,R} ], \\
 \Phi_\text{s}  \!&= \!\frac{1}{2} [\tilde{\phi}_{1,L}  \!-\! \tilde{\phi}_{1,R}  \!-\! \tilde{\phi}_{2,L} \!+\! \tilde{\phi}_{2,R} ],\\ 
 \Phi_\text{sf}\! &=\! \frac{1}{2} [\tilde{\phi}_{1,L} \! -\! \tilde{\phi}_{1,R}  \!+\! \tilde{\phi}_{2,L}  \!-\! \tilde{\phi}_{2,R} ].
\end{aligned}
\label{eq:rotations_to_new_field}
\end{equation}
These fields have clear physical meanings: $\Phi_\text{c}$ and $\Phi_\text{s}$ describe charge and spin variations of the dot, while $\Phi_\text{cf}$ and $\Phi_\text{sf}$ describe charge and spin transfer between the edges.
Specifically, by considering the fermionic density operators 
$$\rho_\alpha = \psi^\dagger_\alpha \psi_\alpha,$$ 
we see that 
$$\partial_x \Phi_\text{c} = \frac{\pi}{K} ( \rho_{1,L} + \rho_{1,R} + \rho_{2,L} + \rho_{2,R} ) =  \frac{\pi}{K} \rho,$$ 
i.e., it is, up to a prefactor, equal to the total charge density $\rho$.
Similarly, the spin field is related to the $z$ component of spin density operators defined in Eq.~\eqref{eq:spin_density_ops}. We define the total spin density $$s^z \equiv s_1^z + s_2^z,$$ 
and then 
$$ \partial_x \Phi_\text{s} = \rho_{1,L} - \rho_{1,R} - \rho_{2,L} + \rho_{2,R} = \frac{\pi}{K} s^z. $$
The definitions of the charge and spin density operators, in combination with Eq.~\eqref{eq:rho_operator}, imply that $t_c = 2K$ for charge density, and $t_s = 2/ K$ for the spin density.

The charge fields $\Phi_\text{c}$ and $\Phi_\text{cf}$ obey the commutation relations
\begin{equation}
\begin{aligned}
\left[ \Phi_\text{c}(x),  \Phi_\text{c} (x') \right] \! =\! \left[ \Phi_\text{cf}(x),  \Phi_\text{cf} (x') \right] \!=\! \frac{i\pi}{K} \text{sgn} (x \!-\! x') ,
\end{aligned}
\label{eq:com_c}
\end{equation}
while the spin fields $\Phi_\text{s}$ and $\Phi_\text{sf}$ obey
\begin{equation}
\begin{aligned}
\left[ \Phi_\text{s}(x),  \Phi_\text{s} (x') \right] \!=\! \left[ \Phi_\text{sf}(x),  \Phi_\text{sf} (x') \right]\! =\! i\pi K \text{sgn} (x \!-\! x').
\end{aligned}
\label{eq:com_s}
\end{equation}
Similar to the chiral fields in Eq.~\eqref{eq:chiral_fields}, the four new fields are independent, and all commutators between different fields vanish.

In terms of the new fields, the edge Hamiltonian becomes
\begin{equation}
\begin{aligned}
H_1 + H_2 
& = \frac{u}{4\pi } \int dx \left[K(\partial_x \Phi_\text{c})^2 + K(\partial_x \Phi_\text{cf})^2 \right. \\
& \ \ \ \ \ \ \ \ + \left. \frac{1}{K}(\partial_x \Phi_\text{s})^2 +\frac{1}{K} (\partial_x \Phi_\text{sf})^2\right].
\end{aligned}
\label{eq:ham_after_rotation}
\end{equation}
Notably, the interaction between fields in the charge sector act inversely compared to those in the spin sector. As a consequence, for attractive electronic interactions $K>1$, the bosonic fields $\Phi_\text{c}$ and $\Phi_\text{cf}$ experience effectively repulsive interactions. In comparison to the general results in Sec.~\ref{sec:stat_and_ops}, we see that $\lambda_\text{c} = 1/K$ for charge (c) and charge flavor (cf) fields, while $\lambda_\text{s} = K$ for spin (s) and spin flavor (sf) fields.
This fact, in combination with Eqs.~\eqref{eq:rho_operator} and \eqref{eq:charge_qj}, indicates that vertex operators $\exp[i\Phi_\text{c}/2]$ and $\exp[i\Phi_\text{s}/2]$  remove a charge and a spin, respectively, from the system. Similar analysis can be done for the charge flavor and spin flavor fields. 
The tunneling Hamiltonians~\eqref{eq:direct_tunneling} and~\eqref{eq:ham_kondo} are expressed in terms of the new fields in the next Section, where we compute the delta-$T$ noise in the QSH tunneling models.

\section{Delta-\textit{T} noise in quantum spin Hall tunneling}
\label{sec:Results}
\subsection{Direct tunneling}
\label{sec:direct_tunneling}
In the bosonized language~\eqref{eq:bosonization_convention}, with the fields~\eqref{eq:rotations_to_new_field}, the single particle tunneling contribution in~\eqref{eq:direct_tunneling} becomes
\begin{equation}
 \frac{t}{2\pi a}\! \left[ F_{1R} F_{2L} e^{i(-\Phi_\text{cf}+ \Phi_\text{sf})}\! +\! F_{1L} F_{2R}  e^{- i(\Phi_\text{cf} + \Phi_\text{sf})} \!+\! \text{H.c.}\right].
\label{eq:direct_tun_boson}
\end{equation}
The scaling dimensions of all terms in Eq.~\eqref{eq:direct_tun_boson} are 
$$\Delta_t=(K + 1/K)/2 \geq 1$$ 
and are therefore irrelevant, unless $K= 1$, where they are marginal [see Eq.~\eqref{eq:renormalizationEquation}]. By using $\Delta_t$ in Eq.~\eqref{eq:c2}, we see that single electron tunneling generates positive delta-$T$ noise, in agreement with Sec.~\ref{sec:FreeElectrons} and Ref.~\cite{RechMartinPRL20}. 

Next, we consider the coherent electron-pair tunneling terms in \eqref{eq:second_order_tunneling}, which we write as
\begin{equation}
\begin{aligned}
&   t_1' \psi^{\dagger}_{1,L} \psi^{\dagger}_{1,R} \psi_{2,L} \psi_{2,R} + t_2'\psi^{\dagger}_{1,L} \psi^{\dagger}_{2,L} \psi_{1,R} \psi_{2,R} + \text{H.c.\,} \\
& =  \frac{2t_1'}{(2\pi a)^2}\cos(2 \Phi_\text{cf}) + \frac{2t_2'}{(2\pi a)^2}\cos(2 \Phi_\text{sf})
\end{aligned}
\label{eq:second_order_tunneling}
\end{equation}
upon omitting Klein factors (they are not important at this stage). Here, the operator $\propto t_1'$ describes coherent tunneling of two electrons (with opposite spins) from one edge to the other. The operator proportional to $t_2'$ describes the coherent exchange of two electrons (with opposite spins) between the edges. 

To relate these expressions to those in Sec.~\ref{sec:stat_and_ops}, we take the vertex operator $\exp(2i\Phi_\text{cf})$ as an example, and express the charge-flavor field as
\begin{equation}
    2\Phi_\text{cf} = \frac{1}{K} \left( \tilde{\phi}_{1,L} + \tilde{\phi}_{1,R} - \tilde{\phi}_{2,L} - \tilde{\phi}_{2,R} \right),
\label{eq:vertex_decomp}
\end{equation}
in terms of the fields defined in Eqs.~\eqref{eq:chiral_fields}. With these fields, the Hamiltonian Eq.~\eqref{eq:ham_before_rotation} and the commutators Eq.~\eqref{eq:chiral_fields} follow the standard form presented in Sec.~\ref{sec:stat_and_ops}. The corresponding density operator
\begin{equation}
\begin{aligned}
    \rho= -\frac{1}{2\pi} \partial_x (\tilde{\phi}_{1,L} + \tilde{\phi}_{1,R} - \tilde{\phi}_{2,L} - \tilde{\phi}_{2,R}),
\end{aligned}
\end{equation}
involves the density difference between two edges, and is of the form of Eq.~\eqref{eq:rho_operator}.
In this case, the vertex operator with the field~\eqref{eq:vertex_decomp} can be considered within the standard form of Sec.~\ref{sec:stat_and_ops} with $\lambda_i = K$, $m_{1,L} = m_{1,R} = 1/K$, and $m_{2,L} = m_{2,R} = -1/K$. By using Eq.~\eqref{eq:primary_field_scaling}, we identify the scaling dimension of $2\cos(\Phi_\text{cf})$ as $\Delta_\text{cf}=2/K$.
The corresponding statistical angles for the tunneling quasiparticles are
$$\Theta_A = \Theta_B = 0 < \pi/2.$$ 
Indeed, in the present case, Eq.~\eqref{eq:theta_ab} yields  the total exchange phase for a product of two vertex operators for counterpropagating modes in each of the subsystems A and B alike. As a result, it involves a difference $$\pi\left(\lambda_{1,R} m_{1,R}^2-\lambda_{1,L}m_{1,R}^2\right)=\pi/K-\pi/K=0$$
for $\Theta_A$, and similarly for $\Theta_B$.
The direct tunneling model is thus an example where a negative delta-$T$ noise accompanies a boson-like tunneling operator.

The scaling dimension of the charge flavor field can also be analysed directly with the Hamiltonian~\eqref{eq:ham_after_rotation} and the commutators~\eqref{eq:com_c}, leading to the same result. In a similar manner, the second term in Eq.~\eqref{eq:second_order_tunneling} involves only the spin flavor field $\Phi_\text{sf}$, which has scaling dimension $\Delta_\text{sf}=2K$. In the absence of interactions, $K=1$, all coherent pair tunneling operators have scaling dimension $2$, and are thus less relevant than the direct tunneling \eqref{eq:direct_tun_boson}, which for $K=1$ has scaling dimension $1$.

However, for sufficiently strong interactions, $K<1/4$ or $K>4$, $t_1'$ or $t_2'$ dominate the tunneling at low enough temperatures. In this case, negative delta-$T$ noise emerges in the parameter regimes $K<1/4$ and $K>4$. We emphasize here that $K$ is not renormalized for point tunneling~\cite{KaneFisherPRB92}.
Noteworthily, although a negative delta-$T$ noise only emerges in the presence of sufficiently strong interactions, two-electron tunneling is always boson-like (regardless of the interaction strength), as it involves the coherent tunneling of two electrons. This indicates that boson-like tunneling does not guarantee a negative delta-$T$ noise, although negative delta-$T$ noise is always due to boson-like quasiparticle operators. 

It is instructive to analyze other possible tunneling operators that could contribute to the delta-$T$ noise. First, we note that descendants of tunneling operators (in the conformal field theory sense, e.g., derivatives~\cite{Francesco2012}) have scaling dimensions larger than one and can only produce positive delta-$T$ noise. Such operators can therefore be neglected in comparison to terms involving the tunneling operators considered above. Similarly, we neglect density operators, which can be considered as first descendants of bosonic operators.

Second, we consider generic multi-particle tunneling operators. Such an operator can be written as
\begin{equation}
\begin{aligned}
    \mathcal{T}_\text{generic} &= (\psi^\dagger_{1,L})^{n_{1,L}} (\psi^\dagger_{2,L})^{n_{2,L}} (\psi^\dagger_{1,R})^{n_{1,R}} (\psi^\dagger_{2,R})^{n_{2,R}}\\
    &\times (\psi_{1,L})^{m_{1,L}} (\psi_{2,L})^{m_{2,L}} (\psi_{1,R})^{m_{1,R}} (\psi_{2,R})^{m_{2,R}},
\end{aligned}
\label{eq:tunneling_higher-orders}
\end{equation}
where $n_\alpha$ and $m_\alpha$ are positive integers labelling the number of involved creation respectively annihilation operators of fermions of type $\alpha$.
For later convenience, we define their difference $\mathcal{N}_\alpha = n_\alpha - m_\alpha$. Then, $\mathcal{N}_\alpha > 0$ implies a net creation of particles of type $\alpha$. In contrast, a negative $\mathcal{N}_\alpha$ refers to a net removal of such particles.
From charge and spin conservation, we must however have 
$\sum_\alpha \mathcal{N}_\alpha = 0$, and $\mathcal{N}_{1,L} - \mathcal{N}_{1,R} - \mathcal{N}_{2,L} + \mathcal{N}_{2,R} =0$.
In addition, we can discard the cases $\mathcal{N}_{1,L} = \mathcal{N}_{1,R} = 0$ or $\mathcal{N}_{2,L} = \mathcal{N}_{2,R} = 0$, since spin-flip tunneling within the same edge is forbidden on the QSH edge. With these conservation requirements, we can simplify Eq.~\eqref{eq:tunneling_higher-orders} in bosonized form as
\begin{equation}
\begin{aligned}
    \mathcal{T}_\text{generic}&\propto \exp [i (\mathcal{N}_{1,L}+\mathcal{N}_{1,R}) \Phi_\text{cf} \! + i(\mathcal{N}_{1,L}-\mathcal{N}_{1,R}) \Phi_\text{sf}].
\end{aligned}
\end{equation}
From Eqs.~\eqref{eq:Local_Scaling} and~\eqref{eq:primary_field_scaling} we find that this operator has the scaling dimension
\begin{align}
   \Delta_\text{generic} = \frac{(\mathcal{N}_{1,L} + \mathcal{N}_{1,R} )^2}{2K} + \frac{(\mathcal{N}_{1,L} - \mathcal{N}_{1,R} )^2 K}{2}.
\end{align}

As a consistency check, according to this formula, single-particle tunneling operators have $\mathcal{N}_{1,L} = 0$, $\mathcal{N}_{1,R} = \pm 1$ or vice versa. The scaling dimension is then $(K+K^{-1})/2$ as it should. Two-particle tunnelings have $\mathcal{N}_{1,L}, \mathcal{N}_{1,R} = \pm 1$ and the scaling dimension is $2K$ or $2/K$. For higher-order tunnelings, either $|\mathcal{N}_{1,L} + \mathcal{N}_{1,R}|$ or $|\mathcal{N}_{1,L} - \mathcal{N}_{1,R}|$ is larger than or equal three. As a consequence, higher-order operators always have scaling dimensions either larger than $9K/2$ or $9/(2K)$. They are, therefore, always less relevant in comparison to the two-particle operators. We can thus safely focus on only single- and double-electron tunnelings in the following.

\begin{figure*}[ht!]
  \centering
    \includegraphics[width=1\textwidth]{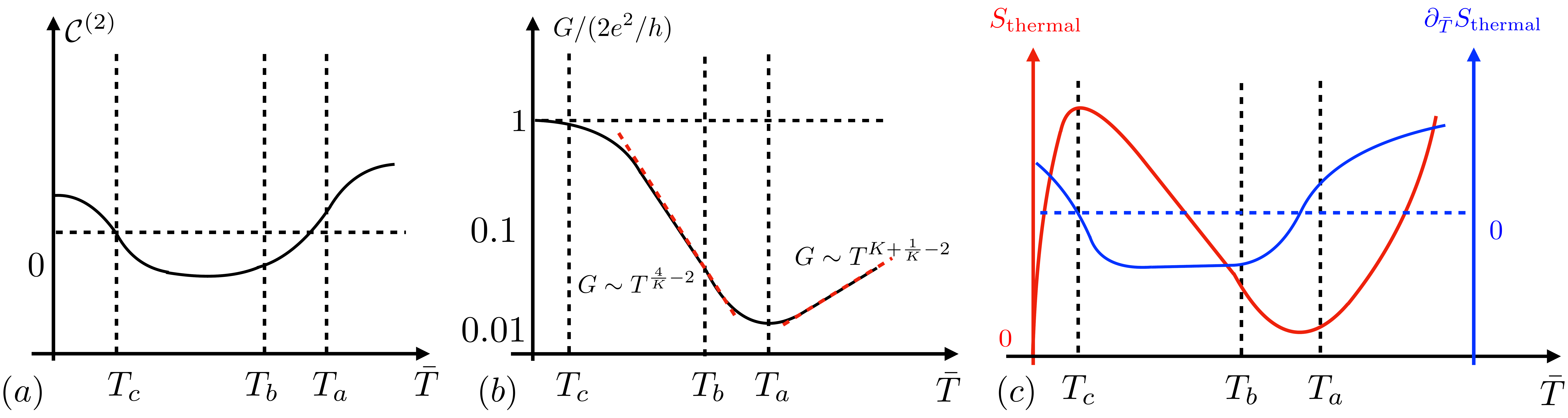}
    \caption{Sketches of (a) the noise coefficient $\mathcal{C}^{(2)}$ [see Eq.~\eqref{eq:c2}] with varying average temperature $\overline{T}$, (b) the linear charge conductance $G$ (in a log-log form), and (c) The thermal noise $S_\text{thermal}$ (red curve) and its derivative over temperature $\partial_{\bar{T}} S_\text{thermal}$ (blue curve).
    Here, $T_a$ represents the temperature with the lowest conductance, or the transition temperature between regimes with dominating single-electron and coherent two-electron tunnelings.
    The other two temperatures $T_b$, and $T_c$ denote the characteristic temperature scales where, with decreasing temperature,  conductance power laws begin to be determined by coherent pair tunneling and backscattering, respectively.
    The sketches are drawn for strongly attractive interactions, $K>4$.}
    \label{fig:schematics_former}
\end{figure*}

To support our argument of possible negative noise from two-electron tunneling, we next compare the delta-$T$ noise with the equilibrium charge tunneling conductance $G=2g_{\mathcal{T}} e^2/h$ [here, $g_{\mathcal{T}}$ is obtained from Eq.~\eqref{eq:gt} using for $\mathcal{T}$ the coherent electron-pair tunneling operators in Eq.~\eqref{eq:second_order_tunneling}], as depicted in Fig.~\ref{fig:schematics_former}. We consider a strongly attractive system with $K>4$. Then, we have $\Delta_\text{cf}<1/2$ and the dominant coherent pair tunneling generates negative delta-$T$ noise according to Eq.~\eqref{eq:c2}, with $\Delta_{\mathcal{T}} = \Delta_\text{cf}$. However, at high temperatures, where $t\gg t_1'$, this negative contribution is smaller than the positive delta-$T$ noise from direct tunneling $\propto t$. We thus anticipate a sign change of $\mathcal{C}^{(2)}$ [high-temperature regime of Fig.~\ref{fig:schematics_former}\textcolor{blue}{(a)}] with decreasing $\overline{T}$.

This can be compared to the zero-voltage-bias (for $T_1 = T_2 = \bar{T}$) charge conductance $G$, sketched in 
Fig.~\ref{fig:schematics_former}\textcolor{blue}{(b)}, which acts as a benchmark for the non-trivial negative delta-$T$ noise in the strongly attractive situation $K>4$.
At high temperatures $\bar{T} >T_a$, with $T_a$ a characteristic temperature scale for dominating direct tunneling, the conductance scales as $G \sim T^{K+1/K - 2}$. Adding a  small $\delta T$, the delta-$T$ noise is positive in this regime.

The conductance starts to deviate from the high-temperature power-law, when $t_1'$ increases during the RG flow. At temperatures $\bar{T}\sim T_a$, $t_1'$ becomes comparable to the direct tunneling, and $G$ takes its minimum value. The value of $t_1'$ keeps growing with decreasing temperature and changes the sign of $\mathcal{C}^{(2)}$ for $T_b<T<T_a$. The coherent pair tunneling starts to dominate at the temperature scale $\overline{T}\sim T_b$, where the conductance scaling changes to $G\sim T^{4/K-2}$.
Finally, at sufficiently low temperatures, $\overline{T}\sim T_c$, the system approaches the strong-tunneling limit, where the leading tunneling operator becomes irrelevant. The temperature window with a negative delta-$T$ noise then closes at $T_c$

The thermal noise, related to the conductance through Eq.~(\ref{eq:sth}), is another possible benchmark, as shown in Fig.~\ref{fig:schematics_former}{\color{blue}(c)}.
Indeed, the temperature derivative of the thermal noise, i.e., $\partial_{\bar{T}} S_\text{thermal}$ has a structure quite similar (although not identical)~\cite{Khrapai} to that of the delta-$T$ noise in Fig.\,\ref{fig:schematics_former}(a). 
In particular, we observe that the zeros of the delta-$T$ correspond to extrema of the thermal noise as a function of temperature. The nonmonotonous $T$ dependence of the thermal noise in Fig.~\ref{fig:schematics_former}{\color{blue}(c)} can be traced back to LL renormalization of the effective transmission coefficient. In the presence of interactions, it depends on both energy and temperature, and is governed by different scaling exponents in different temperature ranges.  
We emphasize that scaling dimensions determine both ``emergent statistics'' of the dominant operators and power-law renormalization of transmission. The thermal noise reveals the latter, while the delta-$T$ noise reveals both.
The difference and connection between the thermal noise temperature derivative and the delta-$T$ noise are discussed in more detail in Appendix~\ref{app:NJ}.

As already mentioned, alternatively, negative delta-$T$ noise is also expected to emerge in strongly repulsive systems $K<1/4$. However, in this case, the noise is instead correlated with a spin tunneling conductance, and is thus experimentally less accessible.
From the above analysis, we conclude that, in the direct tunneling setup, negative delta-$T$ noise can be generated by coherent pair tunneling for sufficiently strong interactions. In Table~\ref{tab:Noise_Table}, we list the leading relevant operator couplings and the sign of the delta-$T$ noise in various regimes governed by the interaction strength $K$.

\begin{table}[h!]
\centering
\begin{tabular}{ | c | c | c | c | c | c | }
 \hline
  & $K<\frac{1}{4}$ & $\frac{1}{4}\!<\!K\!<\!\frac{1}{\sqrt{3}}$ & $\frac{1}{\sqrt{3}}\! <\!K\! <\! \sqrt{3}$ & $\sqrt{3}\! <\! K\! <\! 4$ & $K>4$  \\ 
  \hline
LO& $t_2'$ & $t_2'$ & $t$ & $t_1'$ &  $t_1'$  \\  
 \hline
sign & $\pm$ & +   & + & + & $\pm$ \\ [1ex] 
 \hline
\end{tabular}
\caption{Leading-operator couplings (LO) and the sign of the delta-$T$ noise (sign) as functions of the interaction $K$. Negative delta-$T$ noise is possible for interaction ranges with $\pm$ entries. The points $K = \sqrt{3}$ or $ 1/\sqrt{3}$ are those where the scaling dimensions of coherent pair tunneling and single-particle tunneling coincide, i.e. they are the solutions of $(K + 1/K)/2 = 2/K$ respectively $(K + 1/K)/2 = 2K$.}
\label{tab:Noise_Table}
\end{table}

\subsection{Kondo tunneling}
\label{sec:konod}
Let us now turn to the analysis of delta-$T$ noise for the setup that involves a Kondo dot between the helical edges, as shown in Fig.~\ref{fig:qsh_kondo}{\color{blue}(b)}.
After bosonization with Eq.~\eqref{eq:bosonization_convention} and in the basis of physical fields~\eqref{eq:rotations_to_new_field}, the Kondo Hamiltonian \eqref{eq:ham_kondo} reads
\begin{equation}
\begin{aligned}
& H_{\text{Kondo}} \! =\! 
\frac{J_1^z S_z}{\pi} \partial_x \Phi_\text{s}+ H_\mathcal{T} \\
& \!+\!\frac{J_{1}^{\perp}}{2\pi a} \left[ F_{1,L} F_{1,R} e^{-i(\Phi_\text{s} + \Phi_\text{sf})} S_-  + \text{H.c.}\right]\\
& \!+\! \frac{J_{1}^{\perp}}{2\pi a}\! \left[ F_{2,R} F_{2,L} e^{-i(\Phi_\text{s} - \Phi_\text{sf})} S_-  + \text{H.c.}\right]\\
& \!+\! \frac{J_{2}^{\perp}}{2\pi a}\! \left[ F_{1,L} F_{2,L} e^{-i(\Phi_\text{s} + \Phi_\text{cf})} S_-  + \text{H.c.}\right]\\
& \!+\! \frac{J_{2}^{\perp}}{2\pi a}\! \left[ F_{2,R} F_{1,L} e^{-i(\Phi_\text{s} - \Phi_\text{cf})} S_-  + \text{H.c.}\right]\\
& \!+\! \frac{J_2^z}{2\pi a} \!\left[ F_{1R} F_{2L} e^{i(\Phi_\text{cf} - \Phi_\text{sf})} \!+ \!F_{2R} F_{1L} e^{i(\Phi_\text{cf} + \Phi_\text{sf})}\! +\! \text{H.c.}\right]\! S_z 
.
\end{aligned}
\label{eq:h_kondo}
\end{equation}
Here, we have decomposed the dot spin as $S_\pm = S_x+iS_y$ and we have further distinguished the perpendicular components $J^{\perp}_{1,2}$ (coupling to $S_\pm$) from the $z$-components $J_{1,2}^z$ (coupling to $S_z$), since the latter terms scale differently under the RG flow [see Eq.~\eqref{eq:complete_RG} below]. As we will show next, the inter-edge Kondo operators ($\propto J^{\perp}_2, J_2^z$) and the direct tunneling of $H_\mathcal{T}$ together determine the sign of the delta-$T$ noise. The backscattering terms $\propto J^{\perp}_1$ do not involve charge transport between the edges, and therefore do not contribute to the delta-$T$ noise.

In contrast to the direct tunneling terms~\eqref{eq:direct_tun_boson} and \eqref{eq:second_order_tunneling}, operators proportional to $J^{\perp}_1$ and $J^{\perp}_2$ involve the spin field $\Phi_\text{s}$ and thus couple to the dot spin fluctuations. Initially, the inter-lead Kondo exchange ($\propto J^{\perp}_2$) has the same scaling dimension, $(K + 1/K)/2$, as that of the direct tunneling, and generates positive delta-$T$ noise at high temperatures. However, since there are only two spin states in the dot, the spin field $\Phi_\text{s}$ gradually loses its dynamics during the RG flow, leading to an increasingly relevant inter-edge Kondo exchange.

To derive RG equations of the QSH Kondo model to second order in $J_{1,2}^{\perp}$ and first order in $t$, we use the Coulomb-gas RG technique~\cite{SchillerIngersent97} (see Appendix~\ref{sec:AppendixCoulomb} for details). We find the following set of RG equations:
\begin{equation}
\begin{aligned}
& \frac{d\tilde{J}_1}{d\ell}  = \left[ 1 - \frac{K}{2} - \left( 1 - 2\tilde{J}_1^z \right)^2 \frac{K}{2}  \right] \tilde{J}_1 + 4\tilde{J}_2 \tilde{J}_2^z,\\
&\frac{d\tilde{J}_2}{d\ell}  = \left[ 1 - \frac{1}{2K} - \left( 1 - 2\tilde{J}_1^z \right)^2 \frac{K}{2}  \right] \tilde{J}_2 + 4\tilde{J}_2 \tilde{J}_1^z,\\
& \frac{d}{d\ell}  \left( 1 - 2\tilde{J}_1^z \right) = -2  \left( 1 - 2\tilde{J}_1^z \right) (2 \tilde{J}_1^2 + 2 \tilde{J}_2^2), \\
& \frac{d \tilde{t}}{d\ell} = \left( 1 - \frac{1}{2K} - \frac{K}{2} \right) \tilde{t},\\
& \frac{d \tilde{J}_2^z}{d\ell} = \left( 1 - \frac{1}{2K} - \frac{K}{2} \right) \tilde{J}_2^z + 4 \tilde{J}_1\tilde{J}_2.
\end{aligned}
\label{eq:complete_RG}
\end{equation}
Here, we omitted the flow of $t_1'$ and $t_2'$ as we find them to always be less relevant than the Kondo couplings. In Eq.~\eqref{eq:complete_RG}, we have defined dimensionless coupling parameters $\tilde{J}_1 = J_1^{\perp} /J_0$, $\tilde{J}_2 = J_2^{\perp}/J_0$, $\tilde{t} = t/J_0$, $\tilde{J}_1^z = J_1^z /J_0$, and $\tilde{J}_2^z = J_2^z/J_0$, which are all normalized by $J_0= \pi u$.

From Eq.~\eqref{eq:complete_RG}, we see that at high energies, the intra-edge ($\propto \tilde{J}_1$) and inter-edge  ($\propto \tilde{J}_2$) Kondo exchanges have scaling dimensions $K$ and $(K+1/K)/2$, respectively, when the $z$-component $\tilde{J}_1^z$ is small.
During the RG flow, $\tilde{J}_1^z$ gradually increases to finally reach the point $\tilde{J}_1^z = 1/2$ [see the third line of Eq.~\eqref{eq:complete_RG}].
We emphasize that this fixed point is valid beyond the second-order perturbative beta-functions of Eq.~\eqref{eq:complete_RG}.
To better manifest this fact, we transform the Hamiltonian with the unitary matrix $U = \exp[i S_z \Phi_\text{s}(0)]$~\cite{EmeryKivelsonPRB92} (see also Refs.~\cite{Maciejko2012} and \cite{Yevtushenk2018} for a discussion in the context of helical edges) to remove the $\Phi_\text{s}$ field from vertex operators. As a side effect, this unitary transformation introduces an extra quartic contribution. In combination with the first line of Eq.~\eqref{eq:h_kondo}, the quartic contribution becomes
$
   ( \tilde{J}_1^z - 1/2)S_z \partial_x \Phi_\text{s}/\pi,
$
which is now the only term that contains the field $\Phi_\text{s}$. Consequently, When $\tilde{J}_1^z = 1/2$, the quartic term vanishes and will not be regenerated from further RG flow.

Crucially, at the fixed point with $\tilde{J}_1^z = 1/2$, the contributions from $\Phi_\text{s}$ to the operator scaling dimensions vanish. This ``freezing'' of $\Phi_\text{s}$ indicates that the dot spin at low temperatures becomes fully screened. Similar freezing mechanisms have been reported before in, e.g, Luttinger liquid quantum wires~ \cite{KaneFisherPRB92,FurusakiNagaosaPRB93,SchillerIngersent97}, dissipative resonant level models~\cite{DongPRB14, HuaixiuPRB14}, as well as topological Kondo systems \cite{HerviouPRB16,MichaeliPRB17}.
The $z$-component of the inter-edge Kondo exchange $\tilde{J}_2^z$ always has scaling dimension $(K+1/K)/2$, regardless of the value of $\tilde{J}_1^z$.
The RG irrelevance of $\tilde{J}_2^z$ and the initial-value-independent fixed-point value of $\tilde{J}_1^z$ indicate further the irrelevance of anisotropy when $\tilde{J}_1^z$ reaches the the fixed point. This explains why we are allowed to present Eq.~\eqref{eq:ham_kondo} in an isotropic form.

\begin{figure*}[ht]
  \centering
    \includegraphics[width=1\textwidth]{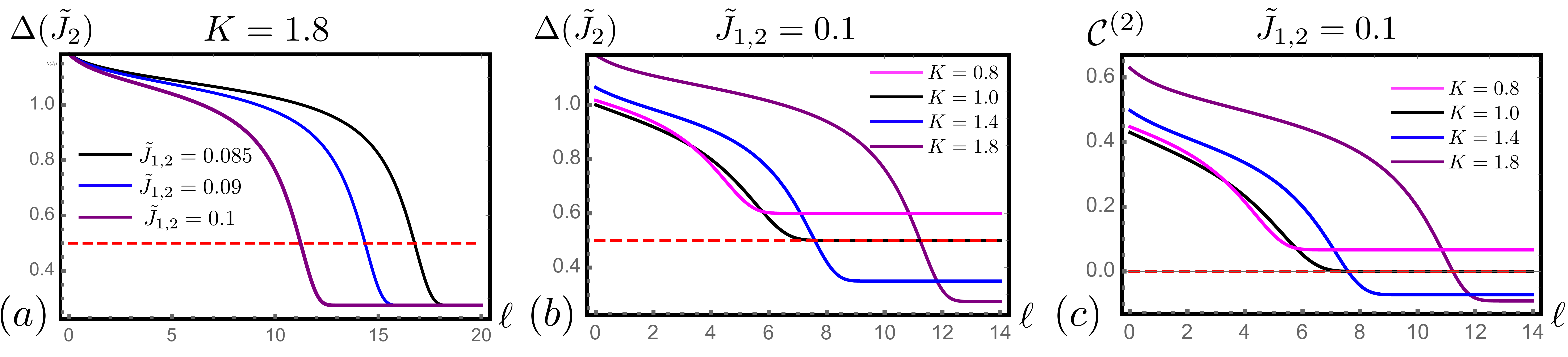}
    \caption{Renormalization of the scaling dimension of the inter-edge Kondo coupling [denoted $\Delta(\tilde{J}_2)$] with decreasing energy (parametrized by the increasing length scale parameter $\ell$). (a) The interaction strength is fixed at $K=1.8$, and the solid, colored curves correspond to different initial values for $\tilde{J}_{1,2}$ [see Eq.~\eqref{eq:complete_RG}]. The red dashed line, $\Delta(\tilde{J}_2)=1/2$, marks the transition between positive and negative delta-$T$ noise [see panel (c)]. (b) The initial Kondo coupling strengths are taken as $\tilde{J}_{1,2} =0.1$. The solid colored curves show how $\Delta (\tilde{J}_2)$ renormalizes for different values of $K$. (c) The change in the noise coefficient $\mathcal{C}^{(2)}$ [see Eq.~\eqref{eq:c2}], with renormalized $\Delta(\tilde{J}_2)$ from (b). For $K>1$,  $\mathcal{C}^{(2)}$ changes sign at sufficiently low energy.}
    \label{fig:j2_dimension}
\end{figure*}

Importantly, the freezing of $\Phi_\text{s}$ alters the tunneling, and the renormalized Kondo exchange tunneling now reads
\begin{equation}
\begin{aligned}
& H_{\text{Kondo,renorm.}} \! =\! \frac{J_{1}^{\perp}}{2\pi a} \left[ F_{1,L} F_{1,R} e^{-i\Phi_\text{sf}} S_-  + \text{H.c.}\right]\\
& \!+\! \frac{J_{1}^{\perp}}{2\pi a}\! \left[ F_{2,R} F_{2,L} e^{i\Phi_\text{sf}} S_-  + \text{H.c.}\right]\\
& \!+\! \frac{J_{2}^{\perp}}{2\pi a}\! \left[ F_{1,L} F_{2,L} e^{-i \Phi_\text{cf}} S_-  + \text{H.c.}\right]\\
& \!+\! \frac{J_{2}^{\perp}}{2\pi a}\! \left[ F_{2,R} F_{1,L} e^{i \Phi_\text{cf}} S_-  + \text{H.c.}\right]\\
& \!+\! \frac{J_2^z}{2\pi a} \!\left[\! F_{1R} F_{2L} e^{i(\Phi_\text{cf} - \Phi_\text{sf})} \!+ \!F_{2R} F_{1L} e^{i(\Phi_\text{cf} + \Phi_\text{sf})} + \text{H.c.}\right]\! S_z,
\end{aligned}
\label{eq:h_kondoMod}
\end{equation}
where strictly all prefactors ($J_{1}^{\perp}$, $J_{2}^{\perp}$ and $J_{2}^{z}$) are those after the renormalization. For simplicity we keep their labeling as in Eq.~\eqref{eq:h_kondo}.
Equation~\eqref{eq:h_kondoMod} is obtained by setting $\Phi_\text{s} = 0$ in Eq.~\eqref{eq:h_kondo}, where $\tilde{J}_1^z$ has flown to the value at the fixed point (i.e., $1/2$).
This point is the Emery-Kivelson point~\cite{EmeryKivelsonPRB92} of our system, near which spin operators can be considered as effectively out of dynamics.
Near this fixed point, it is then legitimate to take the impurity correlations as $\langle S_+ S_- \rangle = \langle S_- S_+ \rangle = 1/2$ for free spin operators. A similar approach was
taken, e.g., in Refs.~\cite{KaneFisherPRB92,SchillerIngersent97} when discussing correlations in a resonant level model, and in Ref.~\cite{GanPRB95} for a two-impurity Kondo system.
We have also dropped the $H_\mathcal{T}$ term since it is less relevant than the other terms.
Since $\Phi_\text{sf}$ and $\Phi_\text{cf}$ are interacting in Eq.~\eqref{eq:ham_after_rotation}, the vertex operators in Eq.~\eqref{eq:h_kondoMod} have the same dynamics as ``Luttinger hyperfermions''~\cite{FendleyLudwigSaleurPRB95} in FQH systems, although they carry an integer number of charges.
These vertex operators have the statistical angle at the maximum values bounded by the scaling dimension Eq.~\eqref{eq:Theta_Delta_bound}. In contrast to the direct tunneling model, boson-like tunneling in the Kondo model appears after the freezing of the charge field.

By neglecting the contributions from $\Phi_\text{s}$,
the scaling dimension of the intra-edge coupling changes to $K/2$, and that of the inter-edge becomes $1/2K$. By contrast, the scaling dimensions of the direct tunneling ($\propto t$) and the Kondo $z$-component ($\propto \tilde{J}_2^z$) do not change during the RG flow, as they do not involve the spin field. They are thus less relevant in comparison to intra and inter-edge Kondo couplings.

Consequently, the final two candidates as the leading relevant operator are intra- ($\propto \tilde{J}_1$) and inter-edge ($\propto \tilde{J}_2$) Kondo exchange operators. The most dominant operator determines the system's ground state. In more detail, if $K<1$, it is the intra-edge Kondo exchange $\propto \tilde{J}_1$ that dominates at low energies, and the dot spin is overscreened by both edges, driving the system towards the two-channel Kondo fixed point (see Ref.~\cite{MitchellSelaPRB12} for details).
Near this fixed point, we expect positive delta-$T$ noise from Eq.~\eqref{eq:c2}, since the leading charge-tunneling operator ($\propto\tilde{J}_2$) in this case has scaling dimension $1/(2K)>1/2$. In strong contrast, for attractive interactions $K>1$, the inter-edge Kondo exchange $\propto \tilde{J}_2$ dominates, with scaling dimension $1/(2K)<1/2$. In view of Eq.~\eqref{eq:c2}, this operator gives negative delta-$T$ noise.
Of this situation, the statistical angle $\Theta_A < \pi/2$ [following Eq.~\eqref{eq:bound_of_phase}] is also boson-like.
Indeed, it equals $\Theta_A = \pi/(2K)$ following Eq.~\eqref{eq:vertex_decomp} that relates $\Phi_\text{cf}$ to the original fields.

In Fig.~\ref{fig:j2_dimension} we plot the scaling dimension of the inter-edge Kondo coupling, denoted $\Delta(\tilde{J}_2)$ as well as the value of $\mathcal{C}^{(2)}$ as functions of $\ell$. The plots are obtained by solving Eq.~\eqref{eq:complete_RG} numerically. Figure~\ref{fig:j2_dimension}\textcolor{blue}{(a)} shows the change in the scaling dimension of $\tilde{J}_2$ when $K = 1.8 >1$. Here, we considered three different initial Kondo exchange strengths $\tilde{J}_{1,2}$. The red dashed line labels the critical scaling dimension $1/2$ below which the delta-$T$ noise becomes negative. Clearly, the scaling dimension of $\tilde{J}_{2}$ decreases during the RG flow, independently of the initial value. Meanwhile, all three curves flow below the critical line $\Delta(\tilde{J}_2)=1/2$, for sufficiently large $l$. In this sense, the bare Kondo coupling strengths are not deterministic: they only change the transition temperature below which delta-$T$ noise becomes negative~\cite{BKT}. 

In Fig.~\ref{fig:j2_dimension}\textcolor{blue}{(b)}, we instead fix the bare values of $\tilde{J}_{1}$ and $\tilde{J}_{2}$, and investigate the scaling dimension $\Delta(\tilde{J}_2)$ for different values of $K$. In contrast to the comparatively trivial effect of $\tilde{J}_{1,2}$, $K$ determines the low-temperature scaling dimension of $\tilde{J}_2$ and the sign of the delta-$T$ noise. Importantly, $\Delta(\tilde{J}_2)$ falls below $1/2$ for $K>1$, in agreement with our previous analysis. Following the flow of the scaling dimension Fig.~\ref{fig:j2_dimension}\textcolor{blue}{(b)}, we calculate the corresponding flow curve of $\mathcal{C}^{(2)}$ in Fig.~\ref{fig:j2_dimension}\textcolor{blue}{(c)}.
In agreement with Fig.~\ref{fig:j2_dimension}\textcolor{blue}{(b)}, $\mathcal{C}^{(2)}$ decreases during the RG flow. At sufficiently low temperatures, $\mathcal{C}^{(2)}$ becomes negative for $K>1$, but remains positive for $K<1$.
This prerequisite of a negative delta-$T$ noise ($K>1$) does not require too strong interaction, and is thus much more accessible in real experiments, in comparison to that ($K>4$ or $K<1/4$) of the direct tunneling model.

\begin{figure}[t!]
  \centering
    \includegraphics[width=1\columnwidth]{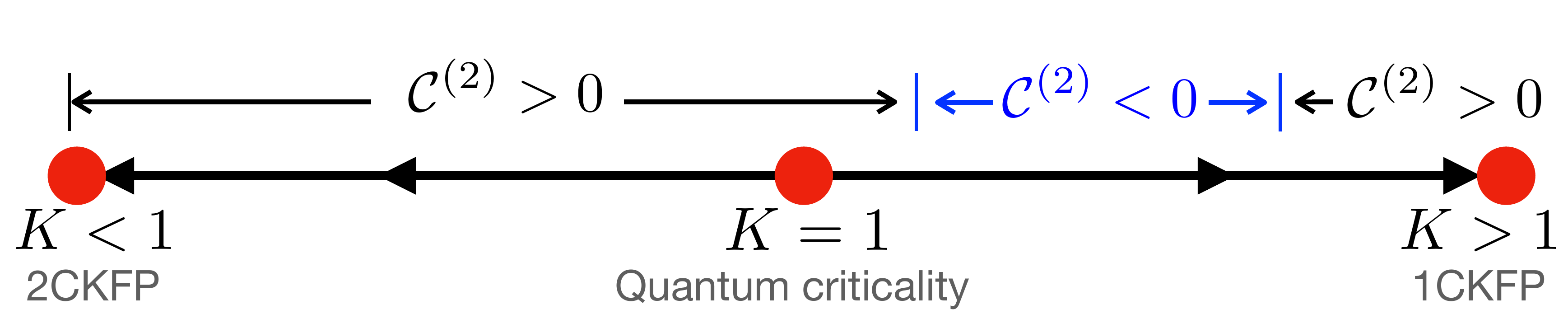}
    \caption{Schematic RG flow of the Kondo tunneling model \eqref{eq:h_kondo} for different interaction strengths $K$. Black arrows depict the flow from the quantum critical point ($K=1$) towards two stable fixed points. For $K<1$, the system flows towards the two-channel Kondo fixed point, while for $K>1$ it flows towards the single-channel Kondo fixed point. In the latter case, we expect a temperature window with negative delta-$T$ noise  ($\mathcal{C}^{(2)} < 0$, in blue) produced by tunneling of emergent quasiparticles.}
    \label{fig:rg_kondo}
\end{figure}

As another important observation, we note that the results presented in this Section suggest that the non-interacting point $K = 1$, known as the quantum critical point between the one-channel and two-channel Kondo fixed points, could be probed by measurements of delta-$T$ noise. If $K>1$, the system flows towards the one-channel Kondo fixed point, and the delta-$T$ noise has a temperature window within which the delta-$T$ noise is negative. By contrast, the noise is always positive for $K<1$, when the system flows towards the two-channel Kondo fixed point (see Fig.~\ref{fig:rg_kondo}).

To briefly summarize this Section, we have demonstrated that both the direct-tunneling model and the QSH Kondo model can exhibit negative delta-$T$ noise. In the direct-tunneling model, the negative delta-$T$ noise is produced when coherent two-particle tunneling, which is always boson-like, starts to dominate over single electron tunneling. In the Kondo model, the negative delta-$T$ noise appears as an interplay of the dot-edge and LL interactions. The combined effect of these interactions causes the Kondo exchange tunneling operator to become boson-like at low energies. Hence, in both models, the emergence of the negative delta-$T$ noise is accompanied by a boson-like nature of the quasiparticles involved in the leading tunneling operator, in accordance with Sec.~\ref{sec:Discussion}.

In this section, we have only analyzed the situation with the impurity spin $S = 1/2$.
More general situations with $S > 1/2$ have been investigated in helical edges only in the context of shot noise~\cite{Maciejko2009,KurilovichPRL19,PashinskyGoldsteinBurmistrovPRB20}.
To the best of our knowledge, an analysis of influence from a spin $S > 1/2$ on the delta-$T$ noise has not yet been carried out in helical-edge systems. Nevertheless, we anticipate that a larger spin would significantly change the features of the delta-$T$ noise. Indeed, when $S > 1/2$, screening of the impurity spin by a single channel alone is not enough, leading to the underscreened Kondo effect~\cite{FradkinNPB90}, where some impurity spin degrees of freedom are left untouched. In this case, the system would prefer the two-channel Kondo fixed point, where the impurity spin is more strongly screened. We expect this conclusion to be valid for a rather wide range of interactions.

\section{Summary and conclusion}
\label{sec:summary}
We have studied delta-$T$ noise, i.e., charge noise at zero bias voltage but finite temperature bias, for interacting electron systems in 1D. 
As our central result, we have showed that for tunneling between two identical, interacting 1D electron systems (e.g., two QSH edges) to second order in $\delta T$, the sign of the delta-$T$ noise is fully determined by the scaling dimension $\Delta_{\mathcal{T}}$ of the leading charge-tunneling operator, see Eq.~\eqref{eq:c2}. The delta-$T$ noise is positive when $\Delta_{\mathcal{T}} > 1/2$, while it is negative when $\Delta_{\mathcal{T}} < 1/2$. For the tunneling between infinite subsystems hosting chiral channels with local interactions, the analysis of the relation of $\Delta_{\mathcal{T}}$ and the statistical phase of the tunneling quasiparticles [see  Eq.~\eqref{eq:Theta_Delta_bound}] implies that the negative delta-$T$ noise is only possible for boson-like tunneling operators. 
This result indicates that the negative delta-$T$ noise is not a property unique to anyons, but can occur generically due to many-body interactions.

We have supported our findings by comprehensively investigating the delta-$T$ noise in two setups involving tunneling between two interacting QSH edges.
In the first setup [Fig.~\ref{fig:qsh_kondo}\textcolor{blue}{(a)}], the edges are coupled by direct tunneling. In Ref.~\cite{RechMartinPRL20}, it was argued that the leading-order operator for such tunneling is RG irrelevant and can only generate positive delta-$T$ noise. However, in Sec.~\ref{sec:direct_tunneling}, we have showed that this conclusion changes when taking into account higher-order coherent tunneling processes. Crucially, if interactions are sufficiently strong, i.e., with Luttinger parameter $K>4$ or $K<1/4$, these processes dominate at low energies, and give rise to negative delta-$T$ noise also in the tunneling of electrons.

In the second setup [Fig.~\ref{fig:qsh_kondo}\textcolor{blue}{(b)}], we have studied an interacting QSH-Kondo model, where electrons can tunnel between the two edges also via Kondo exchange. In Sec.~\ref{sec:konod},  we have demonstrated that the delta-$T$ noise is always positive for repulsive interactions $K<1$, while the negative sign may emerge for attractive interactions $K>1$. In the latter case, the origin of the negative delta-$T$ noise is due to the Kondo exchange favoring boson-like coherent pair-electron tunneling. We have also found that the sign change in the delta-$T$ noise, occuring at $K=1$, coincides with the quantum critical point separating the single-channel and two-channel Kondo fixed points (see Fig.~\ref{fig:rg_kondo}). The interacting QSH-Kondo model is thus an example where quantum criticality can be predicted by measuring delta-$T$ noise.

Returning to the direct inter-edge tunneling model, we may interpret the negative delta-$T$ noise in coherent tunneling of two electrons as mimicking the tunneling of a bosonic Cooper pair. However, in the interacting QSH Kondo model, the emergence of negative delta-$T$ noise is more involved. Still it can also be associated with the tunneling of boson-like particles. Such tunneling is possible close to the single-channel Kondo RG fixed point which requires attractive interactions: $K>1$ (see Fig.~\ref{fig:rg_kondo}).

We have further noted that, since anyons in FQH Laughlin states have scaling dimensions $\Delta_{\rm anyon}<1/2$~\cite{RechMartinPRL20}, the negativity of the delta-$T$ noise naturally follows. However, we want to emphasize that this result, which only involves a single tunneling link, does not involve any actual anyonic braiding. In Appendix~\ref{sec:braiding_delta_t}, we have studied an extended model of a QPC involving two tunneling links. There, we show that, to leading order, i.e., $\mathcal{O}(\delta T^2$),  the contribution to the delta-$T$ noise from anyon braiding vanishes. Hence, the negativity of the delta-$T$ noise is not because anyons have braided, but instead that they renormalize tunneling operators in a nontrivial manner governed by the leading scaling dimension $\Delta_\mathcal{T}=\nu<1/2$, though related to the anyon statistical phase $\Theta=\pi \nu$. As we have shown in this paper, this is true also for other emergent particles in interacting 1D systems. At the same time, the bunching/anti-bunching preference of the tunneling particles is also determined by the statistical phase which thus links it to the delta-$T$ noise.

As a final remark, we note that in LL systems with attached electrical contacts, it is known that the mismatch in interactions between the system and the contacts induces backscattering of the eigenmodes
(see, e.g., Refs.~\cite{KaneFisher1995Contact} and \cite{Polyakov2021} for the case of complex FQH edges). This is not an issue for fully chiral single channels such as those on Laughlin FQH edges. This scattering can however be an issue for edges with counterpropagating channels. Indeed, in a typical setup for measuring the current noise, the electrical currents are measured in the Fermi leads. Therefore, even though the measurements of the delta-$T$ noise avoid applying the bias voltage, effects produced by the interface between the interacting edge and the contact might become important. Understanding such contact effects, in particular, removing the condition of full equilibration within the edge (assumed throughout the paper), deserves further investigation. One way to avoid this complication is to measure the delta-$T$ noise in setups without Fermi contacts---by means of the magnetic field sensing of the local currents~\cite{RothJAppPhys89,WijngaardenPRB96,JOOSSPhysC98,WijngaardenPhysCSC98,JoossRepProPhys02,FeldmannPRB04,DinnerPRB07,MeltzerLevinZeldovPRApp17,Zeldov2020}.

To conclude, we have argued in this paper that the sign of the delta-$T$ noise does not uniquely distinguish anyonic statistics from generic interaction effects in 1D. At the same time, the negative sign of the delta-$T$ noise for tunneling between FQH and QSH edges is always accompanied by the (intrinsic or interaction-induced) boson-like nature of the tunneling quasiparticles. This conclusion is supported by the two case studies for QSH edges, where appropriate choice of the interaction changes the sign of delta-$T$ noise. We have thus demonstrated the potential of delta-$T$ noise as a probe of strongly correlated electron behavior in nanoscale conductors and clarified its relation to quantum statistics.

\begin{acknowledgments}
We thank Matteo Acciai, Igor Burmistrov, Vadim Khrapai, Kyrylo Snizhko, and Oleg Yevtushenko for critical comments on the manuscript and interesting suggestions. We acknowledge support from the Excellence Initiative Nano at Chalmers University of Technology (C.S.), the DFG grants No. MI658/10-1 (C.S.) and MI658/10-2, 
the German-Israeli Foundation through the GIF grant No.~I-1505-303.10/2019, and the Russian Science Foundation (Grant No.~20-12-00147). This project has received funding from the European Union's Horizon 2020 research and innovation programme under grant agreement No 101031655, TEAPOT.

\end{acknowledgments}

\appendix

\section{Energy-dependent transmission and interaction-induced proliferation of low-energy states}
\label{sec:interaction_and_selection}

In Sec.~\ref{sec:FreeElectrons}, we showed that free fermions generate positive delta-$T$ noise if the transmission probability $D$ is energy-independent. In this Appendix, we demonstrate that the delta-$T$ noise of free fermions can be negative if $D$ has a sufficiently fast energy decay. We also show a similar effect for fermions with attractive interactions, where the fermions effectively become more distributed at low energies, i.e., at energies close to the Fermi level, here defined to be zero. This happens because attractive interactions give rise to slowly decaying correlation functions in time. In both cases, low-energy particles dominate the noise generation, which leads to negative delta-$T$ noise. Further, we consider the effect of energy and temperature dependence of the transmission coefficient on the thermal noise in light of this effect on the delta-$T$ noise.

\subsection{Energy-dependent transmission probability}
\label{sec:energy_dependent_trans}

In this Section, we analyze the effect of energy dependence of the transmission probability $D(\epsilon)$ on the delta-$T$ noise for free fermions.
As our starting point, we assume $D(\epsilon)\ll 1$ (as in the main part of the text) and rewrite Eq.~\eqref{eq:correlator_fermion} for $\mu_1=\mu_2=0$ as
\begin{align}
    &S^{(\text{f})} \approx  
    \frac{2e^2}{h} \int d\epsilon D(\epsilon) 
    \left(f_1+f_2-2f_1f_2\right)
    \nonumber
    \\
    &\quad=
    \frac{2e^2}{h}\! \int\! d\epsilon D(\epsilon)\left[ f (T_1,\epsilon) f(T_2,-\epsilon) \!+ \!f(T_1,-\epsilon) f(T_2,\epsilon) \right],
    \label{eq:sf_energy_dependent}
\end{align}
where $f(T,\epsilon)$ is the fermionic distribution function at temperature $T$, and we have used
$f(-\epsilon)=1-f(\epsilon)$. We note that for a symmetric (around the chemical potential) function $D(\epsilon)$, the two terms in Eq.~(\ref{eq:sf_energy_dependent}) are equal if the integration limits are extended to $\epsilon=\pm \infty$. 

We next expand 
\begin{align}
  & f (T_1,\epsilon) f(T_2,-\epsilon) + f(T_1,-\epsilon) f(T_2,\epsilon) 
  \notag\\
  &\qquad \approx 2 f(\bar{T},\epsilon) f(\bar{T},-\epsilon) + A(\epsilon) \frac{\delta T^2}{4}
\end{align}
to leading order in $\delta T$, setting, for simplicity, $\mu=0$. 
Here,
\begin{equation}
\begin{aligned}
    A(\epsilon) & \equiv f(T,-\epsilon)\, \frac{\partial^2 f(T,\epsilon)}{\partial T^2}   + f(T,-\epsilon)\,\frac{\partial^2 f(T,\epsilon)}{\partial T^2}   \\
    &- 2 \frac{\partial f(T,\epsilon)}{\partial T}\,   \frac{\partial f(T,-\epsilon)}{\partial T}
    \\
    & = \frac{\epsilon}{4 T^3 \cosh^4\frac{\epsilon}{2 T}}\left(\frac{\epsilon}{2T} \cosh\frac{\epsilon}{ T}- \sinh\frac{\epsilon}{T}\right),
\end{aligned}
\label{eq:noise_spectrum}
\end{equation}
which can be interpreted as the spectrum of delta-$T$ noise for small $\delta T$.

\begin{figure}[t!]
  \centering
    \includegraphics[width=1 \columnwidth]{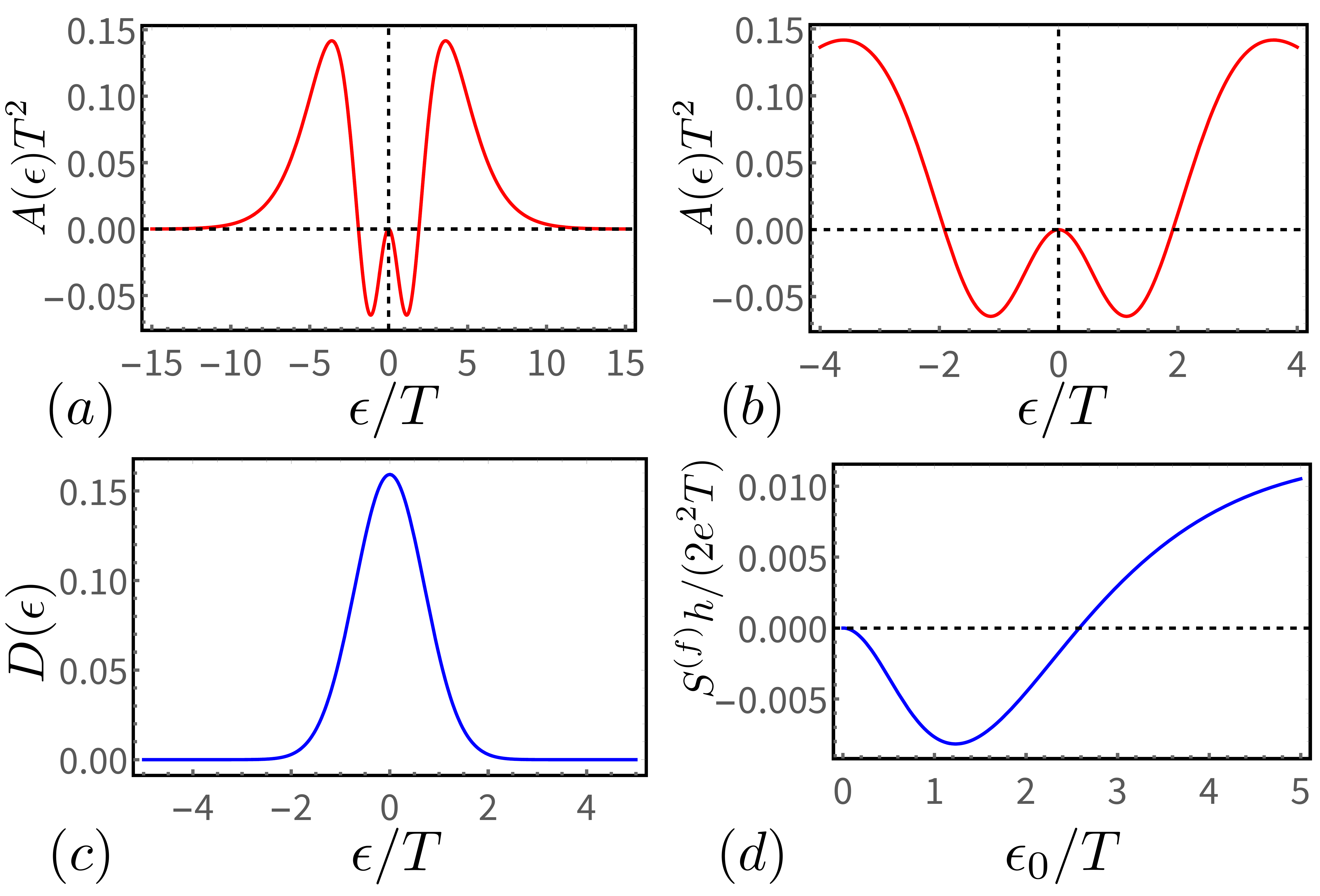}
    \caption{(a) The noise spectrum $A(\omega)$ of free fermions in Eq.~\eqref{eq:noise_spectrum}. The spectrum is positive for most energies, except for a negative contribution at the lowest energies. (b) A zoom-in at low energies. The noise spectrum changes its sign around $|\epsilon| \approx 2 T$. (c) A Gaussian-shape transmission probability $D(\epsilon)$ [Eq.~\eqref{eq:D_Gaussian}], where the width $\epsilon_0$ equals the temperature. (d) The free fermion noise as a function of the width ($\epsilon_0$) of the Gaussian-shape transmission probability, Eq.~\eqref{eq:D_Gaussian}. The delta-$T$ noise changes sign at $\epsilon_0 \approx 2.6 T$.}
    \label{fig:noise_spectrum}
\end{figure}

We plot $A(\epsilon)$ in Fig.~\ref{fig:noise_spectrum}\textcolor{blue}{a} to highlight its features.
Clearly, the spectrum is symmetric around $\epsilon=0$. It is positive at energies $|\epsilon|\gtrsim 2T$, leading to a positive delta-$T$ noise when $D(\epsilon)$ is constant:
\begin{equation}
    \int_{-\infty}^\infty d\epsilon A(\epsilon)=\frac{2\left(\pi^2-6\right)}{9 T^2}>0.
\end{equation}
However, we may infer from Figs.~\ref{fig:noise_spectrum}\textcolor{blue}{a,b}, that negative delta-$T$ noise should emerge if the transmission $D(\epsilon)$ is energy-dependent and is peaked at low energies that contribute negatively to the noise. 

To show this, we choose, as a simple example, a Gaussian-shape transmission probability
\begin{equation}
\label{eq:D_Gaussian}
    D(\epsilon) = \frac{\bar{T} }{2\pi\epsilon_0}\exp(-\epsilon^2/\epsilon^2_0).
\end{equation}
This transmission picks out low-energy fermions within an energy window $\epsilon_0$, see Fig.~\ref{fig:noise_spectrum}\textcolor{blue}{(c)}. The width of the window determines the sign of the delta-$T$ noise as seen in Fig.~\ref{fig:noise_spectrum}\textcolor{blue}{(d)}.
Indeed, when the width is large enough, $\epsilon_0 > 2.6 T$, noise from higher-energy states dominates, leading to a positive delta-$T$ noise. On the contrary, delta-$T$ noise becomes negative if $\epsilon_0 < 2.6 T$, where noise from low energies dominates.

Thus, we conclude that the energy dependence of the elastic transmission probability can be sufficient to induce a negative delta-$T$ noise for free fermions. As such, this sign change can hardly be associated with the particle statistics. However, we will see below that interactions between electrons can lead to the same effect by renormalizing an energy-independent transmission probability. In this situation, the interactions may simultaneously induce a non-trivial statistical angle $\Theta$ for the tunneling quasiparticles.

\subsection{Interaction-induced proliferation of 
low-energy states}
\label{sec:interacting_delta_T}

Following the consideration of the previous section, where the effect of energy dependence of the transmission coefficient was discussed, here we explore the interaction-induced renormalization of $D$ in the context of delta-$T$ noise. 
We focus on tunneling between two semi-infinite spinless LLs [cf. Fig.~\ref{fig:ll_structure}{\color{blue}(b)} and Fig.~\ref{fig:hom}{\color{blue}(a)}]
with the same local interaction (described by the Luttinger parameter $K$), see also Appendix~\ref{sec:LLFormalism} for additional details of the model. 

This is a well-known weak-link problem for tunneling in LLs, as addressed in Ref.~\cite{KaneFisherPRB92}. As discussed in Sec.~\ref{sec:overview},
this setup can be unfolded into two infinite chiral channels, but with \textit{non-local} interactions.
Therefore, the relation of negative delta-$T$ and boson-like statistics, which was made for infinite chiral systems with local interactions, should be reconsidered for this setup.   
Indeed, the above unfolding can be described by means of a proper boundary condition at the ends of the semi-infinite wires.
Since the application of the boundary condition will not modify the statistical phase of tunneling particles (it is still an electron that tunnels through the weak link), a negative delta-$T$ noise in the weak-link setup is not necessarily accompanied by a boson-like statistical phase.

Following Eq.~\eqref{eq:S}, single-particle tunneling across a weak link with the energy-independent tunneling transparency $D\ll 1$ generates noise that can be written in on a form similar to Eq.~\eqref{eq:sf_energy_dependent}:
\begin{equation}
\begin{aligned}
    S_\text{int} & \simeq  \frac{2e^2}{h} D\, \tau_0 ^{2/K - 2}\\
    \times &\int d\epsilon  \left[ a_{K} (T_1,\epsilon) a_{K}(T_2,-\epsilon) + a_{K}(T_1,-\epsilon) a_{K}(T_2,\epsilon) \right],
\end{aligned}
\label{eq:interacting_noise}
\end{equation}
where $D = |t_\mathcal{T}|^2/v^2$ as defined in the main text, and
\begin{equation}
    a_{K} (T,\epsilon) = \frac{1}{2\pi\tau_0^{1/K}} \int d\tau \left\{ \frac{\sinh (i\pi T \tau_0)}{\sinh \left[\pi T ( i\tau_0 + \tau) \right]} \right\}^{1/K}  e^{i\epsilon \tau}
\label{eq:ad}
\end{equation}
is proportional to the lesser Greens function of the particles (with interaction parameter $K$). 
Note that the structure of this expression is analogous to that in Eq.~\eqref{eq:SwrittenOut}, with $\Delta_A=\Delta_B=1/(2 K)$.

\begin{figure*}[t!]
  \centering
    \includegraphics[width=1.0 \textwidth]{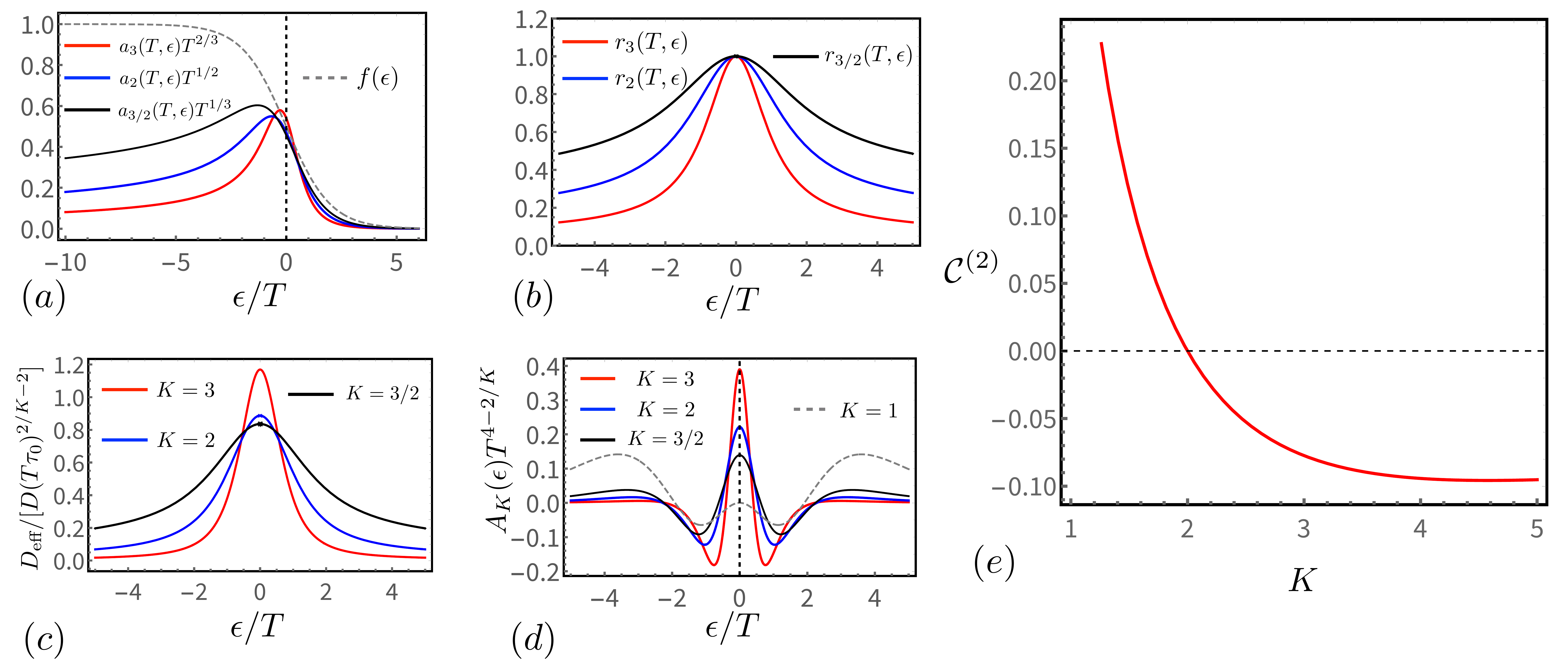}
    \caption{(a) Generalized ``distribution function'' $a_K(T,\epsilon)$ [Eq.~\eqref{eq:ad}] as a function of the dimensionless energy $\epsilon/T$, for several values of the interaction parameters $K$. (b) The normalized ratio $r_K$ [Eq.~\eqref{eq:rk}] as a function of $\epsilon/T$. With increasing attractive interaction $K$, particles become more and more condensed around zero energy, leading to negative delta-$T$ noise. (c) The plot of the effective transmission $D_\text{eff}$ defined in Eq.~\eqref{eq:effective_d}. (d) The noise spectrum $A_K(\epsilon) T^2$ [Eq.~\eqref{eq:interacting_spectrum}] for various interaction strengths $K$. In interacting situations ($K > 1$), the condition of weak transmission requires the bare (UV) value of the transmission probability $D_\text{eff} (\epsilon\sim 1/\tau_0) $ to be extremely small, such that $D_\text{eff} (\epsilon)\ll 1$ for all energies. (e) The coefficient $\mathcal{C}^{(2)}$ determining the delta-$T$ noise as a function of the interaction parameter $K$. The sign change occurs at $K=2$, which corresponds to $\Delta_\mathcal{T}=1/2$. In contrast to FQH edges~\cite{RechMartinPRL20}, where the effective interaction parameter is fixed by topological properties of the system, here arbitrary interaction parameters are possible.}
    \label{fig:normalized_ratio}
\end{figure*}

For non-interacting particles $K = 1$, $a_1(T,\epsilon) = f(T,\epsilon)$, i.e., we recover the Fermi distribution.
When $K\neq 1$, $a_K$ involves the density of states, and plays the role of the effective distribution functions of interacting particles.
Here, we neglect higher-order tunnelings as they always have larger scaling dimensions in a system with fields under the same interaction $K$. This is in stark contrast to the QSH-Kondo setup (see Sec.~\ref{sec:Results}), where charge and spin sectors have inversely proportional scaling dimensions, and the two-particle processes can dominate the tunneling.

When $|\epsilon| \gg T$, 
$a_{K}$ drops to zero for positive energies $\epsilon \gg T > 0$, and decays polynomially as
$$ a_{K} (T,\epsilon) \propto \epsilon^{1/K - 1}$$
for negative energies $\epsilon \ll -T < 0$ and $K > 1$.
In the opposite limit, $T \gg |\epsilon|$, $a_K (T, \epsilon)$ approaches a constant value close to 
$ T^{1/K - 1}/2$. This asymptotic behavior is illustrated in the plot of $a_K$ in Fig.~\ref{fig:normalized_ratio}\textcolor{blue}{(a)}.
These results are manifestations of the zero-bias anomaly in the density of states for the tunneling into the end of a semi-infinite LL wire.

The definition of the noise Eq.~\eqref{eq:interacting_noise} involves an energy-independent prefactor. It reflects the intrinsic UV divergence characteristic of LL systems. Here, we have isolated this piece from the ``generalized distribution function'' Eq.~\eqref{eq:ad}. 
To see the effect of interactions, we next define a normalized function
\begin{equation}
   r_K(T, \epsilon) \equiv \frac{a_{K}(T,\epsilon)}{a_K (T,0)} \frac{f(T,0)}{f (T,\epsilon )}
   \label{eq:rk}
\end{equation}
which can be interpreted as an additional interaction-induced renormalization factor that describes proliferation of particles in low-energy states.
With these functions, noise can instead be presented as
\begin{equation}
\begin{aligned}
    S_\text{int} & \simeq  \frac{2e^2}{h} \int d\epsilon \, D_\text{eff}(\bar{T},\delta T,\epsilon) 
    \notag
    \\
    & \times \left[ f (T_1,\epsilon) f(T_2,-\epsilon) + f(T_1,- \epsilon) f(T_2,\epsilon) \right],
\end{aligned}
\label{eq:interacting_noise_another_form}
\end{equation}
which takes exactly the form of Eq.~\eqref{eq:sf_energy_dependent} with
the effective energy-dependent transmission probability
\begin{equation}
\begin{aligned}
  D_\text{eff} (\bar{T},\delta T, \epsilon) & = 
 D\, \tau_0 ^{2/K - 2}\,  r_K(T_1,\epsilon) r_K(T_2,-\epsilon)\\
 & \times  \frac{a_K(T_1,0) a_K(T_2,0)}{f(T_1,0) f(T_2,0)}.
\end{aligned}
  \label{eq:effective_d}
\end{equation}
Equation~\eqref{eq:effective_d} contains a factor $$\frac{a_K(T_1,0) a_K(T_2,0)}{f(T_1,0) f(T_2,0)} 
\propto (T_1 T_2)^{1/K - 1}$$
that diverges polynomially with decreasing temperatures, for $K > 1$. 

Importantly, for interacting systems, e.g., FQH or Luttinger liquid wires alike, the validity of perturbation theory requires more than just weakness of the bare tunneling transparency $D$. Instead, the dressed tunneling $D_\text{eff} \ll 1$ is required to be small for fixed $\tau_0$, in the whole range of energies, including energies around the two temperatures of interest $T_1$, $T_2$ (see, e.g., Ref.~\cite{FendleyLudwigSaleurPRB95} for details of a related calculation). 
Otherwise, the system approaches the low-temperature fixed point with $D_\text{eff}(\epsilon)\approx \text{const}\sim 1$ for energies in the thermal window. This produces a positive delta-$T$ noise.

To illustrate the above construction, we plot $r_K$ and $D_\text{eff}$ for attractive interaction values $K = 3,\  2$ and $3/2$ in Figs.~\ref{fig:normalized_ratio}\textcolor{blue}{(b)} and \ref{fig:normalized_ratio}\textcolor{blue}{(c)}, respectively.
We see that when $K$ increases, both $r_K (T,\epsilon)$ and the effective transmission $D_\text{eff}$ generate an increasingly narrow peak around zero energy. 
Particles with smaller energies then have a higher transmission probability.
Indeed, attractive interactions induce a slower decay of correlation functions in time~\cite{GiamarchiBook}, which in turn leads to an increased contribution of low-energy states to the noise.

Similarly to the non-interacting noise spectrum $A(\epsilon)$ in Eq.~\eqref{eq:noise_spectrum}, we can define
\begin{equation}
\begin{aligned}
    A_K(\epsilon) & \!\equiv \!a_K(T,-\epsilon)\frac{\partial^2a_K(T,\epsilon)}{\partial T^2}  \! +\! a_K(T,\epsilon)
    \frac{\partial^2a_K(T,-\epsilon)}{\partial T^2}\!   \\
    & \!- \!2 \frac{\partial a_K(T,\epsilon)}{\partial T}  \frac{\partial a_K(T,-\epsilon)}{\partial T},
\end{aligned}
\label{eq:interacting_spectrum}
\end{equation}
as the noise spectrum in the presence of interactions $K$. We plot $A_K(\omega)$ in Fig.~\ref{fig:normalized_ratio}\textcolor{blue}{(d)}. Its behavior explains the negative sign of delta-$T$ noise for sufficiently attractive interactions. In this situation, more particles are transmitted at low energies, which contributes with negative weight to the integral in Eq.~\eqref{eq:interacting_noise}. In turn, this generates negative delta-$T$ noise, even for an energy independent transmission, in agreement with the results of Appendix~\ref{sec:energy_dependent_trans}.
We plot the $\mathcal{C}^{(2)}$ coefficient for the delta-$T$ noise in the weak-tunneling limit as a function of the interaction, in Fig.\,\ref{fig:normalized_ratio}{\color{blue}(e)}. Although it is basically the same as that of Ref.~\cite{RechMartinPRL20} after the replacement $K \to 1/\nu$, the plot of Fig.\,\ref{fig:normalized_ratio}{\color{blue}(e)} applies to LLs with a continuous variation of the interaction, but without any intrinsic anyonic statistics.

Following our case studies of QSH systems in Sec.~\ref{sec:Results}, an interaction-induced negative delta-$T$ noise in infinite chiral systems with local interactions corresponds to a boson-like statistical phase from either (i) the two-electron coherent tunneling (of the direct tunneling model), or (ii) the frozen of the charge field after the RG flow (of the Kondo model).
Our analysis in this Appendix also indicates a possible negative delta-$T$ noise for setups with semi-infinite LLs or chiral channels with non-local interactions. However, in these cases a negative delta-$T$ noise does not necessarily accompanies boson-like tunnelings: indeed, it is an electron that tunnels through a weak link (vacuum) in this setup.
At the same time, the condition for the emergence of negative delta-$T$ noise is still expressed 
in terms of the 
scaling dimension of the leading tunneling operator,
$\Delta_\mathcal{T}<1/2$
(see also Appendix~\ref{sec:LLFormalism}).
Thus, this example clearly demonstrates that, in general, the negative sign of the delta-$T$ noise cannot serve a smoking gun of intrinsic anyonic nature of tunneling quasiparticles.

As a final comment, the results in this Appendix further indicate that negative excess noise for non-interacting electrons is possible without bias voltage (i.e., similarly to delta-$T$ noise) even when the transmission is energy independent. This can be achieved by choosing appropriate non-equilibrium distributions in the reservoirs, e.g., by choosing 
$f(\epsilon)=a_K(\epsilon)$. We emphasize that this type of a fully non-equilibrium setup that produces a negative excess noise involves non-interacting fermions and has nothing to do with nontrivial exchange statistics (intrinsic or emergent). Moreover, the negative excess noise in such a setup is not associated with a small scaling dimension $\Delta_\mathcal{T}$ of the tunneling particles. 
This suggests that nontrivial nonequilibrium states may be used to mimic interaction effects and effects of quantum statistics on the transport properties of nanoscale systems, including the current noise \cite{TikhonovSciRep16,TikhonovPRB2020}. In this paper, however, we focus on the tunneling between fully equilibrated states.

\subsection{Equilibrium Nyquist-Johnson noise: similarities and differences}
\label{app:NJ}

\begin{figure*}[t!]
  \centering
    \includegraphics[width=1.0 \textwidth]{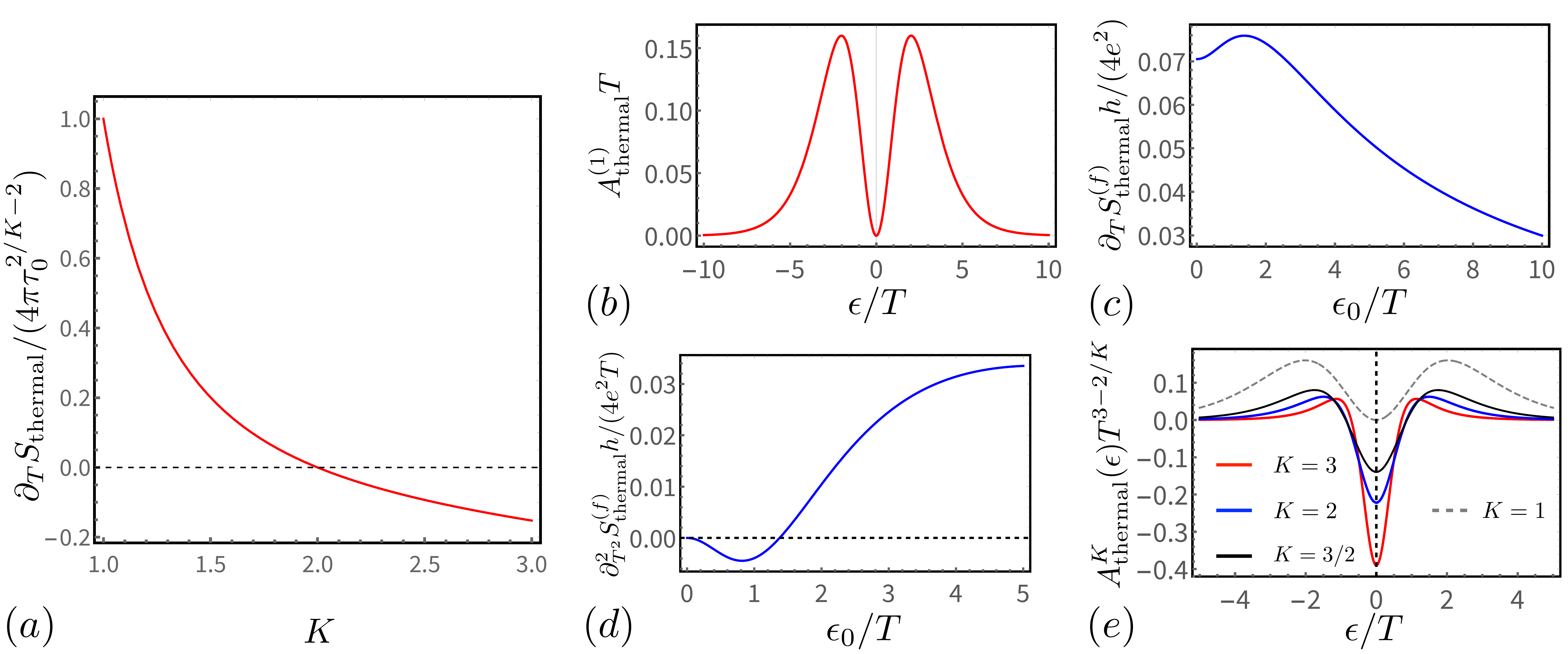}
    \caption{(a) The thermal noise $S_\text{thermal}$ as a function of the interaction parameter $K$. (b) The spectrum  $A_\text{thermal}^{(1)}$ of the Nyquist-Johnson noise [Eq.~\eqref{eq:AthFermions_Appendix}] for free fermions. The function is positive for all energies. (c) Since the spectral function is positive, $\partial_T S_\text{thermal}$ is positive for all $\epsilon_0$, referring to the missing of extreme-value point of $S_\text{thermal}$. (d) The second-order derivative $\partial_T^2 S_\text{thermal}$ can be negative, with the transition value of $\epsilon_0$ close to that of Fig.~\ref{fig:noise_spectrum}(d) for the delta-$T$ noise with an energy-dependent transmission probability. (e) The thermal noise spectrum $A_\text{thermal}^K$ of interacting cases.}
    \label{fig:sth}
\end{figure*}

Here, we briefly discuss equilibrium thermal noise, also called Nyquist-Johnson noise. It is defined generically in Eq.~\eqref{eq:sth} 
and is proportional to the product of the zero-bias dimensionless conductance $g_\mathcal{T}(T)$ given by Eq.~\eqref{eq:gtauGen} and the equilibrium temperature $T$. Equations~\eqref{eq:gtauGen} and \eqref{eq:sth} lead to the following temperature scaling of the thermal noise:
\begin{equation}
S_\text{thermal} \propto 
T^{2 \Delta_\mathcal{T} - 1}.
\label{SNJ-DeltaT}
\end{equation}
Thus, the equilibrium noise increases or decreases with $T$ for $\Delta_\mathcal{T} > 1/2$ respectively $\Delta_{\mathcal{T}}<1/2$.
This is exactly what is shown in Fig.~\ref{fig:sth}\textcolor{blue}{(a)}.
In this sense, the Nyquist-Johnson noise is also potentially capable of capturing the scaling dimension of tunneling operators in interacting systems (cf. Ref.~\cite{KyryloVadimPRB15}).

In more technical terms, the conductance $g_\mathcal{T}$ defined in Eq.~\eqref{eq:gt} contains a differentiation of correlation function of tunneling operators over the bias. The temperature-differentiation of the thermal noise then undertakes twice the derivatives over energies.
This is similar to the derivation of the delta-$T$ noise spectrum in Eq.~\eqref{eq:interacting_spectrum}, which corresponds to a second derivative of the same correlation function 
$\langle \mathcal{T}_{\vec{m}}^{}(t,0)\mathcal{T}_{\vec{m}}^\dagger(t,0) \rangle$ 
with respect to $\delta T$, i.e., also with respect to the variable of dimension energy. In the interacting problem, the characteristic energy scale is given by the temperature. Therefore, based on the simple scaling analysis, one anticipates a qualitative similarity between delta-$T$ noise and the $T$-derivative of the thermal noise. In particular, one observes that if these two functions change sign, they do that at the same value of the scaling dimension $\Delta_\mathcal{T}$ (equivalently, the same value of $K$).

In this Appendix, we explore this similarity. We will show that, although Nyquist-Johnson noise shows sensitivity to the same critical value $\Delta_\mathcal{T} = 1/2$ as the delta-$T$ noise, the thermal noise and the delta-$T$ noise convey different information about the system properties and, therefore, are not interchangeable for probing the system. As an example revealing this difference, we consider the thermal noise of free fermions, $K=1$, in the framework of the scattering formalism, see Sec.~\ref{sec:FreeElectrons} and Appendix~\ref{sec:energy_dependent_trans}. The equilibrium noise is extracted from Eq.~\eqref{eq:correlator_fermion} as
\begin{align}
    S_\text{thermal}^\text{(f)} &
    =  \frac{4e^2}{h} \int d\epsilon\ D(\epsilon) 
    f(\epsilon,T)[1-f(\epsilon,T)]
    \notag
    \\
    &
    =\frac{2e^2}{h} \int d\epsilon\ D(\epsilon)
   \frac{1}{1+\cosh(\epsilon/T)}.
   \label{eq:SthAppFermions}
\end{align}
Note that the term proportional to $D(1-D)$ in Eq.~\eqref{eq:correlator_fermion} does not contribute to the thermal noise at equilibrium, since $f_1=f_2$.
At the same time, as discussed in Sec.~\ref{sec:bunching}, it is this term, which contains the product $f_1f_2$, that is responsible for the effect of statistics (bunching/anti-bunching) on the sign of the delta-$T$ noise.

The temperature derivative of the thermal noise, Eq.~\eqref{eq:SthAppFermions}, is determined by
\begin{equation}
\begin{aligned}
\label{eq:AthFermions_Appendix}
    A_\text{thermal}^{(1)}
    & \equiv \frac{\partial}{\partial T} \frac{1}{1+\cosh(\epsilon/T)} = 
    \frac{\epsilon \sinh(\epsilon/T)}{T^2[1+\cosh(\epsilon/T)]^2}
\end{aligned}
\end{equation}
and is positive at all energies $\epsilon$ [see Fig.~\ref{fig:sth}\textcolor{blue}{(b)}]. Consequently, the positive energy-dependent transmission $D(\epsilon)$ alone cannot lead to the decrease of the thermal noise for free fermions
with increasing temperature. This stands in contrast to the delta-$T$ noise as discussed in Sec.~\ref{sec:energy_dependent_trans}, where an energy-dependent transmission can induce a negative delta-$T$ noise even for free fermions.

To illustrate this consideration, we choose $D(\epsilon)$ as in Eq.~\eqref{eq:D_Gaussian} and compute the $T$-derivative of the Nyquist-Johnson noise, $\partial_T S^\text{(f)}_\text{thermal}$, for free fermions. We plot this quantity as a function of the Gaussian width $\epsilon_0$ in Fig.~\ref{fig:sth}\textcolor{blue}{(c)}. We see that this derivative remains positive for all $\epsilon_0$. 
This is in contrast with the result for the delta-$T$ noise [Fig.~\ref{fig:noise_spectrum}\textcolor{blue}{(d)}], where delta-$T$ noise changes sign for a given width of the transmission amplitude.
Interestingly, the second-order derivative of the thermal noise, $\partial_T^2 S^\text{(f)}_\text{thermal}$, can be negative [Fig.~\ref{fig:sth}\textcolor{blue}{(d)}], and changes sign at a value near the one where the delta-$T$ noise changes its sign in Fig.~\ref{fig:noise_spectrum}\textcolor{blue}{(d)}.
Indeed, $\partial_T^2 S^\text{(f)}_\text{thermal}$
for free fermions is determined by
\begin{equation}
\begin{aligned}
    A_\text{thermal}^{(2)}
    & \equiv \frac{\partial^2}{\partial T^2} \frac{1}{1+\cosh(\epsilon/T)} 
    \\
   & = \frac{\epsilon}{2 T^4 \cosh^4\frac{\epsilon}{2 T}}\left(\frac{\epsilon}{2T} \cosh\frac{\epsilon}{ T}- \sinh\frac{\epsilon}{T}-\frac{\epsilon}{T}\right).
   \label{eq:Ath2Fermions}
\end{aligned}
\end{equation}
which is very similar (but not identical) to Eq.~(\ref{eq:noise_spectrum}).

Similarly, we can also define the thermal noise spectrum
\begin{equation}
    A_\text{thermal}^K \equiv a_K(T,-\epsilon)\frac{\partial a_K(T,\epsilon)}{\partial T} + a_K(T,\epsilon) \frac{\partial a_K(T,-\epsilon)}{\partial T}
\end{equation}
for interacting fermions, 
following that of the delta-$T$ noise Eq.~\eqref{eq:interacting_spectrum}.
The plot [Fig.\,\ref{fig:sth}\textcolor{blue}{(e)}] shows that interaction induces a negative peak of the thermal noise spectrum near zero energy, leading to a negative differentiation of the thermal noise.
However, the thermal noise spectrum of Fig.\,\ref{fig:sth}\textcolor{blue}{(e)} shows clear difference in comparison to that of the delta-$T$ noise [Fig.\,\ref{fig:normalized_ratio}\textcolor{blue}{(d)}]. Indeed, the peak in the spectrum of the latter case is instead positive. A negative delta-$T$ noise of the interacting situation originates from the ``valleys'' near the peak.

Briefly, for interacting cases with an energy-independent transmission $D$, the temperature derivative of the thermal noise and the delta-$T$ noise change sign at the same interaction, corresponding to $\Delta_\mathcal{T}=1/2$. In the weak-link setup in LL wires, the origin of this sign change in both quantities can be traced back to the renormalization of the effective transmission probability, which becomes a function of energy and temperature governed by $\Delta_\mathcal{T}$.
However, despite this similarity, the two quantities bear different information of the relevant quasiparticles. In particular, the emergence of the negative delta-$T$ noise in the HOM-type setup can be traced back to a bunching tendency of quasiparticles, whereas the thermal noise is not sensitive to this effect.  Further, while the delta-$T$ noise can be negative in a non-interacting system with energy-dependent transmission, the thermal noise is monotonous in this case: the spectrum of the temperature derivative of the delta-$T$ noise is always positive in a system with free fermions.

\section{Examples of the bosonization formalism: 
Scaling dimensions and statistical phases}
\label{sec:bosonization_details}

In this Appendix, we outline two important examples of the formalism presented in Sec.~\ref{sec:Discussion}.

\subsection{Tunneling between FQH Laughlin edges}
Here, we consider tunneling between two Laughlin FQH edges at filling $\nu=(2n+1)^{-1}$ for $n$ positive integers. The full Hamiltonian is given in the general form~\eqref{eq:GenBosHam} as
\begin{align}
    H = \frac{v}{4\pi \nu}\int dx\left\{ [\partial_x\phi_1(x)]^2 + [\partial_x\phi_2(x)]^2\right\}.
\end{align}
We take charge vector as $\vec{t}=(1,1)$ and the bosons obey 
\begin{align}
\label{eq:comm_FQH}
\left[ \phi_i(x), \phi_j (x') \right] = i\pi\nu\chi_i \delta_{ij}\, \text{sgn}(x - x'),
\end{align}
with $\chi_1=-\chi_2=1$ and $\lambda_1=\lambda_2=\nu$.
Tunneling is modelled by adding the perturbation
\begin{align}
    H_\mathcal{T}=\int dx \left[\Gamma \delta(x)\mathcal{T}_{\vec{m}}(x)+\text{H.c.\,}\right]
\end{align}
to $H_0$.
Consider first the tunneling operator
\begin{align}
\label{eq:Tqp}
    \mathcal{T}_{\vec{m}}=\mathcal{T}^\text{(qp)} \propto e^{i\phi_1+i\phi_2},
\end{align}
so that $m_1=m_2=1$ according to Eq.~\eqref{eq:TunnelingOperators}. From Eqs.~\eqref{eq:charge_qj} and~\eqref{eq:TotalChargeQ} we see that the net charge vanishes
\begin{align}
    Q = -[1\times 1\times 1 \times \nu+1\times 1\times (-1) \times \nu] =0,
\end{align}
and the particles created and destroyed have charge \begin{equation}
    q=\nu.
\end{equation} 
The particle statistics of the operators $\sim e^{i\phi_{1,2}}$ is found from Eq.~\eqref{eq:theta_j} as
\begin{align}
    \Theta_1=\Theta_2 = \pi\nu.
\end{align}
The scaling dimension of the tunneling operator is obtained from Eq.~\eqref{eq:primary_field_scaling} as
\begin{align}
    \Delta_\mathcal{T} =\nu.
\end{align}
It is the sum of the individual quasiparticle operator scaling dimensions
\begin{align}
    \Delta_\text{qp} = \nu/2. 
\end{align}
We see that, since $\nu<1$, Eq.~\eqref{eq:c2} implies negative delta-$T$ noise for quasiparticle tunneling in all Laughlin states, when the tunneling occurs through the FQH bulk state, in accordance with Ref.~\cite{RechMartinPRL20}.

Next, we consider the alternative tunneling operator
\begin{align}
\label{eq:Te}
   \mathcal{T}_{\vec{m}}= \mathcal{T}^\text{(e)} = \propto e^{i\phi_1/\nu+i\phi_2/\nu},
\end{align}
for which $m_1=m_2=1/\nu$. For this operator, the particles that tunnel have charge $|q|=1$ and statistics angle
\begin{align}
    \Theta_1=\Theta_2=\pi/\nu = \pi.
\end{align}
The latter equality holds because $1/\nu$ is an odd integer and the statistics angle is defined $\text{mod\;}2\pi$. The tunneling operator therefore describes tunneling of electrons. 

The scaling dimension of the electron tunneling operator is
\begin{align}
    \Delta_\mathcal{T} = 1/\nu
\end{align}
and for the tunneling particles
\begin{align}
    \Delta_e = 1/(2\nu).
\end{align}
Since $\Delta_{\mathcal{T}}>1$, Eq.~\eqref{eq:c2} implies positive delta-$T$ noise for electron tunneling in all Laughlin states~\cite{RechMartinPRL20}.
This situation is realized when the tunneling between the FQH edges occurs through vacuum rather than through the FQH bulk state.

\subsection{Tunneling between spinless LL wires}
\label{sec:LLFormalism}

Here, we study tunneling between spinless LL wires. The Hamiltonian of one wire reads
\begin{align}
    H_0 &= \frac{v_0}{4\pi}\int dx\Big\{ [\partial_x\phi_L(x)]^2 + [\partial_x\phi_R(x)]^2\notag \\ &+2u\,\partial_x\phi_L(x) \partial_x\phi_R(x)\Big\}
\end{align}
and describes free $\phi_{R,L}$ bosons which propagate in opposite directions with velocity $v_0$. The bosons obey the commutation relation
\begin{align}
\label{eq:comm_LL}
\left[ \phi_{R,L}(x), \phi_{R,L} (x') \right] =\pm i\pi\, \text{sgn}(x - x').
\end{align}
The associated charge densities are given by
\begin{align}
    \rho_{R,L} = \frac{\partial_x\phi_{R,L}}{2\pi},
\end{align}
implying the charge vector $\vec{t}=(1,1)$ in the free basis. We also see that $u$ in $H_0$ parametrizes density-density between the $L$ and $R$ bosons. Density interactions between $L$-$L$ and $R$-$R$ modes can be incorporated into the velocity $v_0$.

We can put $H_0$ on the form~\eqref{eq:GenBosHam} by introducing chiral fields $\phi_\pm$ according to (see, e.g., Ref.~\cite{GiamarchiBook})
\begin{align}
\label{eq:BasisChange_K}
    \begin{pmatrix}
    \phi_R \\\phi_L\end{pmatrix} = \frac{1}{2K}\begin{pmatrix}
    K+1 & K-1 \\
    K-1 & K+1
    \end{pmatrix} 
     \begin{pmatrix}
    \phi_+ \\\phi_-\end{pmatrix}.
\end{align}
The Hamiltonian then reads
\begin{align}
\label{eq:H0LL}
   H_0 =\frac{v}{4\pi K}\int dx\left( [\partial_x\phi_+(x)]^2 + [\partial_x\phi_-(x)]^2\right).
\end{align}
with the commutation relations
\begin{align}
\label{eq:comm_LL2}
\left[ \phi_{\pm}(x), \phi_{\pm} (x') \right] =\pm i\pi K \text{sgn}(x - x').
\end{align}
Here, the renormalized velocity $v$ and the Luttinger parameter $K$ are given as
\begin{align}
    &v=\frac{2Kv_0}{1+K^2},\\
    &u=\frac{1-K^2}{1+K^2}.
\end{align}
The chiral charge densities are
\begin{align}
    \rho_{\pm} = \frac{\partial_x\phi_\pm}{2\pi}
\end{align}
so that the total charge is preserved during the basis transformation:
\begin{align}
    \rho_{L}+\rho_{R} = \rho_{+}+\rho_{-}.
\end{align}
The charge vector remains the same in the chiral basis: $\vec{\tilde{t}}=(1,1)$. 
As a side note, the chiral fields are related to the conventional canonical ($\theta$, $\varphi$) basis~\cite{GiamarchiBook} as
\begin{align}
    & \phi_\pm=K\theta\mp\varphi.
\end{align}

We next study vertex and tunneling operators. Consider first the vertex operator
\begin{align}
\label{eq:psipm}
    \psi_\pm \propto e^{i \phi_\pm}.
\end{align}
By using Eqs.~\eqref{eq:charge_qj},~\eqref{eq:Delta_jDef}, and~\eqref{eq:theta_j} we see that these operators create and destroy particles with charge $|q|=K$, statistics $\Theta=\pi K$, and scaling dimension $K$.
The operators 
\begin{align}
\label{eq:psipm_e}
    \psi_{\pm,e} \propto e^{i \phi_\pm/K}.
\end{align}
create and destroy particles with charge $|q|=1$
with fermionic statistics $$\Theta=\pi,$$ 
and the corresponding tunneling operator has scaling dimension 
\begin{equation}
   \Delta_\mathcal{T} = 1/K.
\end{equation}

Note that the properties of the operators in Eqs.~\eqref{eq:psipm}-\eqref{eq:psipm_e} are very similar to the vertex operators in Eq.~\eqref{eq:Tqp} and~\eqref{eq:Te} in the FQH example above, upon identification $\nu \leftrightarrow K$. An important difference is that $K$ is not fixed by topology as $\nu$, but depends instead continuously on the interaction strength $u$ via $K$.

The operators $\psi_\pm$ and $\psi_{\pm,e}$ are building blocks for constructing tunneling operators describing scattering off weak and strong impurities in a single LL wire~\cite{KaneFisherPRB92}. Namely, by adding to $H_0$ one of the tunneling Hamiltonians 
\begin{align}
    &H_\text{sbs} = \int \delta(x) \Gamma e^{i(\phi_++\phi_-)} + \text{H.c.\,}\\
    & H_\text{wbs} = \int \delta(x) \Gamma e^{i(\phi_+/K+\phi_-/K)} + \text{H.c.\,},
\end{align}
one can describe either electron tunneling between two semi-infinite LL wires [strong backscattering, see Fig.~\ref{fig:ll_structure}\textcolor{blue}{(b)}], or tunneling of quasiparticles (weak backscattering) between left and right moving chiral fields. 

\textit{We emphasize that our statement on delta-$T$ noise and statistics does not hold for this setup.} Our statement holds for infinite chiral systems with local interactions. Mapping the semi-infinite wires to this configuration introduces non-local and non-translationally-invariant effective interactions in unfolded chiral systems, and statistical angles become ambiguous. At the same time, for tunneling through vacuum in the weak-link setup, the statistical nature of the tunneling particles is clearly fermionic (cf. Appendix \ref{sec:interacting_delta_T}).

Let us now consider the operators $e^{\pm i \phi_{R,L}}$. The properties of the associated excitations are found by first expressing the $\phi_{L,R}$ in the diagonal basis with Eq.~\eqref{eq:BasisChange_K}. Then, application of Eqs.~\eqref{eq:charge_qj},~\eqref{eq:Delta_jDef}, and~\eqref{eq:theta_j} shows that the corresponding particle charges are
\begin{align}
    &q_R=(K+1)/2-(K-1)/2=1\\ &q_L=(K-1)/2-(K+1)/2=-1,
\end{align}
and their statistics angles are
\begin{align}
    \Theta_{L,R} = \pi\frac{K^2+2K+1-(1+K^2-2K)}{4K} = \pi.
\end{align}
The operators $e^{\pm i\phi_{L,R}}$ are the operators consistent with boundary conditions for electron tunneling \textit{between two infinite wires} [see Fig.~\ref{fig:ll_structure}\textcolor{blue}{(a)}]
through vacuum:
these operators describe electron excitations. 

The scaling dimensions for these vertex operators are obtained as
\begin{align}
    2\Delta_{L,R}=\frac{K^2+2K+1+1+K^2-2K}{4K}=\frac{K+1/K}{2}.
\end{align}
By introducing two independent wires on the form~\eqref{eq:H0LL}, with wire indices, $H_a$ and $H_b$, electron tunneling between them is described by adding the point tunneling operator
\begin{align}
    H_{e} = \sum_{j,j'=L,R} \int \delta(x) \Gamma e^{i(\phi_{j,a}+\phi_{j',b})} + \text{H.c.\,}\ .
\end{align}
We then find from Eq.~\eqref{eq:primary_field_scaling} the scaling dimension of each tunneling process
\begin{align}
    \Delta_{\mathcal{T}}=\Delta_L+\Delta_R=\frac{K+K^{-1}}{2}.
\end{align} 
Inserting this into Eq.~\eqref{eq:c2} gives positive noise for all $K$, in accordance with Ref.~\cite{RechMartinPRL20}.

We end this example by re-iterating that the choice of tunneling operators depends crucially on the boundary conditions for the specific setup. While the formula~\eqref{eq:c2} holds for any tunneling process, one may not always connect the sign of the delta-$T$ noise to the statistics of the tunneling particles. 

\section{Anyonic braiding effects on Delta-T noise}
\label{sec:braiding_delta_t}

In this Appendix, we investigate the influence of anyonic braiding in delta-$T$ noise. To be clear, with anyonic braiding we mean here effects that arise from interference between different paths taken by tunneled anyons. We find that the braiding-induced contributions to the delta-$T$ noise vanishes to order $\delta T^2$.

\begin{figure}[t!]
  \centering
    \includegraphics[width=1 \columnwidth]{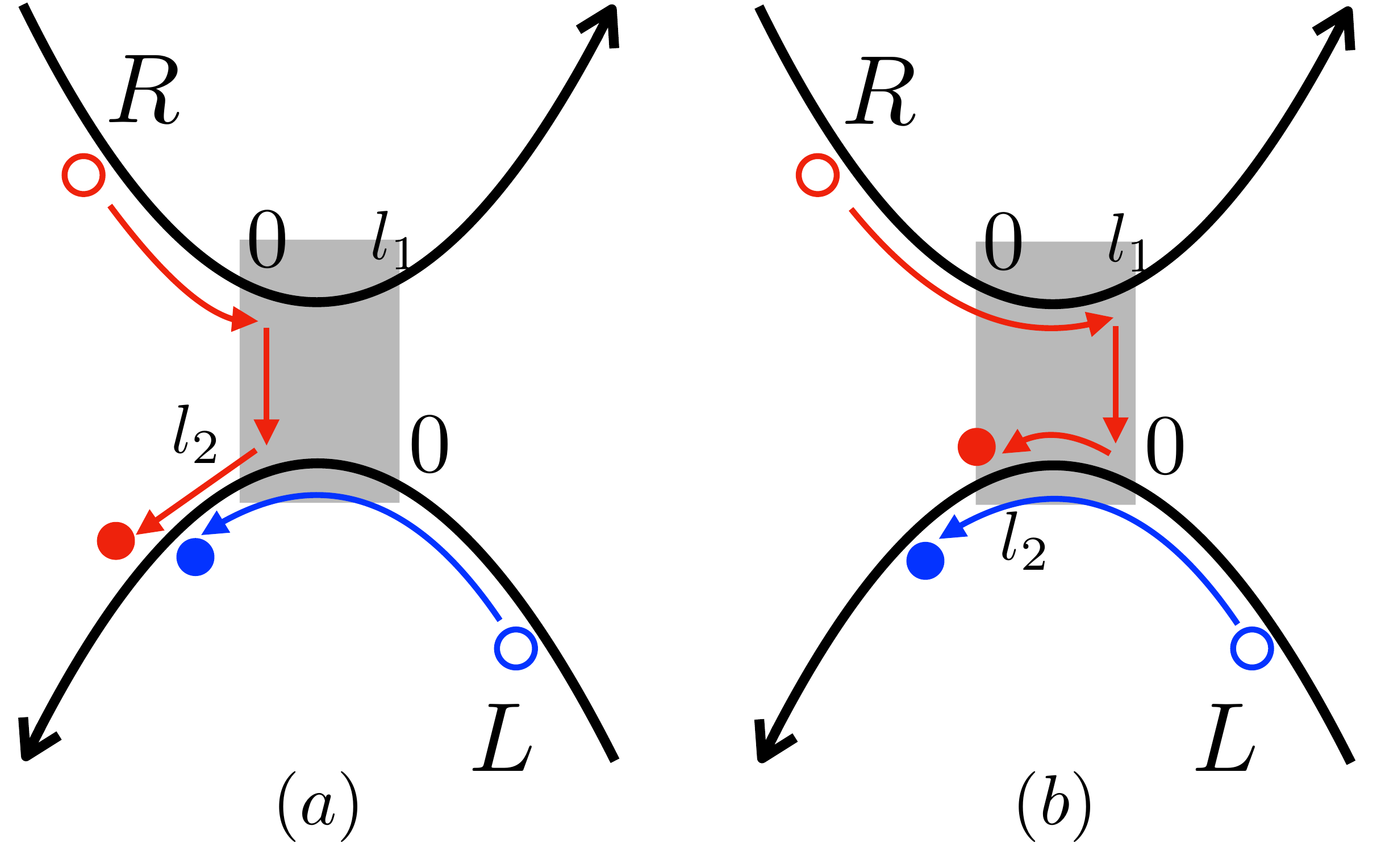}
    \caption{Anyon tunneling between Laughlin edges in a finite size region. The red and blue empty (filled) circles indicate initial (final) positions of two anyons. In panel (a), the top anyon enters the bottom edge from the left side of the scattering area, and propagates in front of the lower anyon. In panel (b), the top anyon instead enters the bottom edge from the right side. Here, the top anyon travels behind the bottom one.}
    \label{fig:finite_scattering_area}
\end{figure}

Consider the setup in Fig.~\ref{fig:finite_scattering_area} where two FQH Laughlin edges (represented by the black arrows) are weakly coupled in a finite size scattering region (gray area). For simplicity, we consider here the case where anyon tunneling occurs at the two boundaries of this region. The scattering segment lengths of the top and bottom edges have lengths $l_1$ and $l_2$, respectively. Due to the spatial distance between the two tunneling points, braiding effects from interference of different tunneling paths are expected. Considering an anyon propagating along the top edge (depicted by red arrows), it travels in front of the bottom anyon (indicated by the blue arrow) by tunneling at the first possible point [Fig.~\ref{fig:finite_scattering_area}\textcolor{blue}{(a)}]. By tunneling instead at the last possible point, it propagates behind the lower anyon [Fig.~\ref{fig:finite_scattering_area}\textcolor{blue}{(b)}]. Taking these two possibilities into account could give rise to signatures in the delta-$T$ noise, which we therefore now compute.

The tunneling Hamiltonian density for the setup in Fig.~\ref{fig:finite_scattering_area} reads
\begin{equation}
    H_\mathcal{T} = \mathcal{T} + \mathcal{T}^{\dagger},
\end{equation}
where
\begin{equation}
\begin{aligned}
\mathcal{T} &= \gamma_1 F_1 e^{i [\phi_R(0,t) - \phi_L(l_1,t)]}\\
&+ \gamma_2 F_2 e^{i [\phi_R(l_2,t) - \phi_L(0,t)]},
\end{aligned}
\label{eq:anyonic_tunneling}
\end{equation}
is the tunneling operator, with the amplitudes $\gamma_1$, $\gamma_2$, and the corresponding Klein factors $F_1$ and $F_2$.
The bosons obey the commutation relations
\begin{align}
    \left[ \phi_{\alpha}(x), \phi (x')_{\alpha'} \right] = i\nu\pi \delta_{\alpha,\alpha'}\text{sgn}(x - x'),
\end{align}
as usual for Laughlin FQH edges at filling factor $\nu$.

In Eq.~\eqref{eq:anyonic_tunneling}, Klein factors obey the state-specific commutation relations \cite{KanePRL03,LawFeldmanGefenPRB06}
\begin{equation}
\begin{aligned}
    &F_1 F_2 = \exp(-2i\pi\nu) F_2 F_1\\
    &F_1 F_2^{\dagger} = \exp(+2i\pi\nu) F^{\dagger}_2 F_1.
\end{aligned}
\end{equation}
These Klein factors are needed to guarantee correct commutation relations, enforced by a locality requirement between the tunneling operators at the edges of the scattering area~  \cite{GuyonPRB02,KanePRL03,LawFeldmanGefenPRB06}.
In contrast to the Luttinger liquid case where Klein factors are associated with fermionic operators, $F_1$ and $F_2$ are here defined with respect to the position of the quantum point contacts~\cite{LawFeldmanGefenPRB06,CampagnanoPRL12,LeeHanSimPRL19}.

For small tunneling amplitudes $\gamma_1, \gamma_2 \ll 1$, the noise can be calculated perturbatively to lowest order as~\cite{HalperinPRL16}
\begin{equation}
  S =  (\nu e)^2 \int_{-\infty}^{\infty} dt \langle \left\{ \mathcal{T}^{\dagger}(0), \mathcal{T}(t) \right\} \rangle.
  \label{eq:noise_anyon}
\end{equation}
With zero bias, the integrand in Eq.~\eqref{eq:noise_anyon} contains two types of contributions. The first type comes from correlations at the same point, e.g., 
\begin{equation}
\begin{aligned}
     2(\gamma_1^2  + \gamma_2^2 ) G_T (0,-t) G_B (0,-t),
\end{aligned}
\label{eq:same_point}
\end{equation}
where $G_{T,B}(x,t)\sim (x-t)^{-\nu}$ are Green's functions for anyons on the top (T) and bottom (B) edges. Eq.~\eqref{eq:same_point} coincides with the results in  Ref.~\cite{RechMartinPRL20}, after substituting the transmission probability of a single point tunneling with $\gamma_1^2 + \gamma_2^2$. Clearly this result does not contain any contribution from anyonic braiding, although information about the anyonic statistics is included in the Green's function exponents.

The second type of terms contributing to the noise involves correlations between different positions
\begin{equation}
\begin{aligned}
    &\gamma_1 \gamma_2 F_1 F_2^{\dagger} G_T (-l_2,-t) G_B (l_1,-t) \\
    &+ \gamma_1 \gamma_2 F_2 F_1^{\dagger} G_T (l_2,-t) G_B (-l_1,-t) \\
   &+\gamma_1 \gamma_2  F_2^{\dagger} F_1 G_T (l_2,t) G_B (-l_1,t) \\
    &+ \gamma_1 \gamma_2 F_1^{\dagger} F_2 G_T (-l_2, t) G_B (l_1,t).
\end{aligned}
\label{eq:braiding_contribution}
\end{equation}
These terms can be calculated by averaging (tracing) over Klein factors. However, following Ref.~\cite{LawFeldmanGefenPRB06}, one has for Laughlin states that
\begin{align}
    \text{Tr}(F_1 F_2^{\dagger}) = \text{Tr}(F_2 F_1^{\dagger}) = \sum_{n=1}^{1/\nu} \exp(2i n\pi  \nu) = 0,
\end{align}
and the leading contribution from braiding, Eq.~\eqref{eq:braiding_contribution}, therefore vanishes. This result has the same origin as the absence of magnetic flux dependence in the second-order contribution of the current in a Mach-Zehnder interferometer \cite{LawFeldmanGefenPRB06}. We therefore conclude that anyon braiding contributions to the delta-$T$ noise may only occur at higher orders than lowest in the tunneling probabilities. Hence, braiding effects are  negligible in comparison to effects from the statistics included in the tunneling operator scaling dimensions. This is consistent with the predictions of Ref.~\cite{LeeHanSimPRL19}, where anyon braiding was shown to induce a non-trivial contribution to the noise to the order $\propto \gamma_1^2\gamma^2_2$.

\section{Coulomb gas RG formalism}
\label{sec:AppendixCoulomb}
In this Section, we briefly describe the Coulomb gas RG technique (discussed in detail in Refs.~\cite{AndersonPRB1970,KaneFisherPRB92}), and sketch the derivation of the flow equations~\eqref{eq:complete_RG}.

Similarly to other momentum-based RG schemes, the purpose of the Coulomb gas RG is to integrate out the physics at large momenta (or equivalently short length scales or high energies), and instead incorporate their influence into a renormalization of couplings at small momenta (long length scales or low energies). As the starting point, the partition function $\mathcal{Z}\equiv\text{Tr}\exp(-\beta H)$ is expanded as
\begin{equation}
\begin{aligned}
\mathcal{Z} &= \sum_n\int_0^{\beta} d\tau_1 \cdots \int_0^{\tau_{i+2}-\tau_c} d\tau_{i+1} \int_0^{\tau_{i+1}-\tau_c} d\tau_{i} \cdots \\
& \Big\langle H_{\text{Kondo}}(\tau_{1}) \cdots H_{\text{Kondo}}(\tau_{i}) H_{\text{Kondo}}(\tau_{i+1}) \cdots \Big\rangle_0
\end{aligned}
\label{eq:partition_function}
\end{equation}
and is demanded to be invariant after each step of the renormalization procedure. In Eq.~\eqref{eq:partition_function}, the subscript ``0'' indicates that  expectation values are evaluated with respect to a quadratic, scale invariant, Hamiltonian [in our case, $H_1 + H_2$ in Eq.~\eqref{eq:disconnected_leads}]. In the language of the Coulomb gas RG, the perturbation Hamiltonian $H_{\text{Kondo}}$ changes the system state (at time steps known as ``kinks'') along an imaginary ``time axis'' periodic in the inverse temperature $\beta$. The crucial assumption is that Eq.~\eqref{eq:partition_function} suppresses the occurrence of two kinks within a short-time (high energy) cutoff $\tau_c$.

The RG procedure consists of three steps: (i) a scaling of the cutoff $\tau_c \to \tau_c + \delta\tau_c$; (ii) integrating out neighbouring kinks within a time interval smaller than the new cutoff $\tau_c + \delta\tau_c$; and (iii) a rescaling of the cutoff back to $\tau_c$. After each step, the coupling constants in $H_{\rm Kondo}$ are adjusted as to keep~\eqref{eq:partition_function} invariant. For an infinitesimal cutoff modification in each step $\delta\tau_c \ll \tau_c$, the infinitesimal changes in coupling constants can be written as continuous equations, known as the RG equations.

As the lowest-order (called tree level) contribution to the RG equations, the re-scaling itself introduces a flow of the coupling, which equals the scaling dimensions of the corresponding operator~\cite{Francesco2012}. In our case, re-scaling of $H_1+H_2+H_{\rm Kondo}$ generates the following tree level RG equations
\begin{equation}
\begin{aligned}
\frac{d\tilde{J}_2^z}{d\ell}& = \left( 1 - \frac{1}{2K} - \frac{K}{2} \right) \tilde{J}_2^z, \\
 \frac{d\tilde{J}_2}{d\ell} & = \left[ 1 - \frac{1}{2K} - \left( 1 - \frac{2J_1^z}{\pi u} \right)^2 \frac{K}{2} \right] \tilde{J}_2, \\
\frac{d\tilde{J}_1}{d\ell} & = \left[ 1 - \frac{K}{2} - \left( 1 - \frac{2J_1^z}{\pi u} \right)^2 \frac{K}{2} \right] \tilde{J}_1, \\
\frac{d\tilde{t}}{d\ell} & = \left( 1 - \frac{1}{2K} - \frac{K}{2} \right) \tilde{t}. 
\end{aligned}
\label{eq:rg_rescaling}
\end{equation}
When all coupling constants are much smaller than unity, the tree level \eqref{eq:rg_rescaling} determines the relevance of operators.

\begin{figure}[t!]
  \centering
    \includegraphics[width=1 \columnwidth]{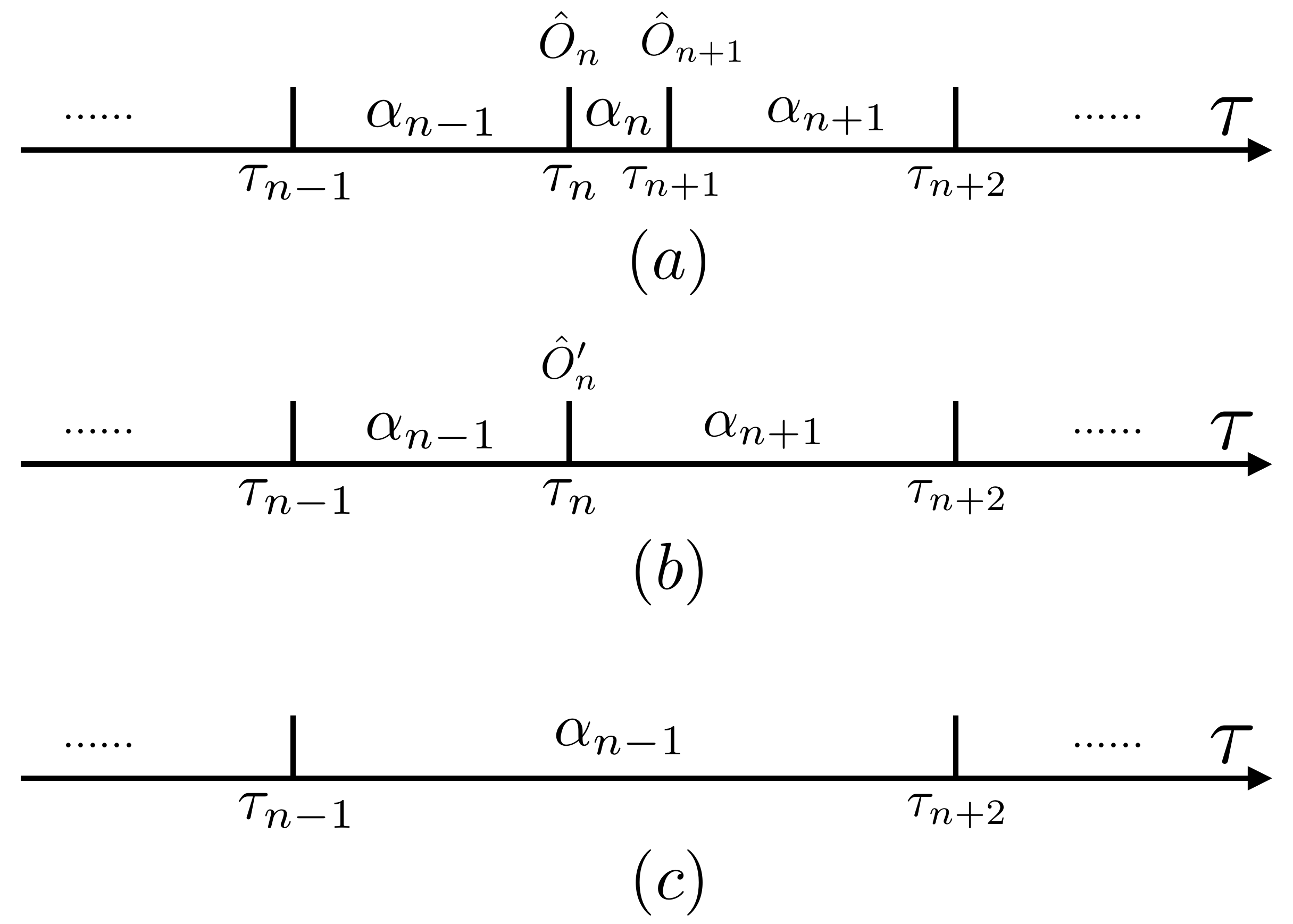}
    \caption{(a) History of the system state $\alpha$ of discrete domains separated by $\tau_n$ (``kinks'') in imaginary time $\tau$. The system is the state $\alpha_n$ for $\tau_n<\tau<\tau_{n+1}$. The discretization is such that $\tau_{n+1} - \tau_n<\tau_c$, implying that this kink pair is to be removed or modified in the current RG step. (b) The non-neutral pair situation, where $\alpha_{n-1} \neq \alpha_{n+1}$. Here the two kinks merge together, adding to the RG flow of the operator $\hat{O}_n' = \hat{O}_n \hat{O}_{n+1}$. (c) The neutral pair situation where $\alpha_{n-1} = \alpha_{n+1}$. In this case, both kinks are to be removed in the current RG step.}
    \label{fig:second_order_rg}
\end{figure}

Next, we consider next order contributions to the RG equations. We take the system histories in Fig.~\ref{fig:second_order_rg} as an example, which describes the system state $\alpha$, changing along the imaginary time axis $\tau\le \beta = 1/k_B T$. The system state changes from $\alpha_{n-1}$ to $\alpha_n$ at $\tau = \tau_n$, under influence of an operator $\hat{O}_n$. Assuming $\tau_{n+1} - \tau_n <\tau_c$, these two kinks are to be dealt with within the present RG step. Generically, we have to distinguish two different situations. As depicted in Fig.~\ref{fig:second_order_rg}(b), $\alpha_{n-1} \neq \alpha_{n+1}$, which is known as the non-neutral pair situation. In this case, we combine kinks at $\tau_n$ and $\tau_{n+1}$ into a single kink, with the effective operator $\hat{O}_n' = \hat{O}_n \hat{O}_{n+1}$. The amplitude of $\hat{O}_n'$ is enhanced by the product of two original operators after the this RG step. Considering the Hamiltonian Eq.~\eqref{eq:h_kondo}, the non-neutral pairs contribute to the flow of parameters as
\begin{equation}
\begin{aligned}
& d \tilde{J}_2^z \to 4 \tilde{J}_1 \tilde{J}_2 ,\ \  d \tilde{J}_2 \to 4 \tilde{J}_1 \tilde{J}_2^z \\
& d\tilde{J}_1 \to 4 \tilde{J}_2 \tilde{J}_2^z,\ \  d \tilde{t} \to 0.
\end{aligned}
\end{equation}
The non-neutral contribution to $\tilde{t}$ is neglected since, strictly speaking, this correction occurs at higher order than the second. 

The second situation, known as the neutral pair situation, is depicted in Fig.~\ref{fig:second_order_rg}(c). In contrast to the non-neutral pair situation, here $\alpha_{n-1} = \alpha_{n+1}$ (or equivalently, $\hat{O}_{n+1} = \hat{O}^{\dagger}_n$). In this case, we neglect the kinks at $\tau= \tau_n$ and $\tau = \tau_{n+1}$. The contribution from this kink pair, $\sim |\hat{O}_n|^2 (\tau_{n+1} - \tau_n)$ is included into the flow of the field dynamics. More specifically, in our case, the neutral pairs introduce the RG contribution
\begin{equation}
d  \left( 1 - \frac{2J_1^z}{\pi u} \right) \sim -2  \left( 1 - \frac{2J_1^z}{\pi u} \right) (2 \tilde{J}_1^2 + 2 \tilde{J}_2^2).
\label{eq:rg_j1z}
\end{equation}
Apparently, $1 - 2J_1^z/(\pi u)$ decreases during the RG flow. Taking into account all terms above, we arrive at the complete set of RG equations,
Eq.~\eqref{eq:complete_RG} in the main text.

\bibliographystyle{apsrev4-1-titles}

\begin{thebibliography}{133}
\bibitem [{Fie()}]{FieldRotation}%
  \BibitemOpen
  \href@noop {} {}\bibinfo {note} {Without the reflection in
  Eq.~\eqref{eq:chiral_fields}, commutators between left-going fields acquire
  an extra minus sign (from the chirality). The rotated fields of
  Eq.~\eqref{eq:rotations_to_new_field} are then not independent.}\BibitemShut
  {Stop}%
\bibitem [{BKT()}]{BKT}%
  \BibitemOpen
  \href@noop {} {}\bibinfo {note} {As predicted in Ref.~\onlinecite{DongPRB14},
  Kondo couplings can decrease to zero before becoming relevant, if their bare
  (non-renormalized) values are sufficiently small. We do not address this
  possibility here, and assume that the bare couplings are strong
  enough.}\BibitemShut {Stop}%
\end{thebibliography}

\end{document}